\documentclass[preprint]{aastex}
\begin{document}


\title{HOT POPULATIONS IN M87 GLOBULAR CLUSTERS}

\author{Sangmo T. Sohn\altaffilmark{1,2},  Robert W. O'Connell\altaffilmark{1}, Arunav Kundu\altaffilmark{3}, 
        Wayne B. Landsman\altaffilmark{4},\\
         David Burstein\altaffilmark{5}, Ralph Bohlin\altaffilmark{6}, 
	Jay A. Frogel\altaffilmark{7,8}, and James A. Rose\altaffilmark{9}}
\email{tonysohn@kao.re.kr, rwo@virginia.edu}

\altaffiltext{1}{Department of Astronomy, University of Virginia, P.O. Box 3818, Charlottesville, VA 22903-0818}
\altaffiltext{2}{Korea Astronomy and Space Science Institute, 61-1, Hwaam-dong, Yuseong-gu, Daejeon 305-348, Korea}
\altaffiltext{3}{Department of Physics and Astronomy, Michigan State University, East Lansing, MI 48824}
\altaffiltext{4}{NASA Goddard Space Flight Center, Code 681, Greenbelt, MD 20771}
\altaffiltext{5}{Department of Physics and Astronomy, Box 871504, Arizona State University, Tempe, AZ 85287}
\altaffiltext{6}{Space Telescope Science Institute, 3700 San Martin Drive, Baltimore, MD 21218}
\altaffiltext{7}{Association of Universities for Research in Astronomy, Inc, 1200 New York Avenue, NW, Washington, DC 20005}
\altaffiltext{8}{Visiting Investigator, Department of Terrestrial Magnetism, Carnegie Institution of Washington}
\altaffiltext{9}{Department of Physics and Astronomy, CB 3255, University of North Carolina, Chapel Hill, NC 27599}

\begin{abstract}

To explore the production of UV-bright stars in old, metal rich
populations like those in elliptical galaxies, we have obtained
HST/STIS far- and near-UV photometry of globular clusters in four
fields in the gE galaxy M87.  To a limit of $m_{FUV} \sim 25$ we
detect a total of 66 globular clusters (GCs) in common with the deep
HST optical-band study of Kundu et al.\ (1999).  Despite strong
overlap in V- and I-band properties, the M87 GCs have UV/optical
properties that are distinct from clusters in the Milky Way and in
M31.  M87 clusters, especially metal-poor ones, produce larger hot HB
populations than do Milky Way analogues.  In color plots including the
NUV band, the M87 clusters appear to represent an extension of the
Milky Way sequence.  Cluster mass is probably not a factor in these
distinctions.

The most metal-rich M87 GCs in our sample are near solar metallicity
and overlap the local E galaxy sample in estimated Mg$_2$ line
indices.  Nonetheless, the clusters produce much more UV light at a
given Mg$_2$, being up to 1 mag bluer than any gE galaxy in $(FUV-V)$
color.   The M87 GCs do not appear to represent a transition between
Milky Way-type clusters and E galaxies.  The differences are in the
correct sense if the clusters are significantly older than the E
galaxies.  

Comparisons with Galactic open clusters indicate that the hot
stars lie on the extreme horizontal branch, rather than being blue
stragglers, and that the EHB becomes well populated for ages $\ga 5$
Gyr.  Existing model grids for clusters do not match the
observations well, due to poorly understood giant branch mass loss or
perhaps high helium abundances.  

We find that 43 of our UV detections have no optical-band
counterparts.  Most appear to be UV-bright background galaxies, seen
through M87.  Eleven NUV variable sources detected at only one epoch
in the central field are probably classical novae.  Two recurrent
variable sources have no obvious explanation but could be related to
activity in the relativistic jet.

\end{abstract}

\section{Introduction}

\subsection{Hot Stars in Elliptical Galaxies}

Ultraviolet observations from space have shown that old stellar
populations are prolific producers of hot stars.  Elliptical galaxies
and large spiral bulges exhibit an unexpected ``ultraviolet-upturn''
or ``excess'' (UVX), a sharp rise in flux at wavelengths shorter than
2000 \AA, which is produced by stars with temperatures of $\sim 22000$
K (Code 1969, Burstein et al.\ 1988, Greggio \& Renzini\ 1990, Brown
et al.\ 1997, O'Connell\ 1999, and references therein).

The UV upturn is the most variable photometric feature of old
populations:  UV to optical flux ratios vary by a factor of 100, a
much larger range than for optical-IR colors (Faber 1983; Burstein et
al.\ 1988; Dorman, O'Connell, \& Rood\ 1995, hereafter DOR95).  The
first large survey of the UVX in nearby galaxies, based on
observations with the {\it International Ultraviolet Explorer} (IUE),
appeared to show a good correlation between its strength and metal
abundance as measured by the Mg$_2$ index at 5170 \AA\ .  A more
extensive recent survey from the {\it Galaxy Evolution Explorer}
(GALEX, Martin et al.\ 2003) by Rich et al.\ (2005) confirms the
enormous range in UV:optical flux ratios but finds no correlation
between UV properties and metallicity.

This remarkable variation in UV light has led to the expectation that
the UVX could be a uniquely sensitive probe of the star
formation and chemical enrichment histories of elliptical galaxies.
If it can be calibrated, the UV upturn would be an important test, for
instance, of the formation times of distant elliptical galaxies, since
preliminary models \citep{Greggio90,Bressan96,Yi99} imply that it will
only appear in systems 5 Gyr or older.  The {\it Hubble Space
Telescope} (HST) and GALEX have already detected a UVX at
redshifts of $0.3< z < 0.6$ \citep{Brown98,Brown00a,Brown03} and $z
\sim 0.1$ \citep{Lee05a}, respectively.

We believe we know the {\it kinds} of stars responsible for the
UVX (reviewed in O'Connell 1999).  UV imaging and spectroscopy 
from the {\it Astro} missions provided good evidence that old, low-mass 
stars on the ``extreme'' horizontal branch (EHB), with $T_{\rm eff} > 
15000$ K, and their hot descendents produce the upturns.  
We also understand the basic physics of the hot evolutionary phases.  
The EHB stars have very small envelope masses ($\la 0.05 {\rm
M}_{\odot}$), and smaller envelopes produce hotter objects.
Differences of as little as 0.01 M$_{\odot}$ can have large effects.
Hence, the net UV output of an EHB population is governed mainly by
envelope mass loss processes on the red giant branch and by the helium
abundance.  In turn, mass loss efficiency is strongly linked to age
(older stars have smaller envelopes to lose) and metal abundance (mass
loss is likely enhanced at higher metallicites by radiation pressure
on grains and molecules).  Less than 20\% of
the evolving population in the galaxies need to pass through the hot
phase to match the observations. 

What we do not yet understand is how the parent population's global
characteristics (age, helium abundance, metal abundances, dynamics)
determine the distribution of EHB temperatures and hence the strength
and shape of the upturns.  Observations of globular clusters (GCs)
undoubtedly offer the best means of clarifying these global effects.
GCs are the best understood stellar systems from the standpoint of
evolution, and they are (mostly) simple stellar populations with a
small internal dispersion in age and abundance.

Unfortunately, the available UV data on GCs are limited and hard to
interpret, and the Galactic GC sample does not overlap the galaxies in
metallicity as measured by Mg$_2$ (summarized by DOR95).  Some
globulars have total UV-star fractions (as indicated by their $FUV-V$
colors) that rival those of any elliptical galaxy.  However, the UV
fractions are largest for moderately metal-poor clusters (with [Fe/H]
$\sim -1.5$), such as $\omega$ Cen \citep{Whitney94} and NGC 6752
\citep{Landsman96}.  There is large scatter at all metallicities and
no strong trend with metallicity (although the UV sample is far from
complete).  Meanwhile, the recent GALEX observations by Rey et al.
(2005) show that the far-UV properties of M31 GCs are similar to those
of MWGCs.  By contrast, the UV fractions among nearby galaxies in the
IUE sample appear to increase strongly with metallicity.  Such
differences in behavior could be linked to differences in age or
helium abundance or to other factors influencing the EHB, such as the
dynamical environment (e.g.\ Paresce et al.\ 1991, Fusi-Pecci et al.\
1993, Bailyn 1995, Sweigart 1997). 

There is a critical gap in metallicity between the most metal-rich of
the observed Milky Way globulars and the bulk of elliptical galaxies.
This corresponds to Mg$_2$ values in the range $\sim$ 0.1--0.3 (see
Figure 1 of DOR95 or Figure \ref{f:fuvv_mg2} below).  An obvious next
step is to fill in this gap with clusters of higher metallicity.  Such
objects are rare in our Galaxy and also difficult to study because
they tend to be preferentially situated in regions with large UV
extinction.  Where data have been obtained on metal-rich Galactic
globulars, there is a large dispersion: 47 Tuc has a definite but
small EHB population 
\citep{OConnell97} while NGC 6388 and 6441 have rich EHB populations
\citep{Rich97}.  A small EHB component is present in the high
metallicity open clusters NGC 188 and NGC 6791 \citep{Liebert94,
Landsman98}.  But overall the Milky Way is not a fertile terrain for
investigating the UV behavior of metal-rich clusters. 

To enlarge the sample of metal rich objects that can link globular
cluster populations to those in galaxies, we must instead turn to the
large cluster systems of giant elliptical galaxies.

\subsection{The M87 Globular Cluster System}

Giant elliptical galaxies possess enormous populations of GCs, many
of which have high metal abundances.  Hundreds of such globulars in
elliptical galaxies in the Virgo cluster have $V \sim 20-23$.  These
systems are essentially free of foreground extinction, with $E(B-V)
\lesssim 0.02$.  Among the Virgo galaxies, M87 has the largest and
best studied GC system.  Numerous, faint point sources around M87 were
first identified as globular clusters by \citet{Baum55}.  Early
photographic studies showed that the GCs in M87 are on average bluer
than the galaxy background \citep{Racine68} and that the mean color of the
clusters has a radial trend such that clusters become bluer as radial
distance increases \citep{Strom81}.  The radial dependence of mean color 
was later confirmed by the CCD observations of \citet{Lee93} and 
\citet{Cohen98}.

Kundu et al.\ (1999; hereafter K99) studied the GC system in
the inner region of M87 using images taken with the HST WFPC2 camera.
Their sample, over a thousand clusters, allowed them to investigate
the luminosity function, the optical color distribution, and structural
information.  An important result is the presence of bimodality
in the optical color distribution.  Bimodality of GC
systems in early-type galaxies was first (statistically) studied by
\citet{Zepf93}.  Recent studies of HST archival data (Kundu \& Whitmore
2001a, Larsen et al. 2001) have shown that bimodal color distributions
are present in at least 30\% and up to 60\% of early-type GC systems.
Several models have been proposed to explain the observed bimodal
cluster groups: major mergers of late-type galaxies \citep{Ashman92,Ashman98},
two bursts of in situ star formation \citep{Forbes97}, and hierarchical
formation \citep{Cote98}.

The goal of this study is to determine how the M87 metal-rich clusters
compare to the Galactic globulars and the elliptical galaxies in their
UV color properties.  Among the observational questions of particular
interest are these: (1) Does the M87 cluster system overlap in UV
properties with the Galactic system or is it distinct? (2) Does
the M87 cluster system overlap with the E galaxies? (3) Does the
M87 system represent a transition between Milky Way-type clusters and
the galaxies?  (4) Are secondary dependencies found in UV properties,
e.g.\ a dependence on galactocentric radius at a given metallicity? (5)
Do the UV properties of the two M87 cluster subsystems differ?

We have obtained near- and far-UV photometry of M87 globular clusters
using the imaging mode of the HST Space Telescope Imaging Spectrograph
(STIS) to answer these questions.  In the next section we describe our
observations and reductions and our far-UV detection rate.  Subsequent
sections discuss interpretation of the resulting photometry,
comparison to the globular clusters in the Milky Way and M31, the
absence of radial trends in our photometry, comparison to elliptical
galaxies, comparison to existing model grids for clusters, possible
selection effects, and the surprisingly large sample of ``UV-only''
sources we detect.

\section{Observations and Data Reduction}

\subsection{Observations}

\begin{figure}
\epsscale{1.0}
\plotone{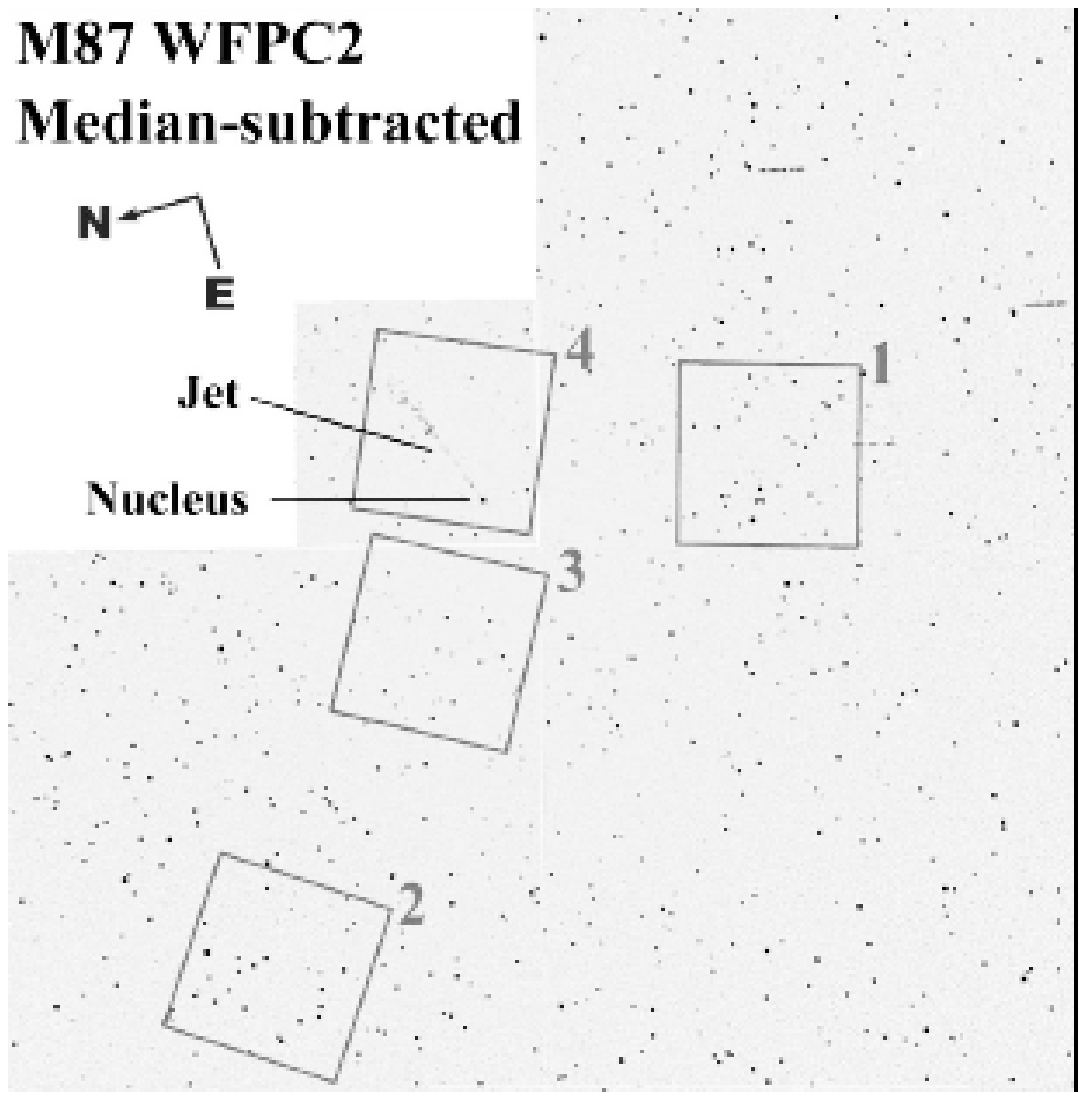}
\caption{ The four HST/STIS far-UV fields that were observed
	  superimposed on a WFPC2 image of the inner region of M87.
	  To remove the diffuse background, a median filtered image
	  was subtracted from the original WFPC2 image.
	  The entire field of view of the image is
	  160\arcsec$\times$160\arcsec.  Locations of the nucleus and
	  optical jet of M87 are indicated. Most of the point sources
	  shown in the figure are globular clusters. The centers of
	  Fields 1, 2, 3, and 4 are located at angular distances of
	  38, 73, 22, and 10 arcseconds, respectively, from the
	  nucleus of M87.  \label{f:fuvpointings} }

\end{figure}

The data used in this study consist of HST/STIS far- and near-UV images of
four different fields in the inner ($r < 1\farcm 5$) region of M87.
At our adopted distance modulus of $(m-M) = 31.12$ \citep{Whitmore95} for
M87, this region corresponds to $r < 7.3$ kpc.  Figure
\ref{f:fuvpointings} shows the location of the far- and near-UV fields
superposed on the median-subtracted WFPC2 image of K99.  The fields
are labeled from 1 to 4, and we use these designations throughout this
paper.

Fields 1, 2, and 3 were observed with the HST/STIS far-UV camera in
HST Program GO-8643 (PI: R. O'Connell).  This camera has a field of
view of 25\arcsec$\times$25\arcsec with 0.024\arcsec pixels. Total
exposure time was 10,480 s for each field.  The far-UV camera employs
a multianode micro-channel array (MAMA) which has no readout noise and
little sensitivity to cosmic rays.  Since the long wavelength cutoff is
determined by the rapidly decreasing detector sensitivity longward of
2000\AA , red leaks are negligible.  We used the F25SRF2 longpass
filter (1270 -- 2000 \AA, pivot wavelength 1460 \AA) to extend the
bandwidth of our images as far to the blue as possible while avoiding
the strong geocoronal Lyman-$\alpha$ emission at 1216 \AA .  The
FUV-MAMA/F25SRF2 combination admits the dayglow emission lines of
\ion{O}{1} (1302, 1356 \AA ), which create a relatively bright sky
background.  However, the \ion{O}{1} features are strongly variable in
both time and geographic latitude, and their strength cannot be
confidently predicted in advance.  Therefore, instead of restricting
our observing periods to orbital night by using the SHADOW TIME
special requirement, we observed in TIMETAG mode.  In this observing
mode, the data consist of a stream of time-tagged photon events and
X/Y coordinates. Periods of bright dayglow emission can be rejected
during the subsequent data reduction, thereby making best use of
available low dayglow periods.

We also obtained three optical band images centered on the same
positions as the far-UV frames of Fields 1, 2, and 3, using the STIS
CCD and F28X50LP (5500 -- 11000 \AA, pivot wavelength 7230 \AA) filter.
These images were obtained mainly to verify pointings of our targets.
Each CCD image has a field of view of 50\arcsec$\times$50\arcsec ~of
which only the central 28\arcsec$\times$50\arcsec ~is usable due to
the size of the filter.  All three far-UV fields fall on the usable
area of the CCD.  The exposure time for the CCD observations was 518
seconds for each field.  

Observations of Field 4 were taken from the STScI/HST
Archive\footnote{\url{http://archive.stsci.edu}}.  These included a total of
four far-UV and sixteen near-UV images.  The relevant observations are
those of HST Programs GO-8048, -8140, -8780, and -9461 (PI: J.
Biretta).  All downloaded images were observed in ACCUM mode (normal
integration mode) either using FUV-MAMA/F25SRF2 in the far-UV or
NUV-MAMA/F25QTZ (1480 -- 3500 \AA, pivot wavelength 2360 \AA) in the
near-UV.  Each observation has four near-UV exposures with integration
times of $\sim 600$ seconds.  Far-UV images were available only in the
GO-8140 observation, whose field coverage is shown in Figure
\ref{f:fuvpointings}.  Four FUV-MAMA exposures with integration times
ranging from 850 to 910 seconds were taken.  The orientations of
GO-8048, -8780, and -9461 observations slightly differ from one
another, but the field coverage is not much different from that of
Field 4 in Figure \ref{f:fuvpointings}.  We denote these three fields
respectively as Field 4a, 4b, and 4c when we need to distinguish
them.  All of these observations were taken as part of the ongoing
investigation of the variability of the M87 nonthermal jet (e.g.\
Perlman et al.\ 2003).  Field 4 is subject to the bright diffuse background
that increases rapidly toward the M87 nucleus.

A summary of all observational material is given in Table \ref{t:obslog}.

\subsection{Image Reductions}

The STIS CCD images for Fields 1, 2, and 3 were calibrated using the
standard IRAF.CALSTIS\footnote{IRAF(the Image Reduction and Analysis 
Facility) is distributed by the National Optical Astronomy Observatories.} 
pipeline.  Each CCD image went through the steps
of data quality initialization, overscanning, bias subtraction, cosmic
ray rejection, dark subtraction, flat-fielding, and geometric
correction.  The final calibrated CCD images still show a considerable
number of remaining hot pixels not removed from the dark subtraction
stage.  We used the IMGCLEAN procedure in IDL\footnote{The Interactive Data Language
is distributed by Research Systems Incorporated (\url{http://www.rsinc.com}). 
IMGCLEAN and other applications routines are available through the 
IDL Astronomy User's Library (\url{http://idlastro.gsfc.nasa.gov/homepage.html})
maintained by W. Landsman}.
to eliminate these hot
pixels.  IMGCLEAN was originally designed to remove cosmic ray hits
from WFPC images, but extensive tests showed that it works very
well on our STIS CCD images.  The CCD images are only used for
astrometry of sources in this study, so any photometric bias
introduced by the IMGCLEAN routine has no impact.

Because the far-UV images of Field 1, 2, and 3 were obtained in the
TIMETAG mode, our first step was to convert the TIMETAG tables into
2-D images.  For each data set we plotted the total counts in each
five second bin as a function of time.  Since the total count is
dominated by the sky background, higher total count rates indicate
higher background counts.  To reject periods of bright skyglow, we can
simply reject periods of high total count rates.  We note that
exposures {\it o6be02ceq} of Field 1 and {\it o6be04z4q} of Field 2
were made entirely in dark background periods.  The time periods are
input to the INTTAG task in the IRAF/STSDAS package to produce 2-D
images.  In practice, it was straightforward to identify intervals
with serious skyglow.  Once the conversion was done for all of our
exposures, we proceeded with the standard STIS imaging calibration
procedures as described below.

The dark current in the STIS far-UV camera is known to have a low but
spatially and temporarily variable enhancement in the upper left
quadrant of the image, which is called the {\it glow}.  It is known
that the glow correlates well with the detector temperature.  While
the temperature of the detector itself is hard to measure, the
charge-amplifier temperature (CAT), which is recorded in the image
headers, is a reliable proxy for the detector temperature.  The glow is
negligible for CAT lower than 35 K but above that temperature the
global count rates for dark frames increase with increasing CAT.  Most
of our far-UV frames have CAT higher than 35 K so we performed
subtraction of the glow pattern in the following manner.  First, a
constant dark component was subtracted by the CALSTIS routine.  Next,
to deal with the glow component on the upper left-hand quadrant of
some images, we used the ``glow only'' image as a template which is
available online.\footnote{\url{http://www.stsci.edu/hst/stis/performance/background}}
While the glow profiles of our far-UV frames are well represented by
this image, the intensity of the glow varies image by image.  We have
generated projected mean profiles along the X and Y axes of those
images and used them as references.  Then we scaled and subtracted the
glow profile from the images affected by the glow and generated
profiles.  This step was iterated until the subtracted profile closely
matched the reference profile.  Each glow-subtracted image was
visually inspected for any anomaly caused by this technique.  It
turned out that this method works reasonably well except for Field 3,
which has a strong stellar background gradient due to its proximity to
the M87 nucleus.  Since the background dominates for this field, we
did not attempt to subtract the dark glow.  Finally, we flagged all of
the hot pixels present in our science images.

The four individual images of Fields 1 and 2 have slight positional
offsets. By inspecting the locations of the bright sources in these
images, we derived and applied the integer shifts in X and Y
directions for registration.  (Note that a one pixel shift is a small
fraction of the 70\% encircled energy radius for the point spread
function, which is 8 pixels.)  The images were then co-added.  For
Field 3, there was an interval of 4 months between the two visits and
the orientations differ by approximately 170 degrees.  We performed a
rotation on the earlier epoch images using the ROT function in IDL
with the CUBIC parameter set to $-0.5$.  The cubic convolution
interpolation method used in this function closely approximates the
optimum sinc interpolation function by adopting cubic polynomials.  To
ensure that introducing rotation does not affect the object
photometry, we have compared the photometry (explained below) done on
several sources in both the pre- and post-rotated images.  The
differences in all cases were negligible.  After rotating the images,
integer shifts were applied using the same approach as in Fields 1 and
2.

The MAMA images of Field 4 were also calibrated with the
standard CALSTIS pipeline.  We did not attempt to subtract the dark
glow of the Field 4 far-UV images because of the high galaxy diffuse
background.  The registration and co-adding for each observation of
Field 4 were done as described above, resulting in
one combined FUV-MAMA image and four combined NUV-MAMA images.

In Figure 2, we show the combined STIS FUV-MAMA and
STIS CCD images for Field 1, 2, and 3.  Images and identifications
in Field 4 are shown and discussed in \S 9.2.

\subsection{Photometry and Flux Conversion}

We derived astrometric solutions for all of our STIS images (CCD,
FUV-MAMA, and NUV-MAMA) using the WFPC2 image as a reference to aid
the identification of sources.  The IRAF task TFINDER was used for
doing the astrometry.  The offsets between new solutions and old
header records are less than 1\arcsec in most cases for all STIS
images, and the uncertainties for the solutions are less than 0.02\arcsec.

Sources in the STIS CCD (optical-band) frames of Fields 1, 2, and 3
were identified using the automated DAOFIND routine (Stetson 1987).
Since the K99 observations go deeper than ours, all sources found in
our CCD images are also listed in the photometry of K99.\footnote{Both paper
and electronic versions of K99 only include one hundred objects
closest to the center of M87.  Throughout this paper, we use an
updated complete photometry list made by A. Kundu but refer to it as
K99 photometry.}

For the far-UV and near-UV images, we did not use an automated
detection algorithm since the sources have poorly defined point spread
functions (PSFs).  Instead, we centroided on each object detected by
eye and cross-identified them with objects listed in K99 photometry.
When detected far-/near-UV sources did not match any of the objects in
K99 photometry, we labeled them as ``UV-only sources.''  Similarly,
sources not found in our far-/near-UV images but listed in K99
photometry within our field of view were labeled as ``optical-only
sources.''  The NUV-MAMA images of Field 4 covered four different
epochs separated from five months to as long as two years.  We
tabulate identifications separately for each epoch in Table
\ref{t:catalog3} to study possible variable sources.  We note that in
Field 4, all UV-only sources in the far-UV frame are also found in one or
more of the near-UV frames.

Aperture photometry on the UV sources was performed using a modified
version of the IDL APER routine, an implementation of the DAOPHOT
\citep{Stetson87} PHOTOMETRY routine.  The dark current is so low in the
MAMA detectors that the background levels are quantized to just a few
values.  The median or mode is not a good estimation of the background
when dealing with such signals.  According to \citet{STISHANDBOOK},
the best way to estimate the background is to identify all hot pixels
in the local background of each source, and then to use a {\it mean}
for the remaining pixels.  Our modified APER routine fully
accommodated this approach.  The net flux of each source was measured
through an aperture of 8 pixels in radius, corresponding to 0.20
arcsec and using a  local sky background measured within an annulus of 20
to 30 pixels in radii.  In some cases, the sources are so close to
each other that the background measurements are affected by the nearby
sources.  For those, we masked the nearby sources with apertures of
appropriate size prior to doing photometry.  The conversion of MAMA
counts into magnitudes in the STMAG system was done using the expression:
\begin{eqnarray}
m_{\scriptscriptstyle STMAG} = -2.5\log_{10} {\bigg({\rm {counts
\times PHOTFLAM} \over {EXPTIME} }\bigg)} - 21.10, 
\end{eqnarray} 
where PHOTFLAM defines the inverse sensitivity in units of ergs
s$^{-1}$ cm$^{-2}$ 
\AA$^{-1}$/(counts s$^{-1}$) and EXPTIME is the exposure time in seconds. 
The version of the CALSTIS routine we used (v2.13, 26-April-2003) does
not take into account the time dependent sensitivity changes for MAMA
imaging modes.  The PHOTFLAM values in the image headers all default
to $3.808\times10^{-17}$ ergs s$^{-1}$ cm$^{-2}$ \AA$^{-1}$/(counts
s$^{-1}$) for the FUV-MAMA and $5.922\times10^{-18}$ ergs s$^{-1}$
cm$^{-2}$ \AA$^{-1}$/(counts s$^{-1}$) for the NUV-MAMA, regardless of
the date of observation.  To calculate the correct PHOTFLAM values for
our observing periods, we used the BANDPAR task in IRAF STSDAS.SYNPHOT
package with the observation Julian dates as inputs.  

The final
integration times and PHOTFLAM values used are tabulated in Table
\ref{t:photflam}.  For UV-only and optical-only sources, we determined
upper limits in $V$ and far-/near-UV fluxes using the nominal flux
plus $3 \sigma$ when the measured fluxes are positive and $3 \sigma$
when the measured fluxes are negative.  In the case of the UV-only sources,
we re-measured the original K99 image frames at the coordinates of the
UV identifications, applying the K99 photometric procedures.
We attempted to construct
far-/near-UV growth curves for a few of our bright sources to derive
aperture corrections.  However, because even the brightest sources
have a poorly defined PSF, it is not possible to derive reliable
aperture corrections.  Therefore, we rely on the aperture corrections
for point sources tabulated in Table 8 of 
\citet{ISRSTIS2003-01}.

The foreground reddening towards M87 estimated from diffuse 100$\mu$ dust
emission is $E(B-V) = 0.022$
\citep{Schlegel98}.  Using this value and the extinction law of
\citet{Cardelli89}, we derived extinctions of $A_{\scriptscriptstyle
FUV} = 0.183$ mag and $A_{\scriptscriptstyle NUV} = 0.178$ mag,
implying $E(FUV-V) = 0.117$ and $E(NUV-V) = 0.112$.  All photometry was
corrected for extinction using these values.  Foreground $E(B-V)$
estimated from HI emission is zero (Burstein
\& Heiles 1984),  but even if we had adopted the latter value, the large
color offsets discussed below between the M87 and Milky Way clusters
would still persist. 

Our photometric catalogs are presented in Tables \ref{t:catalog1} and
\ref{t:catalog2}.  For each object we list the following parameters:
identification number, coordinate offsets from the center of M87,
radial distance from the center, far- and near-UV magnitudes and
errors in the STMAG system, and dereddened UV-optical colors.  Details
are given in the footnotes to the tables. 

\subsection{Far-Ultraviolet Detection Rate}

We use the WFPC2 $V$ photometry of K99 to examine the detection rate
of clusters in our fields.  Our limiting far-UV magnitude is
$m_{\scriptscriptstyle FUV} \sim 25$, corresponding to
$M_{\scriptscriptstyle FUV} \sim -6$.  We detect in the FUV a total of
69 objects in Fields 1, 2, and 3, of which 50 are globular clusters
identified by K99.  There are 127 optically-identified clusters in the
K99 sample in these fields to $V_0 = 26$.  (We discuss the non-K99
identifications later in \S 9).  In Field 4, 7 and 37 objects are
detected in the far- and near-UV images, respectively.  In Figure
\ref{f:fuvdetection}, we show our far-UV detections with respect to
optical detections in Fields 1, 2, and 3.  (Field 4 has been excluded
from this plot since the short exposure time and the strong background
gradient make the detection rate low and not comparable to the other
fields.)  For $V_{0} \leq 22.0$, all but two K99 clusters are detected
in the far-UV, resulting in a detection rate of 94\% with respect to
optical detections. The mean $(V-I)_{0}$ color for these 29 clusters
is $1.09 \pm 0.02$.  K99 show that the completeness of their cluster
selection for $V_{0} \leq 22.0$ is nearly 100\%.  For $V_{0} \leq
23.0$ and 24.0, our detection rate drops to 63\% and 48\%,
respectively.  Overall, we detect 39\% of the K99 cluster sample.  We
return to this subject and consider possible selection effects in \S 8. 

As seen in Figure \ref{f:fuvdetection}, the clusters detected in the
far-UV are at the bright end of the optical luminosity function for
the M87 GC system (also see Figure 5 of K99).  The two peaks in the
optical color distribution found for M87 clusters by K99 lie at
$(V-I)_{0} = 0.95\pm 0.02$ and $(V-I)_{0} = 1.20\pm 0.02$.  The
histogram in Fig.\ 3 shows that most of the clusters we detect in the
far-UV belong to the bluer [$(V-I)_{0} = 0.95$] subpopulation.

\section{Interpretation of UV-Optical Colors}

In the subsequent sections, we discuss our results for the M87 system
in the context of the available UV photometry samples of Milky Way
globular clusters and early-type galaxies.  Most of the existing data
are from the {\it International Ultraviolet Explorer} (IUE), but
several other experiments also contributed data (see Appendix A and
DOR95).  The recently-launched GALEX mission will rapidly enlarge 
these samples.

The far-UV band is sensitive only to hot stars.
In purely old populations, these exist in significant numbers only on
the horizontal branch.  The far-UV flux therefore depends mainly on
the number of HB stars hotter than about 10000~K, with a modest
dependence on their temperature distribution.  On the other hand, such
stars contribute almost nothing to the optical bands (here $V$ and
$I$), which are instead dominated by main sequence and giant light
(DOR95, Schiavon et al.\ 2002).  Thus, we take the $(FUV - V)$ color
to be a measure of the {\it fraction} of the stellar population that appears
in the form of hot HB stars. 

The situation in the near-UV ($\sim 2400$ \AA) band is more
complicated (Burstein et al.\ 1988).  In the absence of HB light, the
near-UV band will be dominated (over 90\% contribution) by stars near
the main sequence turnoff.  For a given age, there will be a strong
dependence of color on metallicity, estimated by Dorman, O'Connell, \&
Rood (2003) to be $\partial\, (NUV-V) /
\partial \log Z \sim 1.8$ mags/dex.  But hot HB stars do often have a
major impact in the NUV.  Dorman et al.\ (2003) estimated
that hot HB stars contribute 40--80\% of the NUV light in a sample of
nearby gE galaxies, and comparable numbers should also apply to those
globular clusters with bluer $(FUV - V)$ colors.  

The $(FUV-NUV)$ color for populations with comparable ages and
metallicities is in principle a measure of the temperature of the hot
HB stars.  The simple models of DOR95 indicate that the FUV light of
objects with $(FUV-NUV) < -0.3$ is dominated by HB populations that
extend to the EHB range ($T_{eff} > 15000$~K).

\section{Comparison of M87 to Local Group Globular Clusters}

Figure \ref{f:colormag} shows the extinction-corrected $(FUV-V,
M_{\scriptscriptstyle V})_0$ and $(FUV-V, M_{\scriptscriptstyle
FUV})_0$ color-absolute magnitude diagrams (CMDs) for M87 clusters.
We do not include the
FUV limits for the optical-only sources.  For comparison, we also plot
the Milky Way GCs as {\it open triangles}.  The Milky Way data were
taken from DOR95 but updated with the new $E(B-V)$ values listed in 
\citet{Harris96} (see Appendix for details). The figure shows that we
have been able to sample in M87 almost the full V-band luminosity
range present in the Milky Way UV-detected sample, although the
average M87 detection is more luminous than in the Milky Way.  There
are three M87 clusters that are brighter than the most massive Milky
Way GC, $\omega$ Cen, in $M_{\scriptscriptstyle V}$ and two in
$M_{\scriptscriptstyle FUV}$.  (While $\omega$ Cen and M54 have
similar brightnesses in $V$, $\omega$ Cen is brighter in the far-UV by
1.2 mags.)

The M87 clusters in Figure 4
extend to a significantly bluer $(FUV-V)$ color than the Milky Way GCs.
Figure \ref{f:colorhist2_15v} shows the $(FUV-V)_{0}$ color
distributions for all M87 clusters plotted in Figure \ref{f:colormag}
({\it upper panel}) and for the clusters brighter than $V_{0} = 22$
({\it lower panel}).
We also plot the histogram of the color distribution for Milky Way
GCs.  The mean color of the M87 clusters in Figure \ref{f:colorhist2_15v}a
is $(FUV-V)_{0} = 1.50 \pm 0.08$ (the error being the standard error
of mean); and for clusters in Figure \ref{f:colorhist2_15v}b, the mean
color is $(FUV-V)_{0} = 1.72 \pm 0.10$. Both M87 histograms have peaks
at $(FUV-V)_{0} = 1.30 
\pm 0.10$ (the error being half of the bin size).  The mean and peak
for the Milky Way GCs are at $(FUV-V)_{0} = 2.35 \pm 0.16$ and
$(FUV-V)_{0} = 2.30 \pm 0.10$, respectively.  Thus, the M87 GC sample
we observed is significantly bluer than the MW sample.  The two-sided
Kolmogorov-Smirnov (K-S) test shows that the probability of the Milky
Way and M87 GC color distributions being drawn from the same
parent distribution is only $6.3\times 10^{-4}$.  Only a few MW clusters,
including $\omega$ Cen and NGC 6752, have $(FUV-V)_{0}$ colors
comparable to the peak of the M87 distribution.  

To compare our M87 results to those for M31 clusters from Rey et al.\
(2005), we must first convert the GALEX photometric system (defined in
terms of flux per unit frequency) to our STMAG system (defined in terms
of flux per unit wavelength).  The conversion is $(FUV-V)_{\scriptscriptstyle STMAG}  =
(FUV-V)_{\scriptscriptstyle GALEX} - 2.82$.  The detected M31 GCs
lie in the range $(FUV-V)_{0} \sim 1.2$--2.2, within the blue envelope of
the Milky Way clusters and not as extreme as the M87 sample.

In Figure \ref{f:colorcolor}, we show the $(V-I, FUV-V)_{0}$
color-color diagram for all data plotted in Figure \ref{f:colormag}.
The $(V-I)$ colors corresponding to the centers of the two cluster
subpopulations as defined by K99 are indicated by arrows.  Most of our
detections belong to the bluer group, although they tend to lie on the
red side of the color centroid.  They also correspond in $(V-I)_0$
color to the average of the Milky Way sample.  Seventeen of the
UV-detected M87 clusters have $(V-I)_0 > 1.1$.  These evidently belong
to the red subpopulation.  They are redder in $(V-I)_0$, and presumably more
metal-rich, than any Milky Way sample cluster.  On average, these objects
have redder $(FUV-V)_0$ colors than the other M87 GCs, but they
are considerably bluer than metal-rich Milky Way clusters like 47 Tuc
and NGC 6388.  Because of the selection effects (see \S 8) evident in
Figure 
\ref{f:fuvdetection}, the true mean color of the metal-rich M87 
clusters must be redder than suggested by the figure.  
However, it is clear that M87 clusters with metallicities considerably 
in excess of those in the Milky Way are capable of producing 
large EHB star populations.

The M87 clusters show significant scatter in Figure \ref{f:colorcolor}
but form a rough diagonal sequence along which
clusters bluer in $(V-I)$ color are also bluer in
$(FUV-V)$.  The slope of this sequence
is measured to be ${\Delta}(FUV-V)/{\Delta}(V-I) \sim 4.5$. Since
$(V-I)$ colors are widely used as a proxy for metallicity (assuming a
modest age dispersion), this trend in turn suggests that more
metal-poor GCs in our M87 sample produce larger fractional hot HB
populations.  The Milky Way sample is too small and inhomogeneous to
establish such a metallicity-UV dependence, but it not inconsistent
with the M87 trend. 

Except for the two bluest Milky Way clusters in $(FUV-V)$, $\omega$
Cen and NGC 6752, {\it there is no overlap} between the M87 and Milky
Way clusters in the color-color diagram.  At a given $(V-I)$, the M87
GCs are significantly bluer in $(FUV-V)$ than the Milky Way
clusters. 

Combining both M87 and Milky Way data, there is a hint in Figure
\ref{f:colorcolor} of three parallel sequences in the diagram with
offsets in $FUV-V$: M87 GCs with $(FUV-V)_{0}<2$, Milky Way GCs
(unlabeled) plus M87 GCs with $(FUV-V)_{0}>2$, and a sequence of
UV-faint Milky Way GCs (labeled).  The latter includes 47 Tuc. However,
the inhomogeneous photometry and limited Milky Way sample preclude any
definite conclusion, nor is there an obvious physical explanation for
such a separation.

If we consider the $(V-I)$ color as a proxy for metallicity again, the
pronounced scatter in $(FUV-V)$ colors at a given $(V-I)$ in Figure
\ref{f:colorcolor} is an extreme manifestation of the ``second
parameter effect'' (see Lee, Demarque, \& Zinn 1994 and references
therein). 

The main conclusion from this section is that despite strong overlap
in V- and I-band properties, the M87 GCs have UV/optical properties
that are distinct from clusters in the Milky Way and in M31.  M87
clusters produce larger hot HB populations than do their Milky Way
analogues, and metal poor GCs are better at this than are metal rich
GCs.  Because we have detected a large fraction of all cluster
candidates in our fields (see \S 8), we do not believe this result is
strongly biased by preferential selection of an unrepresentative
extreme UV-bright tail of the M87 cluster population.  M87 metal rich
clusters can produce significant hot HB populations.

\section{Radial Dependence of Luminosities and Colors}

Figure \ref{f:radial_mag_color} tests for a variation of the absolute
far-UV magnitude, $M_{FUV}$, and the $(m_{\scriptscriptstyle FUV} -
V)_{0}$ colors with radial distance $r$ from M87 nucleus.  Since Field
4 has a different exposure depth than the other fields, we separated
this field with dashed lines in both plots.  Studies 
of radial gradients in the optical color distribution of GCs in M87 
by \citet{Strom81}, \citet{Lee93}, \citet{Cohen98}, and \citet{Harris98} show
that on average,  GC optical colors become bluer as radial
distance increases.  
\citet{Harris98} found that the color distribution is flat for $r <
1\arcmin$ (which encompasses 3 of our 4 fields) and then becomes bluer
with radial distance.  \citet{Ohl98} studied the integrated
far-UV color gradient in M87, which is dominated by the diffuse
background, and found that the $FUV-B$ color is flat out to a radial
distance of $\sim 20$\arcsec.  Beyond that, the color becomes redder
with increasing radial distance.  In our case, we find no evidence for
radial trends in the colors of individual GCs in Figure
\ref{f:radial_mag_color}, although our radial coverage is not large.

\section{UV Colors and Metallicities in Clusters and E Galaxies}

Here we present population diagnostic diagrams similar to those in
DOR95, including data for both the M87 and Milky Way globular cluster
samples, for two Galactic open clusters,  and for early-type galaxies.

\subsection{Metallicity Indicators}

To place the M87 clusters in the context of other old populations with
UV data, we employ the Lick Mg$_{2}$ narrow-band absorption line index
as a metallicity indicator.  The Mg$_{2}$ index is widely used in
population studies, and it is available for many Milky Way GCs and
nearby early-type galaxies.  UV color-Mg$_{2}$ diagrams were presented
for early type galaxies in Figure 1 of \citet{Burstein88} and for
galaxies and Galactic globular clusters in Figure 1 of DOR95.

Unfortunately, neither Mg$_{2}$ nor any other spectroscopic measures
are available for our M87 clusters.  To place them in the diagram, we
have estimated Mg$_{2}$ values from their $(V-I)_{0}$ colors using the
color-[Fe/H]-Mg$_{2}$ relations derived in Appendices A and B.  These
are based on color and spectroscopic data for GCs in the Milky Way,
the gE galaxy NGC 1399, and a small sample of GCs on the outskirts of
M87 (see the Appendices for details).  In Appendix B we show that the
three cluster systems lie on a common $(V-I)_{0}$-[Fe/H] relation.

The inferred Mg$_{2}$ values for the M87 clusters must obviously be
treated with caution.  However, the color range in this calibration
encompasses all of our UV-detected sample, and our basic results do
not depend on the details of the conversion unless there are strong
nonlinearities not evident in the data sets discussed in the
Appendices.  

An alternative that would avoid whatever uncertainty resides in this
conversion would have been to use $(V-I)_{0}$ colors, which are known
to correlate well with metallicity, instead of Mg$_{2}$.
Unfortunately, most of the early type galaxies with existing far-UV
observations do not presently have published $(V-I)_{0}$ colors.  

Another potential difficulty for interpretation is that it is now well
established that the abundances of the light metals in old
populations, including Mg, do not change in lockstep with the iron
peak (e.g.\ Worthey 1998; Thomas, Maraston, \& Bender 2003; Schiavon
2005; and references therein).  Trager et al.\ (2000) estimate that
the light metals are enhanced by $\sim 0.2$ dex with respect to the
Fe-peak in typical luminous elliptical galaxies.  However, it is the
light metals, especially O, that dominate the overall metal
abundance.  They therefore have a dominant influence on the evolution
of galaxy spectral energy distributions, although the Fe-peak is also
important through its effects on opacity.  Mg is probably a better
indicator of metallicity than Fe.  The relatively small enhancements
found by Trager et al.\ (2000) should not distort the metallicity
scale much with respect to one based on the Fe-peak.  In particular,
they should not importantly affect the ranking of objects by
metallicity, which is the main issue in the following discussion.

\subsection{Far-UV/Mg$_2$ Correlation}

In Figure \ref{f:fuvv_mg2}, we have added the M87 clusters to the
far-UV color-Mg$_{2}$ diagram for Milky Way GCs and nearby early-type
galaxies presented by DOR95.  The galaxy measurements refer to the
central $10\arcsec\times20\arcsec$ region covered by the IUE aperture. 

The lowest inferred Mg$_{2}$ indices for the M87 clusters shown in the figure
coincide with the lowest values for Milky Way
clusters.  But the M87 sample extends to considerably higher values
than does the Milky Way sample.  The most metal-rich M87 GC is
inferred to have solar metallicity.  As in Figure \ref{f:colorcolor},
the M87 sample is significantly bluer in $(FUV-V)$ at a given Mg$_{2}$
than the Milky Way clusters.

The metal-rich M87 clusters have filled the gap between the Milky Way
cluster and galaxy sequences in the diagram, as we had hoped in
planning the observations.  However, the clusters are still clearly
separated from the galaxies in this diagram.  Many
of the M87 GCs are $\gtrsim 1$ mag bluer than the bluest
ellipticals in $(FUV-V)$.

About ten metal-rich M87 clusters overlap with the lower metallicity
end of the galaxy sequence.  (The S0 galaxy M85 shows evidence for
star formation and internal extinction [Kinney et al.\ 1993], so we
exclude this object from further discussion.)  The lowest metallicity
galaxy in the sequence is the Local Group dwarf M32.  Many studies have
indicated that M32 has approximately solar abundance (Worthey 2004 and
references therein).  Our UV-detected metal-rich M87 GC sample is
$\sim 2$--3 mags bluer than M32.  In fact, it has an average
$(FUV-V)_0$ color bluer than 90\% of the entire galaxy sample.

Because of selection effects, we cannot compare M32 to a complete
sample of M87 clusters with comparable metallicity.  However, an age
difference is likely to be a contributing factor to the large color
discrepancy between the available metal-rich sample and M32.  The
luminosity-weighted age of M32 is known to be $\sim$3--5 Gyr
\citep{OConnell80,Rose84,Trager00,Schiavon04,Worthey04}, whereas the
clusters are probably 8--12 Gyr old.  Younger systems are
expected to have a smaller UVX (other things being equal) because
the envelope masses of their HB stars are larger, yielding lower HB
temperatures (e.g.\ Yi et al.\ 1999, O'Connell 1999 and references
therein).

Rose (1994) and Rose \& Deng (1999) had already demonstrated
significant population differences, probably involving age, between
M32 and the metal rich Milky Way cluster 47 Tuc.  They made a careful
analysis of spectral features in the 2600-4400 \AA\ region, which
originate mainly on the main sequence and giant branches.  The
differences on the hot HB revealed in the far-UV by Figure
\ref{f:fuvv_mg2} are much more conspicuous.  Using the simple
formulation for treating the UV spectra of old populations described
in Dorman et al.\ (2003), we find that populations like the bluest of
the M87 metal rich clusters could contribute at most about 6\% of the
V-band light in M32.  

As mentioned in the Introduction, the recent GALEX survey of quiescent
early-type galaxies (Rich et al.\ 2005) exhibits a large range of
UV/optical color but no correlation of color with metallicity, unlike
the IUE sample shown in Figure \ref{f:fuvv_mg2}.  The differences
between the two samples are not understood and are still being
explored.  However, after taking account of the differences in
definition between the GALEX and STMAG systems (\S 4), we find that
roughly 40\% of the GALEX sample is bluer than NGC 1399, the bluest
elliptical in the nearby IUE sample.  The bluest GALEX objects
correspond to $(FUV-V) \sim 0$, which is 2 magnitudes bluer than NGC
1399.  This situation is not unphysical for an old population because
DOR95 showed that if all horizontal branch stars evolved through the
hottest EHB channel, such colors could result.  However, it does
signify rather different conditions in the GALEX sample than in nearby
systems.  Whatever these are (e.g.\ contamination by recent star
formation, environmental effects leading to larger ages), they may
have concealed the trends with metallicity found by Burstein et al.\
(1988).  

The key result from Figure \ref{f:fuvv_mg2} is that the metal rich
subset of M87 globular clusters does indeed overlap the Mg$_2$ values
found on the elliptical galaxy sequence but the UV-detected clusters
produce much more far-UV light than do the galaxies at a given
Mg$_2$.  The metal-rich clusters of M87 do not appear to represent a
transition between MW-type clusters and E galaxies.  This comparison
is tentative because of incompleteness in both samples and the
uncertainty in the inferred Mg$_2$ values for the M87 clusters.
However, the effect is in the correct sense if the detected clusters
have significantly older light-weighted ages than the galaxies.

\subsection{Comparison With Galactic Open Clusters: Ages and Blue
Stragglers}

We have marked in Figure \ref{f:fuvv_mg2} the approximate locations of
two well-known Galactic open clusters, NGC 188 and NGC 6791.  These
have estimated ages of 6 and 7 Gyr, respectively (Dinescu et al.\ 1995;
Kaluzny \& Rucinski 1995).  Integrated far-UV fluxes were based on
Ultraviolet Imaging Telescope observations made by Landsman et al.\
(1998).  V-band integrated fluxes were taken from Lata et al.\
(2002).  To estimate Mg$_2$, we used the synthesis predictions for
Lick indices by Thomas et al.\ (2003).  We assumed the ages quoted
above, [$\alpha$/Fe] $= 0$, and adopted [Fe/H] values from Friel et
al.\ (2002).  (We did not use the empirical calibration for globular
clusters discussed in the Appendices because these objects are
significantly younger than the globulars.)

The positions of these clusters are rough because of lack of 
completeness, uncertainties in the Mg$_2$ conversion, and stochastic
effects in the (small) number of hot stars responsible for the UV
light in these clusters.  

Both clusters have appreciable far-UV contributions, and, as
anticipated by Landsman et al.\ (1998), their $(FUV-V)_0$ colors are
comparable to the bluest elliptical galaxies.  They fall together with
the metal-rich M87 clusters in Figure \ref{f:fuvv_mg2} and have much
bluer colors than elliptical galaxies with similar Mg$_2$ values.  NGC
6791 is as blue as the bluest Milky Way globulars. 

Theoretical models for the UV upturn (e.g.\ Yi et al.\ 1999) predict
that its strength is a step function in age, with populations older
than $\sim 5$ Gyr having a significant upturn (i.e.\ a rich EBH).
Therefore, even if the locations of NGC 188 and 6791 in Figure
\ref{f:fuvv_mg2} are accurate, we cannot conclude that the surrounding
M87 metal rich clusters have intermediate ages like those of the open
clusters.  The open cluster observations can also be taken as evidence
that the transition age is younger than 6-7 Gyr.  If so, Figure
\ref{f:fuvv_mg2} indicates that the light of most of the galaxies is
dominated by populations younger than about 6 Gyr.  

The far-UV in these two open clusters is produced by extreme HB stars,
just as we believe is the case in the elliptical galaxies.  Blue
straggler stars (BSS) are not important in the far-UV in either case
(Landsman et al. 1998).  However, under certain circumstances, BSS can
dominate the UV light.  The open cluster M67 is the best example of
this.  Landsman et al. (1998) showed that a single, unusually hot BSS
(F81, with $T_e \sim 12700$ K) contributes over 90\% of the far-UV
flux in M67.  The predicted integrated $(FUV-V)_0$ is $\sim 1.5$,
comparable to that of NGC 6791.  However, the more typical BSS stars,
with $T_e < 8500$ K, make very little contributions to the light below
1600 \AA\ (e.g.\ see Fig. 3 of Landsman et al.\ 1998).

Therefore, it is possible for BSS-rich populations to match the far-UV
colors of the metal-rich M87 globulars or the bluer gE galaxies.  But
we do not think BSS components are important in those systems for
several reasons (see Landsman et al.\ 1998).  First, massive BSS
capable of affecting the integrated far-UV light will appear only for
a narrow range of population age, near that of M67 (4 Gyr).  They are
not important in clusters like NGC 188 and 6791, only a few Gyr older,
because they become too cool.  In younger populations, the UV
influence of BSS is overwhelmed by the light of main sequence turnoff
stars.  Second, their importance in M67 has probably been enhanced by
dynamical evaporation of lower mass stars, which is not a factor in
galaxies and is probably less important in massive globular clusters.
Finally, it is thought that many BSS are binaries produced by stellar
collisions (Bailyn 1995), but the dynamical environment of galaxies is
much less favorable to this process than is the case in clusters.  

In the near-UV (2000--3200 \AA) the influence of BSS can be
considerably larger.  They probably affect the NUV integrated spectra
of clusters with significant BSS components, even where there is
strong competition from horizontal branch stars.  However, for the
last two reasons discussed above, it is doubtful that BSS have much
effect on the NUV spectra of galaxies.

\subsection{Near-UV Color-Color Correlations}

Among our data, only Field 4 has near-UV observations.  Since the
far-UV observations in Field 4 are not as deep as the near-UV, we have
$(FUV-NUV)$ colors for only seven clusters (see Table
\ref{t:catalog2}).

The $(NUV-V)$ versus Mg$_{2}$ diagram is shown in Figure
\ref{f:nuvv_mg2}.  This is the NUV analogue to Figure
\ref{f:fuvv_mg2}.  We plotted only M87 GCs with
$\sigma_{\scriptscriptstyle V-I} < 0.15$.  Because of the large
contribution of main sequence stars to the NUV fluxes, we expect the
$(NUV-V)$ color to become redder as the metal abundance increases
(Burstein et al.\ 1988, Dorman et al.\ 2003 and \S 3).  The M87 GCs
seem to follow this general trend.  However, as in Figure
\ref{f:fuvv_mg2}, they are systematically offset to bluer colors than
the Milky Way GCs for a given metallicity.  The slope of the blue
envelope in Figure \ref{f:nuvv_mg2} agrees roughly with expectations (see \S
3) for a metallicity effect if vertical scatter has been introduced by
a variable hot horizontal branch population.  The five most Mg-rich
galaxies have bluer colors than the rest of the galaxy sample,
probably due to strong contamination of the NUV light by their large
EHB populations (also evident in Figure
\ref{f:fuvv_mg2}). 

Figure \ref{f:fuvnuv_mg2} shows the $(FUV-NUV)$ color versus Mg$_2$ for
the three samples.  The M87 GCs have $(FUV-NUV)$ colors bluer than the
Milky Way GCs and comparable to the average galaxies.  However, in
this diagram the M87 clusters appear to be an extension of the Milky
Way GCs, becoming bluer in $(FUV-NUV)$ with increasing metallicity.
As noted in our discussion above on interpreting UV colors, colors
bluer than $(FUV-NUV) \sim -0.3$ indicate the presence of dominant EHB
populations.  Most galaxies and 6 of the M87 clusters fall in this
category.  M32 is again significantly redder than metal-rich M87 GCs
in this plot, by $\sim 2$ mags in $(FUV-NUV)$. 

The UV color-color diagrams are shown in Figure \ref{f:fuv_nuv_v_colorcolor}.
DOR95 emphasized the separation between Milky Way clusters and galaxies
in similar plots.  Here, the M87 GCs again extend the Milky Way
sequence to the bluer side, and the distinction between the GC
and galaxy sequences is preserved in both plots.  DOR95 suggested that
the galaxies have bluer $(FUV-NUV)$ than clusters because the more
metal rich stars in the galaxies have larger NUV line blanketing.

It is interesting to note that the metal-rich Milky Way cluster NGC
6441, and to a lesser extent NGC 6388, has photometric properties in
Figures 7--11 that are comparable to those of the UV-detected M87 GCs.
NGC 6388 and 6441 are known to have RR Lyrae variables with periods
anomalously larger than any other Milky Way clusters
\citep{Pritzl00,Pritzl01,Pritzl02}.  Further studies of these two
clusters may provide hints on the origin of the bluer $(FUV-V)$ colors
seen in M87 GCs compared to the Milky Way counterparts.

The important empirical results here (for what is admittedly a small
cluster sample) are that the M87 clusters appear to represent an
extension of the Milky Way GC sequence in FUV-NUV-V-Mg$_2$ space and
that the galaxies are found to have stellar populations distinct from
those in either cluster system.  The UV light in the galaxies is
evidently not produced by globular cluster-like subpopulations.

\section{The M87 Globular Clusters in the Age-Abundance Grid}

The bimodal distribution of cluster optical broadband colors seen in many
elliptical galaxies indicates the presence of two different
subpopulations (see \S 1.2). Since optical broadband colors of very
old stellar populations are more sensitive to metallicity than age
(e.g.\ see Yi 2003), the bimodality is normally assumed to reflect the
presence of distinct metal-poor and metal-rich populations.  The
prevailing interpretation is that the bluer, metal-poor population is
older and formed during the earliest phases of galaxy assembly,
whereas the redder, metal-rich population formed later as a byproduct
of gas-rich mergers between systems with more advanced chemical
evolution.  K99 argued that the color and luminosity distributions in
M87 were consistent with the redder population being 3-6 Gyr younger.

However, the interpretation of the red clusters as younger is
controversial (e.g.\ Catelan et al. 1998).  Using their Keck spectra of
150 GCs and the stellar population models of \citet{Worthey94},
\citet{Cohen98} found no significant age different between the two
subpopulations.  Also, \citet{Jordan02} used the Str\"omgren
photometry of 628 clusters and various population synthesis models
\citep{Bruzual93,Worthey94,Maraston98} to conclude that within their
measurement errors, the two subpopulations are coeval.  By contrast,
Puzia et al.\ (2005) recently used Lick absorption line indices to
identify a group of metal-rich, younger (5-8 Gyr old)
GCs in M31.  

Here we investigate the extent to which our UV data can contribute to
the question of M87 cluster ages.  We compare our data
to the most complete fiducial model grid of
Lee, Lee \& Gibson (2002; hereafter LLG02). The advantage of their
models lies in the fact that they include detailed HB modeling and
provide isochrones for a wide range of colors. 
LLG02 and \citet{Yi03} show that, with proper calibration of the
HB phases, colors such as $(FUV-V)$ can be more sensitive
to ages than any optical-band integrated color.
However, 
the LLG02 models do not include the EHB population ($T_{e} > 15000$ K).

Our version of LLG02 Figures 14-(a) and (b)  is shown in Figure
\ref{f:iso}. 
We have plotted the M87 cluster sample using [Fe/H] values transformed
from their $(V-I)_{0}$ colors, as described in \S 6.1 and the Appendices. 
We superpose the isochrones used in LLG02, kindly
provided by H.-C. Lee.  The models exhibit rapid bluing of $(FUV-V)$
colors for a given metal abundance as the age increases past a threshold
that varies with abundance.

Because of their blue $(FUV-V)$ colors, the M87 clusters fall at the
extreme right hand side of the LLG02 isochrone grid, where the models
predict large ages for a given metallicity.  According to K99, the
boundary between the blue and red subpopulations is near $(V-I) \sim
1.1$, which corresponds to [Fe/H] $\sim -0.8$.  If we adopt this as a
dividing line in Figure \ref{f:iso}, we find that the models imply a
large internal age spread of 2--6 Gyr in both the blue and red
subpopulations.  It seems clear that the LLG02 isochrones are not
realistic in the context of these data because 15 M87 clusters lie
beyond the $\Delta t = +4$ Gyr contour and many of the others would be
2--4 Gyr older than Milky Way clusters.  These values are unphysical
because the color-magnitude diagrams for Milky Way clusters show them
to be at the maximum age of $\sim 13$ Gyr allowed by current
cosmological parameters.

The reason for the age discrepancies found above is almost certainly
the lack of a well-established physical prescription for determining
the HB distribution for a population of given age and abundance.
Dependences on core mass and abundance are well understood, but there
is no deterministic theory for the RGB mass loss that fixes the HB
envelope mass.  Higher mass loss near the helium flash on the RGB will
yield a hotter HB for a given age or metallicity.  The effects seen in
Figure \ref{f:iso} could be explained with larger RGB mass loss in M87
than prevails in Milky Way clusters.  In any case, we
cannot use our UV observations to constrain the ages of the M87
clusters without a better understanding of RGB mass loss.

An alternative, or possibly complementary, possibility is that the
M87 clusters are super-helium-rich, as recently proposed for one of the
subpopulations in the Galactic globular cluster $\omega$ Cen by
\cite{Lee05b}.  We explore this possibility in a forthcoming
paper (Kaviraj et al.\ 2005, in preparation).

\section{Selection Effects}

One of our main conclusions in the sections above is that the M87
globular clusters have systematically brighter FUV components than do
their Milky Way counterparts.  For instance, Figure \ref{f:colorcolor}
shows that for a given $(V-I)$, M87 GCs are always bluer in $(FUV-V)_0$
than MWGCs.  To what extent is this result affected by observational
selection?  Unfortunately, selection effects are difficult to judge.
The Milky Way data are from various spacecraft, and target selection
was inhomogeneous, based for instance on apparent magnitude and low
foreground extinction rather than metal abundance or luminosity.  Our
M87 selection is actually much more homogeneous and is limited only by
apparent UV magnitude ($m_{FUV} \lesssim 25$).

An attempt to evaluate selection effects as a function of $(V-I)$
color is shown in Figure \ref{f:selection}. The upper panel of Figure
\ref{f:selection} is the color/apparent magnitude diagram for the M87
optical and UV samples in fields 1-3 taken from Figure 3.  Our UV
sample includes almost all clusters brighter than $V_0 = 22$ but none fainter
than $V_0 = 24.1$.  We have divided the intermediate region into two
bins: 
 (1) $22 < V_{0} < 24$ and 
$(V-I)_{0} \le 1.1$, (2) $22 < V_{0} < 24$ and $(V-I)_{0} > 1.1$.
The vertical dividing line corresponds to the color of the reddest
cluster (47 Tuc) in the Milky Way UV-detected sample. 
Very few of the clusters in region (2) are detected in the FUV, whereas 
59\% of the clusters in region (1) 
are detected.  Overall, we have UV detections for 61\% of the M87 optical
sample of clusters with $(V-I)_0$ colors in the range covered by the
Milky Way GC sample.

In the lower panel, we plot the $(FUV-V)_{0}$ colors or upper limits
for the M87 optical sample and for the UV-detected Milky Way
sample.  The blueward half of this panel makes a good case that the
difference in UV properties between the M87 and Milky Way clusters is
genuine and not a selection effect.  If one takes all the detections
and limits plotted in this region as a hypothetical parent
distribution, it is clear that there is a very low probability of
choosing by chance a sample with the redder $(FUV-V)_{0}$ colors of
the Milky Way subset.  Optical $(V-I)$ colors are known to be good, if
not perfect, indicators of the main sequence and giant branch
populations in clusters, and they are negligibly influenced by the HB
stars that produce most of the FUV light.  The inference is that for a
given pre-HB population, the M87 clusters typically produce larger hot
HB populations than do Milky Way clusters.  

Although the Milky Way sample is incomplete, it does include most of
the well-studied clusters with [Fe/H] $\lesssim -1.5$, and there is no
reason to suppose that a better MW sample would change this situation.
Recall that the GALEX sample of M31 clusters (preferentially luminous
ones) has properties similar to those of the MWGCs (Rey et al.\ 2005),
so M87 is distinct from both of the other two cluster systems explored
in the UV.  

We doubt that cluster mass is a factor here.  Although, based on
their V-band luminosities, the M87 detections are on average more
massive than the MW sample, the UV color distinction persists throughout
nearly the full mass range present in the Milky Way sample (see Figure
4).

Turning to the redder side of the lower panel in Figure \ref{f:selection},
we have UV detections for about
50\% of the M87 clusters in the range $1.1 \lesssim (V-I)_0 \lesssim 1.2$ but
very few of the redder clusters.  We cannot determine typical properties
for the redder M87 clusters, but we can say that some very metal-rich
M87 clusters produce UV-bright populations comparable to those of
metal-poor MW clusters.

\section{Ultraviolet-Only Sources}

\subsection{Far-UV Detections}

A surprising result of our photometry is that in all four fields, we
find sources that are detected in the far- or near-UV frames but {\it
not} in the optical frames.  In Fields 1, 2, and 3, we find 19
``UV-only'' sources (the second digit of the ID number is set to 2 in
Table \ref{t:catalog1}), and in Field 4, we find 24 UV-only sources (Table
\ref{t:catalog3}).  All of the Field 4 UV-only sources are detected 
in the NUV exposures (see \S 2.3), which go deeper than the FUV exposures, so we
consider only those here.

First, we discuss the far-UV-only sources in Fields 1, 2, and 3.
These lie in the range $23.0\lesssim m{\scriptscriptstyle FUV}\lesssim
25.5$.  Close structural inspections show that they are similar to
other sources, but our poor PSFs limit conclusions.  We photometered
these sources independently on all images and found that none show
significant variations beyond photon statistics.  In Figure
\ref{f:fuvonly_colorlimits}, we plot the color limits for these
sources in the CMD.  All sources have colors bluer than 0.1 in
$(FUV-V)_{0}$, compared to the two bluest optically-confirmed clusters
found in Figure \ref{f:colormag}, which have $(FUV-V)_{0} = 0.44$ and
$0.59$.  

There are several alternative explanations for these sources:

\begin{itemize}

\item First, we doubt that many are detector artifacts. MAMA detectors
do not suffer from cosmic rays.  Other detector artifacts should
persist for some period of time; therefore they should appear in the
same location in different fields, but they do not.  No other studies
using the STIS FUV-MAMA have reported artifacts similar to the effect
found in our fields.  Finally, in \S 9.2 below, we show that
there are a number of ``NUV-only'' sources that are detectable at
different epochs at different detector locations.

\item We cannot rule out the possibility that these sources are old
M87 clusters with a large EHB star fraction.  However, DOR95 predicts
that if all RGB stars go through the EHB channel, the $(FUV-V)$ color
cannot be bluer than $\sim 0.1$.  A similar limit was predicted from
the models of \citet{Yi98}.  Most sources in Figure
\ref{f:fuvonly_colorlimits} have colors much bluer than the
theoretical blue limit.  Hot PAGB stars are an alternative source of
UV light, but considering the short lifetime of PAGB stars, it is
unlikely that a sufficiently large population of such objects could be
produced.

\item These sources could also be young star clusters with ages
$\lesssim 1$ Gyr.  According to \citet{Bruzual93}, the $(FUV-V)$
colors of young stellar systems can be as blue as $\sim -4$. Colors in
the observed range of 0 to $-2$ correspond to cluster ages of 30--300
Myr.  However, there is no other evidence for recent star formation in
M87.

\item Some of them could be transient sources in M87 such as novae.  Many
classical novae in M31 are found to reach $M_{\scriptscriptstyle V}
\sim -8.5$ to $-9.5$ at their maximum \citep{DellaValle91}, which is
comparable to integrated magnitudes of globular clusters.
\citet{Baltz03} estimate the nova eruption rate of 100-1000 per year
for M87, following a simple purely theoretical calculation.  Our far-UV
observations span periods from a few hours to several months for
different fields.  Since the maxima of fast novae only last for a day
or so, the UV-only sources detected in Field 3 for which exposures
were split by 4 months are not likely to be of this kind.  We conclude
that only a few of the far-UV-only sources in Fields 1, 2,
and 3 are likely to be nova eruptions caught near their maximum. 

\item UV-bright background galaxies are perhaps the best candidates.
So far, there have been only a few deep UV searches for galaxies.  The
UV (2000 \AA) survey of galaxies in SA57 by the FOCA balloon-borne
camera \citep{FOCA92} showed that many UV-selected late type galaxies
in the redshift range $0<z<0.5$ have $(m_{\scriptscriptstyle 2000}-B)$
colors as blue as $\sim -2$.  They also find a number of sources
showing extreme UV-optical colors (bluer than normal Hubble sequence
galaxies), reaching up to $(m_{\scriptscriptstyle 2000}-B) \sim -4$.
These are thought to be galaxies undergoing intense star formation
with little internal extinction.  The FOCA sample is much brighter
than the objects we detect and has too low a surface density ($\sim
0.03$ arcmin$^{-2}$ down to $m{\scriptscriptstyle 2000} = 18.5$) to be
important in our fields (our combined field coverage is 0.52
arcmin$^{2}$), but there are analogues to these sources at fainter
magnitudes.  We can use the STIS observation of the Hubble Deep Field
South (HDFS; Gardner et al.\ 2000) to estimate the density of the
fainter UV background sources in our fields.  We note that the far-UV
observation of the HDFS was also done using the FUV-MAMA but with a
different filter (F25QTZ; 1450 -- 1900 \AA, pivot wavelength 1590 \AA).
We converted their AB magnitudes to STMAG magnitudes for direct
comparison with our far-UV magnitudes and selected the sources with
$(FUV-50CCD)$ colors or color limits bluer than 0.5.  Then, we counted
the number of sources that have far-UV magnitudes or limits in the
range $23.0\lesssim m{\scriptscriptstyle FUV}\lesssim 25.5$.  We
obtain a surface density of $\sim 104$ sources per arcmin$^{2}$.  This
is in acceptable agreement with $\sim 35$ sources per arcmin$^{2}$, the
number density of the UV-only sources in our fields.

\end{itemize}

We therefore believe that the most likely explanation is that the
far-UV-only sources in Fields 1, 2, and 3 are UV-bright background
galaxies.  We believe we do detect a number of M87 novae in Field 4
(see next section), but the ratio of novae to background galaxies
should decline with radius, so that fewer will be found in Fields 1,
2, and 3.  The statistical uncertainty of these observations is too
large for a meaningful estimate of the UV optical depth of M87, but it
is clearly small.  Interestingly,
\citet{Gardner00} note that they do not find any objects in their UV
images that do not appear on the optical image of the HDFS.  In the
case of the M87 fields, the bright diffuse galaxy background seriously
limits the detection depth at optical wavelengths.  A deeper optical
image of our fields will possibly reveal the nature of these sources.

\subsection{Near-UV Detections in Field 4}

We now discuss the UV-only sources on the near-UV frames of Field 4.
The near-UV frames were taken in four different epochs that span more
than three years in time.  The long span of these observations allows
us to investigate the variability of the UV-only sources.  In Figure
15 we plot the long-term variability of the NUV-only
sources.  The $m_{\scriptscriptstyle NUV}$ magnitudes of each source
are plotted for four different epochs.  For sources not detected in
specific epochs, we provide the 3$\sigma$ upper limits of the
$m_{\scriptscriptstyle NUV}$ magnitudes.  The finding charts for these
sources are provided in Figure 16.  

Based on the variability of the sources, we can divide them into three
different categories; (1) sources showing constant brightness: NUV-01,
-03, -04, -05 , -10 , and -21; (2) sources only detected in one
epoch:  NUV-02, -08, -09, -11, -12, -13, -17, -18, -19, -20, and, -22;
and (3) recurrent sources which appear variable:  NUV-07, and -16.
The sources showing constant brightness are likely to be background
galaxies as discussed above for Fields 1, 2, and 3.  

Sources that fall into category (2) are candidates for nova
outbursts.  Recently, in an attempt to search for microlensing events
in M87, \citet{Baltz03} have monitored in $I$ and broad $R$ bands the field
around the center of M87 over a month-long interval. Out of the seven
variable sources, they identify two as obvious novae: one classical
(PC1-1) and one in a globular cluster (WFC2-6); two as possible novae
(PC1-3 and PC1-3); and one as a microlensing event (WFC2-5).  Shara et
al.\ (2004) also recently reported a classical nova outburst (at optical
wavelengths) in an M87 globular cluster.  

Two of the  \citet{Baltz03}  sources, PC1-1 and PC1-2, lie in our near-UV
fields.  We detected PC1-1 in only one of the epochs (GO-8780: NUV-13)
while PC1-2 was not detected in any.  Coincidentally, the date of the
GO-8780 near-UV observation is approximately 40 days after the date
PC1-1 reached its maximum brightness ($M_{I} \sim -8.9$) in optical
bands.  From the light curve in Figure 6 of \citet{Baltz03}, the optical
luminosity of PC1-1 seems to decline rather rapidly and after 40 days
it is expected to be significantly fainter than it was at maximum
light.  Interestingly, the 53 day OAO-A2 observation of Nova FH
Serpentis 1970 \citep{Gallagher74} showed that the nova became progressively
bluer during the early decline of the visual light curve with the
bolometric luminosity remaining constant.  Our detections of
PC1-1 in the GO-8780 near-UV frames suggest that this behavior was
also shown by PC1-1.  It is likely that most of the other sources in
category (2) are also novae.  

On the other hand, the recurrent variable sources such as NUV-07, and
-16 are difficult to explain.  They do not fit the pattern expected
for novae, and most other kinds of accretion-driven variable stellar sources
are intrinsically faint in the UV.  Because both sources NUV-07 and
-16 lie close to M87's nonthermal jet in projection, a more exotic
alternative is that they are a new class of source induced by activity
in the jet.  Earlier authors have speculated that the continuing,
dramatic activity in the M87 jet might, through an undetermined
mechanism, trigger point-like flares in its vicinity (e.g.\ Sparks et
al.\ 2000, Perlman et al.\ 2003).  However, if triggering occurs, the
observed variability will be strongly influenced by light travel time
effects.  These objects are interesting enough to justify detailed
monitoring.

\section{Summary and Conclusions}

We have obtained deep HST/STIS far- and near-UV photometry of four
fields near the center of the gE galaxy M87 in order to study the
hot-star populations in its globular cluster system.  We made new
FUV-MAMA/F25SRF2 imaging observations in Fields 1, 2, and 3 (25\arcsec\
square each) using TIMETAG data acquisition in order to maximize access
to periods of low sky background.  We extracted additional FUV and NUV
images taken at four different epochs from the HST Archive for a
nuclear field (Field 4).  We performed aperture photometry on all UV
sources detected in the fields and at the positions of all
optically-identified clusters in the deep HST survey of Kundu et al.\
(1999).

Our limiting far-UV magnitude is $m_{\scriptscriptstyle FUV} \sim 25$,
corresponding to $M_{\scriptscriptstyle FUV} \sim -6$.  We detect in
the FUV a total of 69 objects in Fields 1, 2, and 3, including 50 of
the 127 globular clusters identified by K99 to $V_0 = 26$.  Most of
our UV detections appear to belong to the bluer of the two cluster
subpopulations identified by K99.  We did not find any radial trends
of UV luminosities or UV/optical colors of our clusters, although
given its small range of radii our sample is not well suited for
testing such trends.

We compare the UV/optical photometric properties of the M87 clusters
to those of Milky Way clusters (from DOR95) and M31 clusters (from Rey
et al.\ 2005).  Despite strong overlap in V- and I-band properties, the
M87 GCs have UV/optical properties that are distinct from clusters in
the Milky Way and in M31.  The color distribution of the M87 clusters
has a strong peak at $(FUV-V)_{0}= 1.30$ and a mean of 1.50, compared
to corresponding values for the Milky Way sample of 2.35 and 2.30,
respectively.  In the $(V-I, FUV-V)_{0}$ color-color diagram, there is
almost no overlap between the M87 and Milky Way samples, with M87
clusters being significantly bluer in $(FUV-V)$ 
at any $(V-I)$.  M31 clusters have UV properties like the Milky Way
sample and unlike M87.  In color plots including the NUV band (only
seven M87 clusters here), the M87 clusters again are distinct from
Milky Way clusters but appear to represent an extension of the MW
sequence.  NGC 6441 is the closest Milky Way analogue to the M87
clusters.  

Seventeen of the UV-detected M87 clusters are redder in $(V-I)_0$, and
presumably more metal-rich, than any Milky Way sample cluster.  Some
of these have $(FUV-V)_0$ colors as blue as any metal-poor MW
cluster.  We do not sample the more metal rich population as well as
the lower metallicity clusters.  However, there is a suggestion of a
trend in which lower metallicity GCs in our M87 sample have bluer
$(FUV-V)$ colors.

It is difficult to assess selection effects, especially because the
Milky Way UV sample is very inhomogeneous.  Overall, in Fields 1, 2,
and 3 we have UV detections for 61\% of the M87 optical sample with
$(V-I)_0$ colors in the range covered by the Milky Way GC sample.  We
therefore believe our UV detections are representative of the M87 GC
population for metallicities found among Local Group clusters
and that the difference in UV properties between them and M87 GCs
is genuine and not a selection effect. 

We believe the stars producing the far-UV light lie on the extreme
horizontal branch.  Blue stragglers are in general too cool to
contribute to the far-UV, though they could be important in the
near-UV light of some clusters (if not elliptical galaxies).  The
inference is then that for a given pre-HB population (i.e.\ main
sequence and red giant branch), M87 clusters typically produce larger
hot HB populations than do Milky Way clusters.  M87 globulars
with metallicities considerably in excess of those typical of
the Milky Way are capable of producing large hot HB populations.  The
five bluest M87 clusters have a mean value of $(FUV-V)_0$ of 0.62.
This implies, based on the estimate of DOR95 that the bluest possible
color for an old population is $\sim 0.14$, that $\sim 60$\% of the HB
stars in these clusters evolve through the hot EHB channels.  This
sounds like a dramatic concentration, but it is only a factor of
$\sim$2 higher than occurs in well-observed clusters like
$\omega$Cen.

Existing model grids for cluster UV properties (e.g.\ LLG02) do not
match the M87 observations well.  Effects such as larger giant branch
mass loss or perhaps high helium abundances evidently act to enhance
the importance of hot HB stars over the predictions for a given age
and abundance.  

In order to compare our M87 clusters to earlier compilations of
elliptical galaxy data, we have estimated Mg$_2$ line
index values for them from their $(V-I)_0$ colors using a new
calibration based on data for clusters in the Milky Way, the gE galaxy
NGC 1399, and the outer parts of M87.  We find that the three cluster
systems lie on a common $(V-I)_{0}$-[Fe/H] relation.

The most metal-rich M87 GCs in our sample are near solar metallicity
and overlap the E galaxy sample in estimated Mg$_2$ line indices.
Nonetheless, the detected clusters produce much more UV light at a
given Mg$_2$, being up to 1 mag bluer than any gE galaxy in $(FUV-V)$
color and up to 3 mag bluer than the fiducial Local Group elliptical
M32 at the same Mg$_2$ value.  The detected metal-rich clusters have an
average $(FUV-V)_0$ color bluer than 90\% of the entire galaxy sample.
They do not appear to represent a transition between Milky Way-type
clusters and E galaxies.   

We find that two Galactic open clusters, NGC 188 and 6791, have EHB
components that place them among the metal-rich M87 clusters in
the Mg$_2$-$(FUV-V)_0$ diagram.  These have well-determined ages
of 6-7 Gyr.  If, as the theoretical models predict, there should
be a transition age above which the EHB becomes well populated, then
this is younger than 6 Gyr.

Our main conclusion is that insofar as their UV/optical properties are
concerned, the Milky Way clusters, the M87 clusters, and the elliptical
galaxies constitute three distinct types of stellar populations.  More
subtle distinctions in the optical, near-IR, and near-UV among cluster
systems and between clusters and E galaxies have been established for
some time (e.g.\ Frogel, Persson, \& Cohen 1980; Burstein et al.\
1984; Ponder et al.\ 1998).  Cluster mass does not seem to be a factor
in the differences between the Milky Way and M87 cluster systems
because the color effects are detected at all masses.  The very
different dynamical environment of M87 could influence its cluster
populations, for instance by affecting concentration (but this cannot
be measured at the resolution of HST).  Enhanced red giant branch mass
loss or high helium abundances might also account for the large hot HB
populations in M87 clusters.  The UV color differences are in the
correct sense if the detected M87 clusters have significantly older
light-weighted ages than the E galaxies.  Whether or not the trends of
elliptical galaxy UV colors discussed in \S 6 are consistent with the
``old base plus younger frosting'' models discussed by Trager et al.\
(2000) is something that should be explored.  

The large samples from GALEX will help to clarify the behavior of hot
populations in elliptical galaxies.  The existing GALEX sample (Rich
et al.\ 2005) does not exhibit the strong metallicity-UVX correlation
of the older IUE sample (in Figure 8).  However, we find that
40\% of the GALEX sample is bluer than any galaxy in the IUE sample,
which suggests that additional factors such as age or recent star
formation are playing a role.

Surprisingly, we find that 43 of our UV detections have no optical-band
counterparts.  We consider various explanations for these ``UV-only''
sources.  Most appear to be UV-bright background galaxies, seen
through M87, which therefore has low UV optical depth.  A set of 11
NUV variable sources detected at only one epoch in the central field
are probably classical novae.  Two recurrent variable sources have no
obvious explanation but could be related to activity in the
relativistic jet.  


\acknowledgments 
 
This work was supported at UVa in part by NASA grants GO-8643 and
GO-9455 from the Space Telescope Science Institute and Long Term Space
Astrophysics grant NAS5-02052.  STScI is operated by the Association
of Universities for Research in Astronomy, Inc., under NASA contract
NA5-26555.  AK acknowledges support from NASA LTSA grant NAG5-12975
and grant AR-09208.01 from STScI.  JAF thanks Dr. Sean Solomon for his
hospitality at the DTM, Carnegie Institution of Washington.  The
referee, Ruth Peterson, made some important suggestions that
significantly improved the paper.  We are grateful to Robert Rood, Ben
Dorman, and Ricardo Schiavon for helpful conversations and to H.-C.
Lee for providing his model isocrones.

\clearpage

\appendix


\section{Ultraviolet and Mg$_2$ Data for Milky Way Globular Clusters}

DOR95 compiled UV data for Milky Way GCs from various sources and
listed them in their Table 1.  Because the metallicities and reddening
values they used are outdated, we have updated them with the new
[Fe/H] and $E(B-V)$ values in the most recent version of Harris catalog
\citep{Harris96}\footnote{The catalog can be found at
\url{http://physun.physics.mcmaster.ca/~harris/mwgc.dat}}.  
The update was done as following.  For the $(15-V)_{0}$, $(15-25)_{0}$, 
and $(25-V)_{0}$ colors, we reverted to the ``uncorrected'' values using
the old $E(B-V)$ and the reddening law by \citet{Cardelli89}, and then
corrected for reddenings, this time using the new $E(B-V)$ and the
same reddening law.  The OAO far-UV fluxes \citep{deBoer85} were only used
for four clusters: 47 Tuc, NGC 1851, M79, and M5.  For M79, we
averaged the colors from OAO and UIT observations listed in DOR95. The
far-UV observation for $\omega$ Cen is taken from UIT photometry
\citep{Whitney94}.  
To these data, we have added integrated $(V-I)_{0}$ colors and 
$M_{\scriptscriptstyle V}$ magnitudes taken from the Harris Catalog.

The catalog was also updated by adopting Mg$_2$ measurements from the
recent literature.  Specifically, the Lick/IDS Mg$_2$ measurements and
associated errors of (1) \citet{Trager98} were adopted for NGC 5024,
5272, 5904, 6205, 6341, and 7078, and of (2) \citet{Puzia02} were adopted
for NGC 6388, 6441, and 6626.  For clusters that overlap in both
studies (NGC 6356, 6624, and 6637), we have taken error-weighted means.

Although DOR95 used a linear relation for the conversion of [Fe/H] 
to Mg$_2$ for clusters that do not have Mg$_2$ measurements, other
authors (e.g.\ Burstein et al.\ 1984, Brodie \& Huchra 1990) found
nonlinear relationships.  
\citet{Puzia02} noted that the linear relation breaks 
down in the metal-rich regime and provide a  parabolic relation between 
various line indices and [Fe/H].  However, we found that the coefficient 
for the Mg$_2$ versus [Fe/H] listed in their Table 6 is incorrect in the 
sense that the fitted curve in their Fig 7 is not consistent with the 
coefficients provided.  Using the same clusters as \citet{Puzia02}, 
we derived the following relation:

\begin{eqnarray}
{\rm Mg_{2}} = 0.268(\pm 0.006) + 0.204(\pm 0.011){\rm [Fe/H]} + 0.042(\pm 0.004){\rm [Fe/H]^{2}}.
\end{eqnarray}

We used the relation above to derive predicted Mg$_2$ values from 
their [Fe/H] for Milky Way clusters in our updated table 
that do not have Mg$_2$ measurements either in \citet{Trager98} 
or \citet{Puzia02}. 
The adopted values of all parameters for our Milky Way GCs are
presented in Table \ref{t:mwgc_dor95_dat}.

For consistency, we have also updated the Mg$_2$ measurements 
for the galaxies discussed in \S 6 by adopting the Lick/IDS Mg$_2$ 
measurements of \citet{Trager98}.

The Mg$_{2}$ index is, of course, sensitive to age as well as
metallicity (e.g.\ Worthey 1994).  The calibration here is technically
correct only for systems in the age range $\sim 10$-13 Gyr; but the
actual age range in the Milky Way sample evidently produces only a
small effect on the line index.


\section{Color-Metallicity Transformation for M87 Globular Clusters}

Here we investigate the transformation between  $V-I$ colors
and [Fe/H] for the M87 clusters.   A number
of such studies of other cluster systems have been made.
\citet{Couture90} derived an empirical relationship between [Fe/H] and
$V-I$ using the compilation of Milky Way clusters by \citet{Reed88}
and \citet{Zinn85}.  More recently, \citet{Harris00} derived the
dependence of $(V-I)$ on [Fe/H] using the 1999 version of the McMaster
catalog of Milky Way clusters.  However, there are well-known pitfalls
(see e.g. Kissler-Patig et al.\ 1998) in trying to extrapolate results
for Milky Way clusters to higher metallicities such as those
characteristic of many M87 clusters.  
\citet{Kissler98} combined their data for the gE galaxy NGC 1399 
with those of the Milky
Way clusters in order to extend the relation to the higher metallicity
range.  Meanwhile, \citet{Kundu98} pointed out that the difference between
the \citet{Couture90} and \citet{Kissler98} relations are mostly due to the
choice of the independent variable.  They also argue that neither of
the quantities should be used as an independent variable but instead
use a ``bisector'' \citep{Isobe90} to derive the relationship.

To study the color-metallicity relation for M87 clusters, we use ten
M87 clusters in Table 2 of the spectroscopic study by
\citet{Cohen98} that overlap with the $V$ and $I$ photometry of K99.
Using the ID numbers of \citet{Cohen98}, these are 5001, 5002, 5012,
5021, 5028, 649, 697, 746, 750, and 892.  The metallicities of these
clusters are in the range $-1.3 <$ [Fe/H] $< 0.1$.  Here, we derive
our own relation for the combined sample consisting of Milky Way
clusters with $E(B-V) < 0.2$, NGC 1399 GCs, and M87 GCs.  We used a
bisector method \citep{Isobe90} to derive a new empirical relation as
shown in Figure \ref{f:vifeh_conversion}.  The figure shows that, with
the exception of one outlier, the M87 clusters fall on the same linear
color-metallicity relation as the Milky Way and NGC 1399 clusters.
The correlation is good despite what must be a finite dispersion in
ages among the three systems and within each system, although it
would undoubtedly be improved if ages could be assigned to individual
clusters.  

Table \ref{t:vifeh_relation} lists our fitting parameters and compares
them to those of other studies mentioned above.  We adopt this
relation for the discussion of M87 clusters in the age-abundance grid
(\S 7).  We also use this relation in combination with the results of
Appendix A to estimate Mg$_2$ values for our M87 cluster sample (\S
6).  We use this approach in preference to attempting to determine a
color-Mg$_2$ relation for M87 using the spectroscopy of
\citet{Cohen98} because of reservations they expressed about their
Mg indices.



\begin{figure}
\figurenum{2a}
\epsscale{1.00}
\plotone{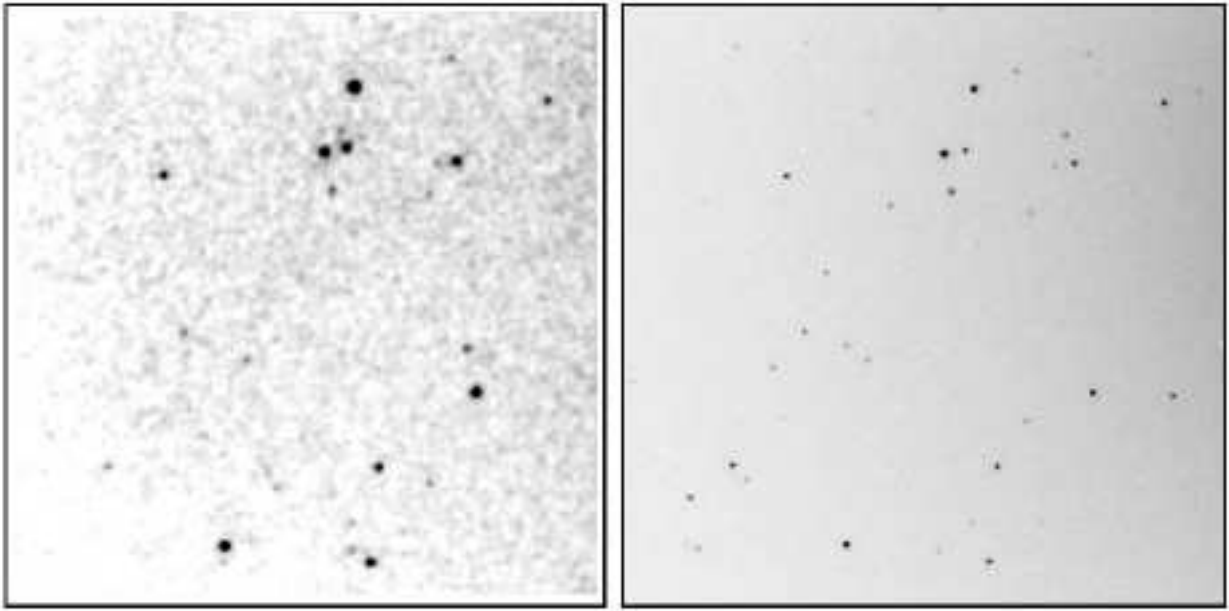}
\caption{ STIS FUV-MAMA ({\it left} panels) and STIS optical-band CCD ({\it right}
	  panels) images of Field 1.  The FUV frames have been smoothed by
          a Gaussian with FWHM = 0.24\arcsec\ (10 pixels) in order to improve 
          the contrast of the resolved sources against the background.  
          The CCD frames, which have
	  larger field of views, were cropped and scaled to match the
	  25\arcsec$\times$25\arcsec\ FUV-MAMA frames.   The resolved objects
          in the optical frame are all globular clusters.  Most ``UV-only''
          sources in the FUV frame are probably background UV-bright galaxies
          (see \S 9). 
          \label{f:field123} }
\end{figure}

\begin{figure}
\figurenum{2b}
\epsscale{1.00}
\plotone{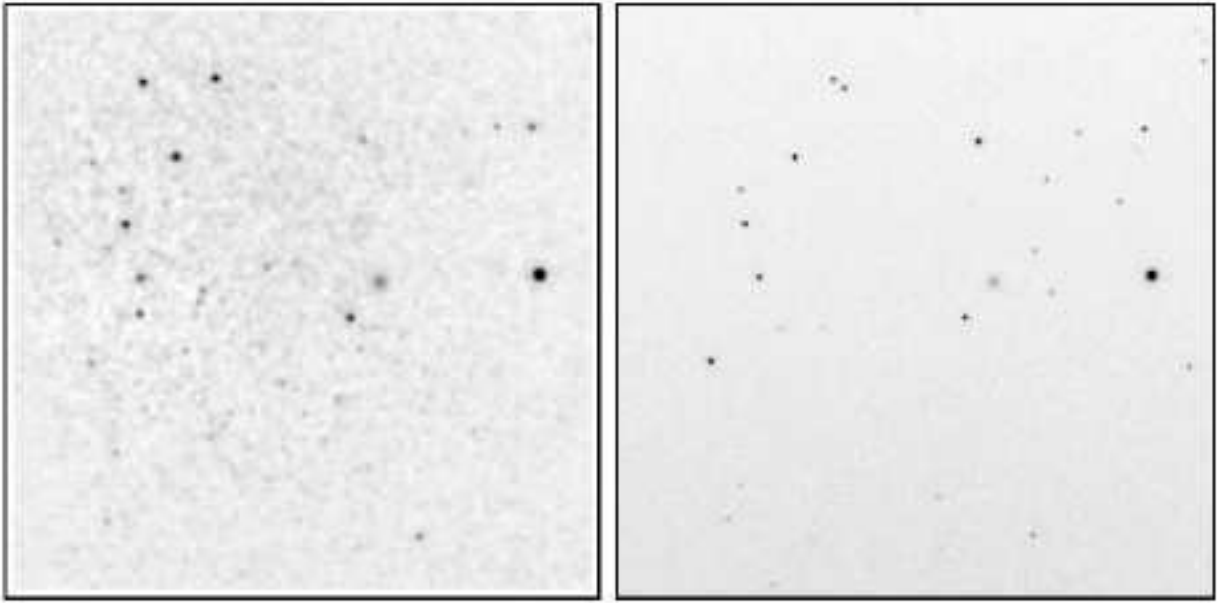}
\caption{ Same as Figure 2a but for Field 2.  A background galaxy is at right
  center.}
\end{figure}

\begin{figure}
\figurenum{2c}
\epsscale{1.00}
\plotone{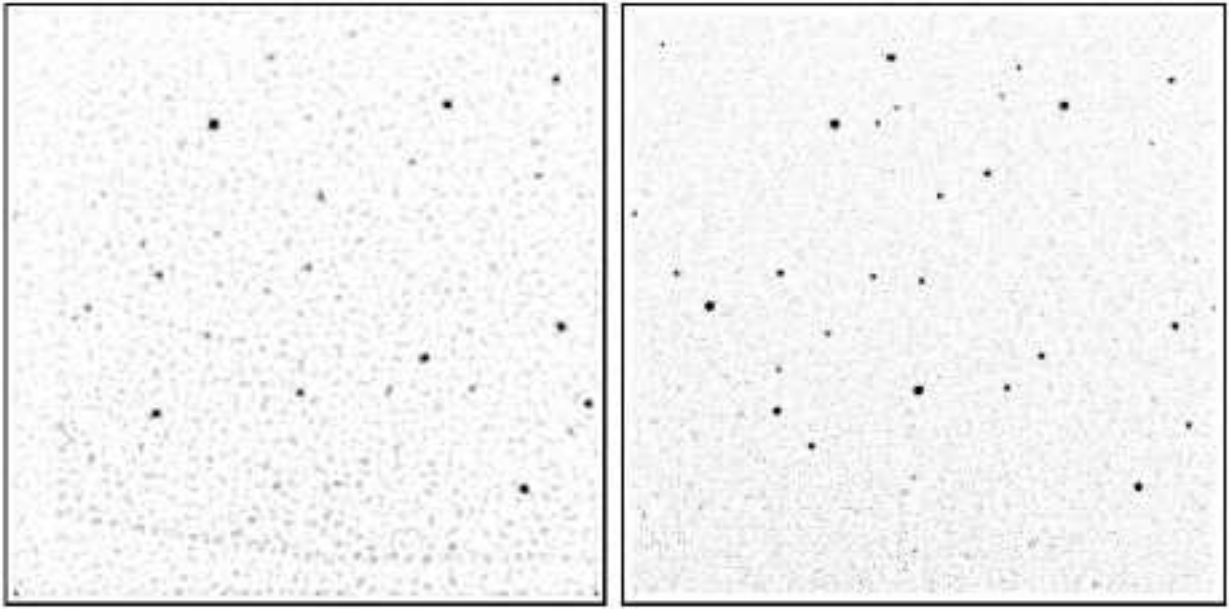}
\caption{ Same as Figure 2a but for Field 3.  Because of the strong background
          gradient in this field, we subtracted a median-filtered image
          from both panels.  The irregular shading
	  in the FUV-MAMA image is a result of co-adding
	  images with two different position angles.}

\end{figure}
\setcounter{figure}{2}


\begin{figure}
\epsscale{0.9}
\plotone{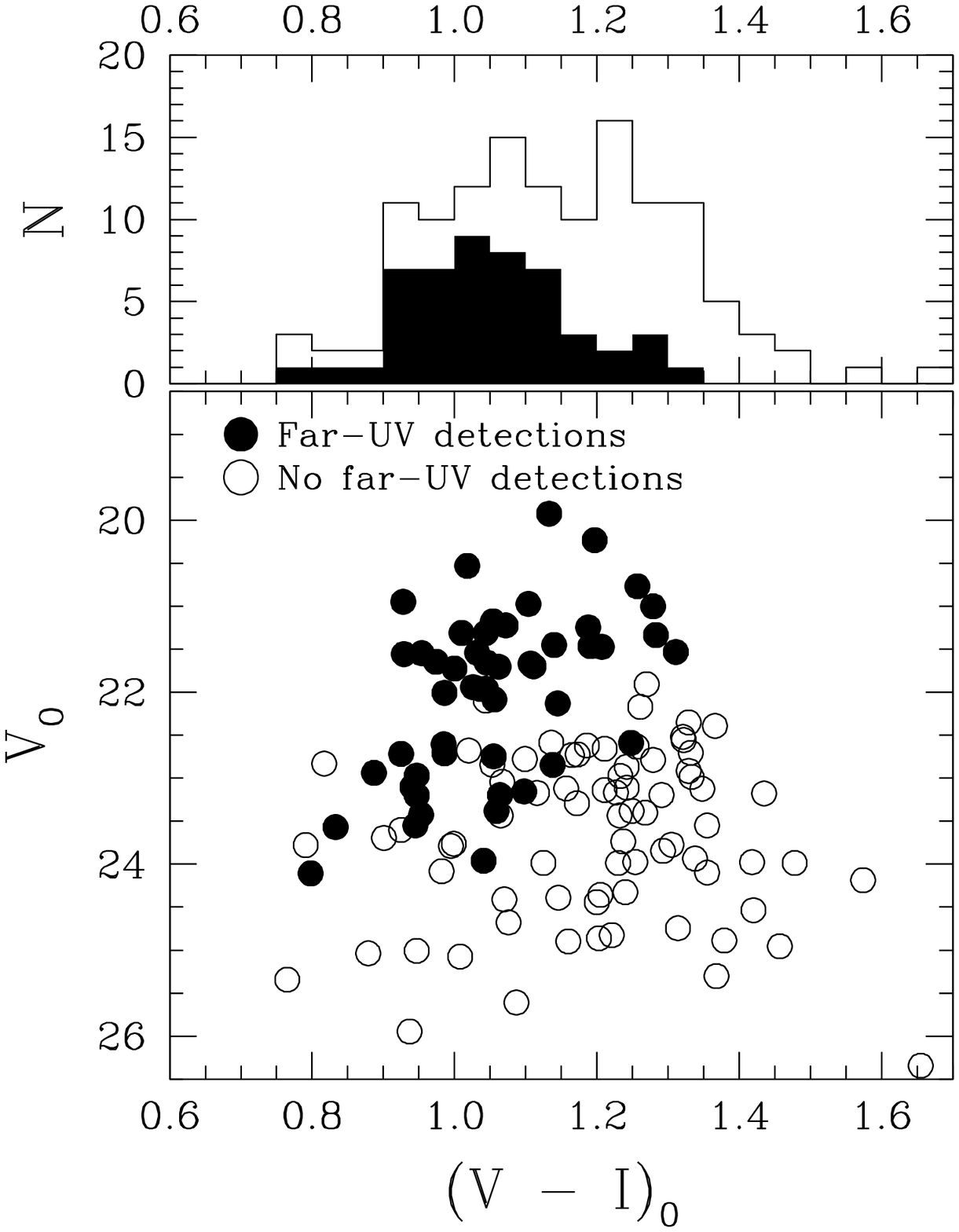} 
\caption{ $(V - I, V)_{0}$ color-magnitude diagram ({\it lower panel})
	  and $(V - I)_{0}$ histogram ({\it upper panel}) of all
	  clusters detected by K99 that lie in Fields 1, 2, and 3.
	  The $V$ and $I$ magnitudes were adopted from the HST/WFPC2
	  photometry of K99.  The {\it open circles} in the lower
	  panel are for sources that are detected only in the WFPC2
	  optical frames while the {\it filled circles} are for
	  sources detected both in STIS/FUV and WFPC2 frames. The {\it
	  non-shaded} and {\it shaded} histograms in the upper panel
	  are respectively for the {\it open} and {\it closed} circles
	  in the lower panel.
          \label{f:fuvdetection} }
\end{figure}

\begin{figure}
\epsscale{1.0}
\plotone{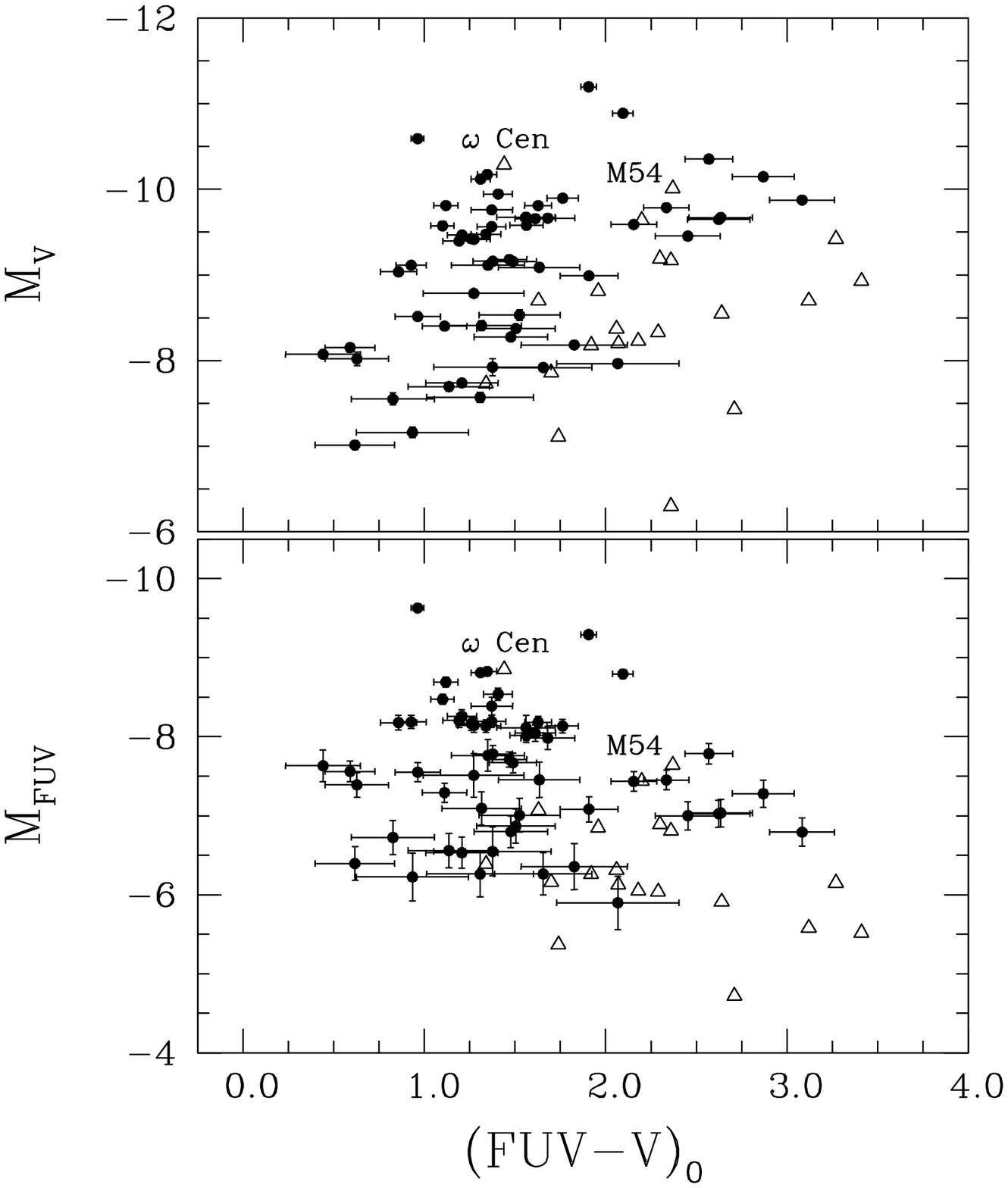} 
\caption{ $(FUV-V, M_{\scriptscriptstyle V})_{0}$ and $(FUV-V,
	  M_{\scriptscriptstyle FUV})_{0}$ color-magnitude diagrams
	  for M87 clusters ({\it filled circles}) and Milky Way
	  globular clusters ({\it open triangles}). 
	  The two brightest Milky Way globular clusters in the far-UV,
	  $\omega$ Cen and M54, are labeled.  They are comparable in
          V-band luminosity to the brightest M87 clusters but have
          redder UV-optical colors than many M87 objects.
          The bluest M31 globular clusters in the GALEX sample of
          Rey et al.\ (2005) coincide with the blue limit for
          Milky Way clusters in this diagram at $(FUV-V)_0 \sim 1.2$.
          Many M87 clusters lie blueward of this limit. 
          Note that 47 Tuc, which
	  has $M_{\scriptscriptstyle V} = -9.42$,
	  $M_{\scriptscriptstyle FUV} = -4.78$, and $(FUV-V)_{0} =
	  4.64$ lies off this plot.  \label{f:colormag} }
\end{figure}

\begin{figure}
\epsscale{1.0}
\plotone{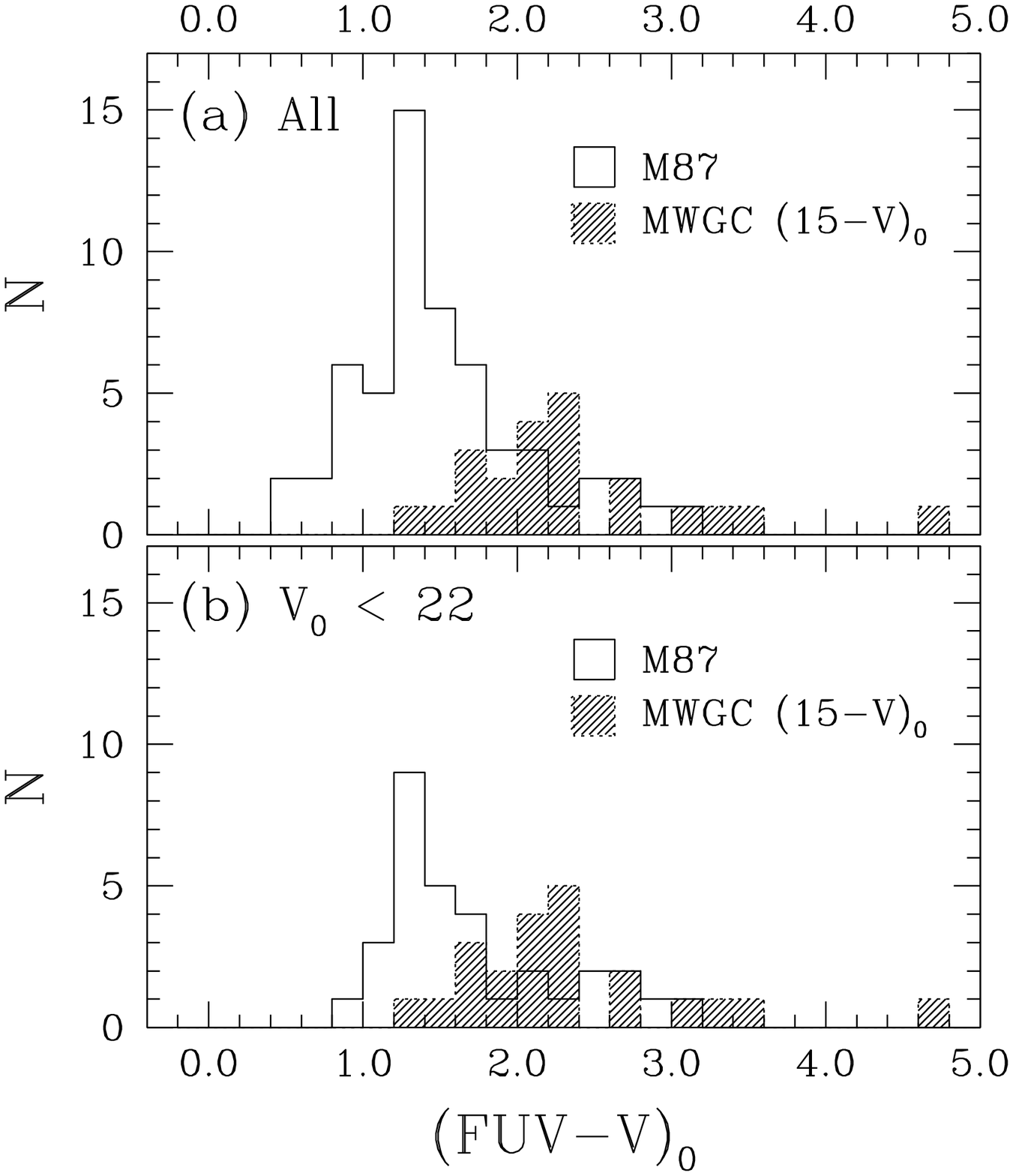} 
\caption{ Dereddened $(m_{\scriptscriptstyle FUV} - V)$ histogram for
	  (a) all M87 clusters detected both in optical and far-UV, and
	  (b) brighter clusters only, with $V_{0} < 22$. For
	  comparison, the $(FUV - V)_{0}$ histogram for Milky Way GCs
	  compiled by DOR95, but updated with the new E($B-V$) values
	  listed in \citet{Harris96}, is plotted with shading. Bin sizes are
	  0.2 mags for all distributions.  47 Tuc is the reddest
          cluster plotted.  Compared to the Milky Way sample, 
          the M87 samples are strongly shifted to bluer colors.
           \label{f:colorhist2_15v} }
\end{figure}

\begin{figure}
\epsscale{1.0}
\plotone{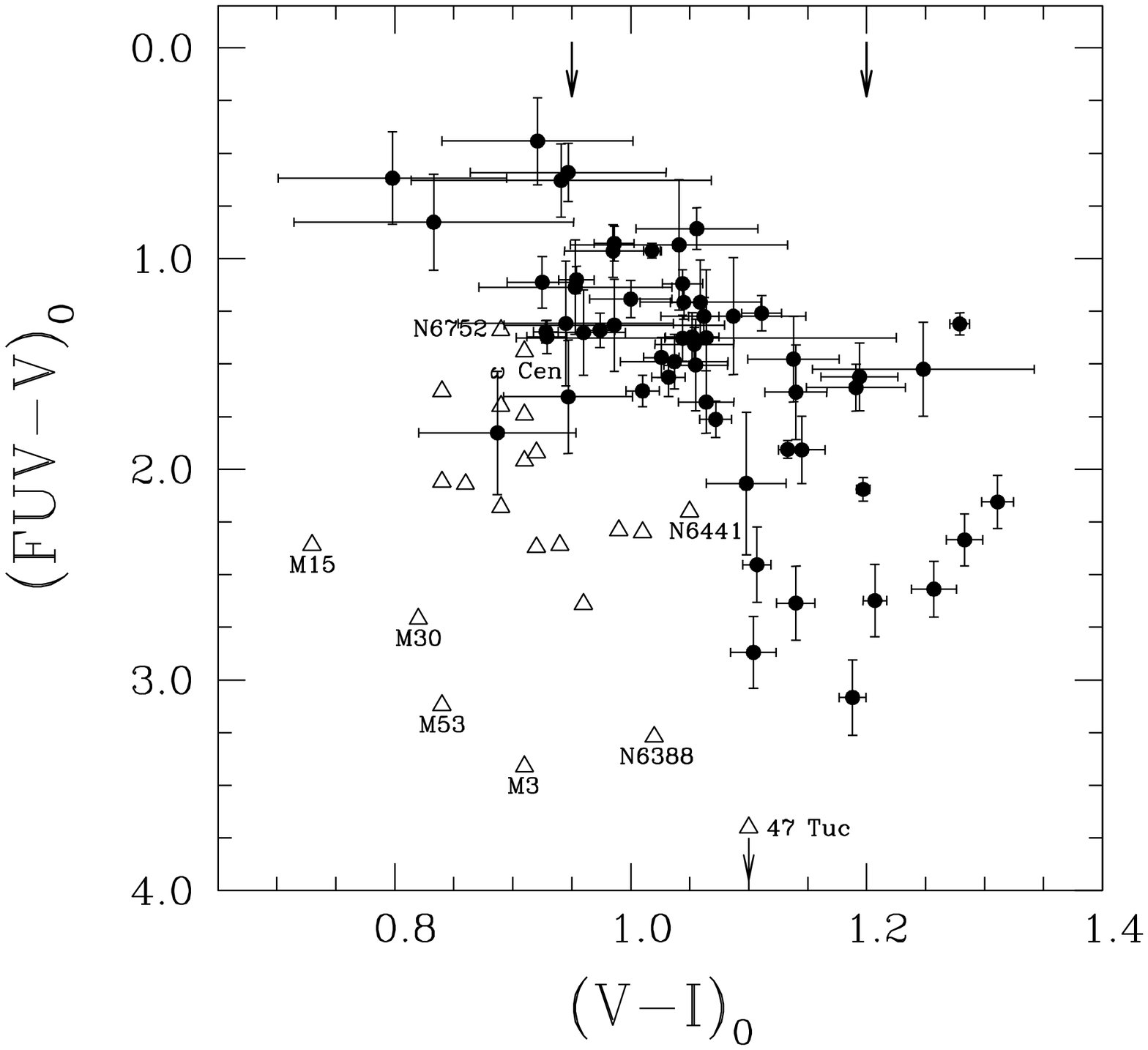} 

\caption{ $(FUV-V)_{0}$ versus $(V-I)_{0}$ color-color diagram for M87
	  clusters ({\it filled circles}) and Milky Way globular
	  clusters ({\it open triangles}).  The downward arrows at
	  $(V-I)_{0} = 0.95$ and 1.20 show the two peaks in the
	  optical M87 color distribution found by \citet{K99}.
	  \label{f:colorcolor} }

\end{figure}

\begin{figure}
\epsscale{1.0}
\plotone{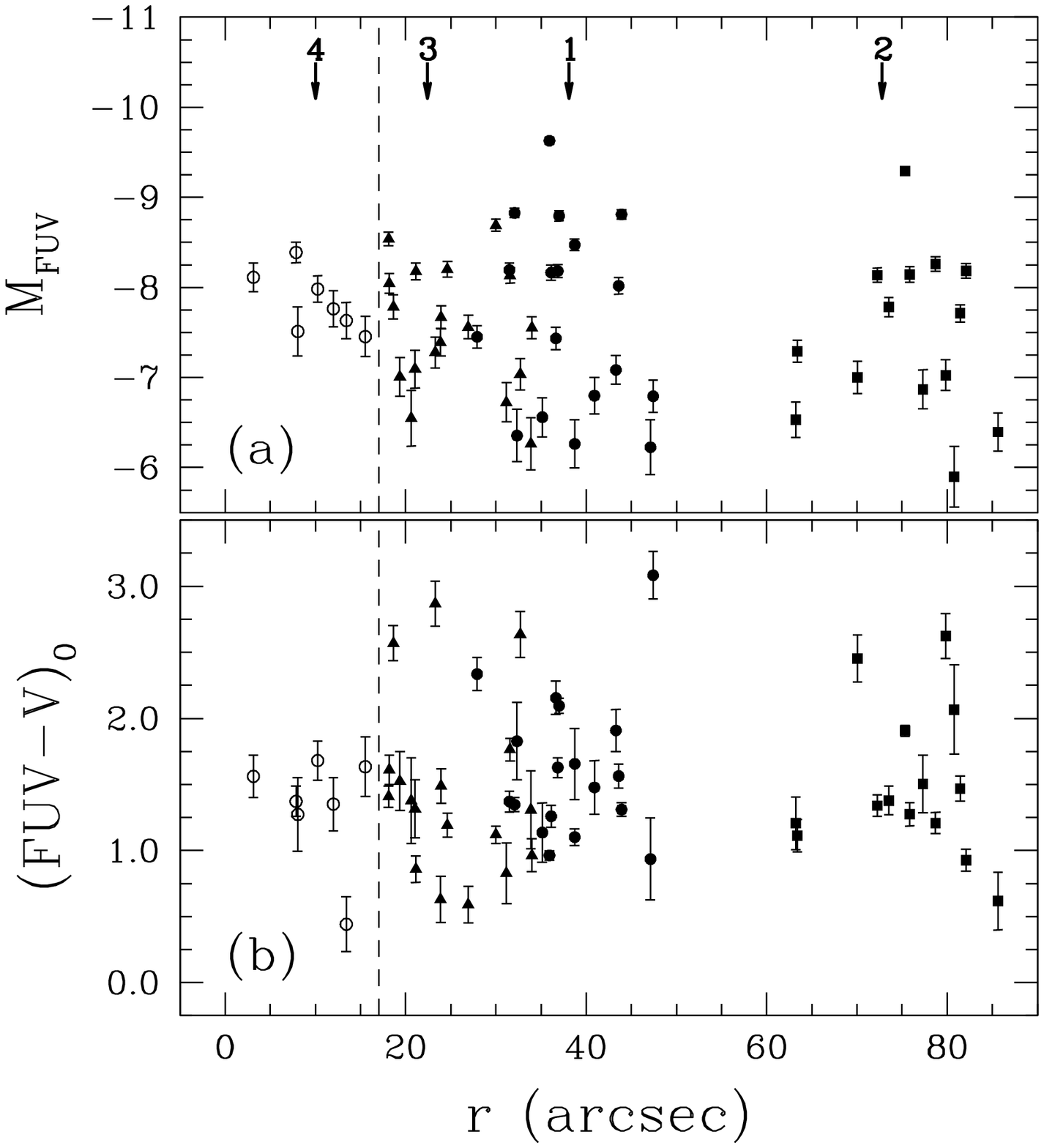} 
\caption{ Radial dependences of (a) $M_{FUV}$ and (b) $(FUV - V)_{0}$
	  color for M87 clusters.  $r$ is the distance from the galaxy
          nucleus.  We use different symbols for
	  clusters in different fields; {\it filled circles} are for
	  Field 1, {\it filled squares} are for Field 2, {\it filled
	  triangles} are for Field 3, and {\it open circles} are for
	  Field 4.  The downward arrows at the top of the plot
	  indicate the field center for each field.  The vertical
	  dashed line at $r = 17$ is drawn to separate the clusters in
	  other fields with those in Field 4, for which the total
	  integration time was considerably shorter.  The magnitude
	  trend in Field 4 is probably a selection effect caused by
	  the bright diffuse background.  \label{f:radial_mag_color} }
\end{figure}

\begin{figure}
\plotone{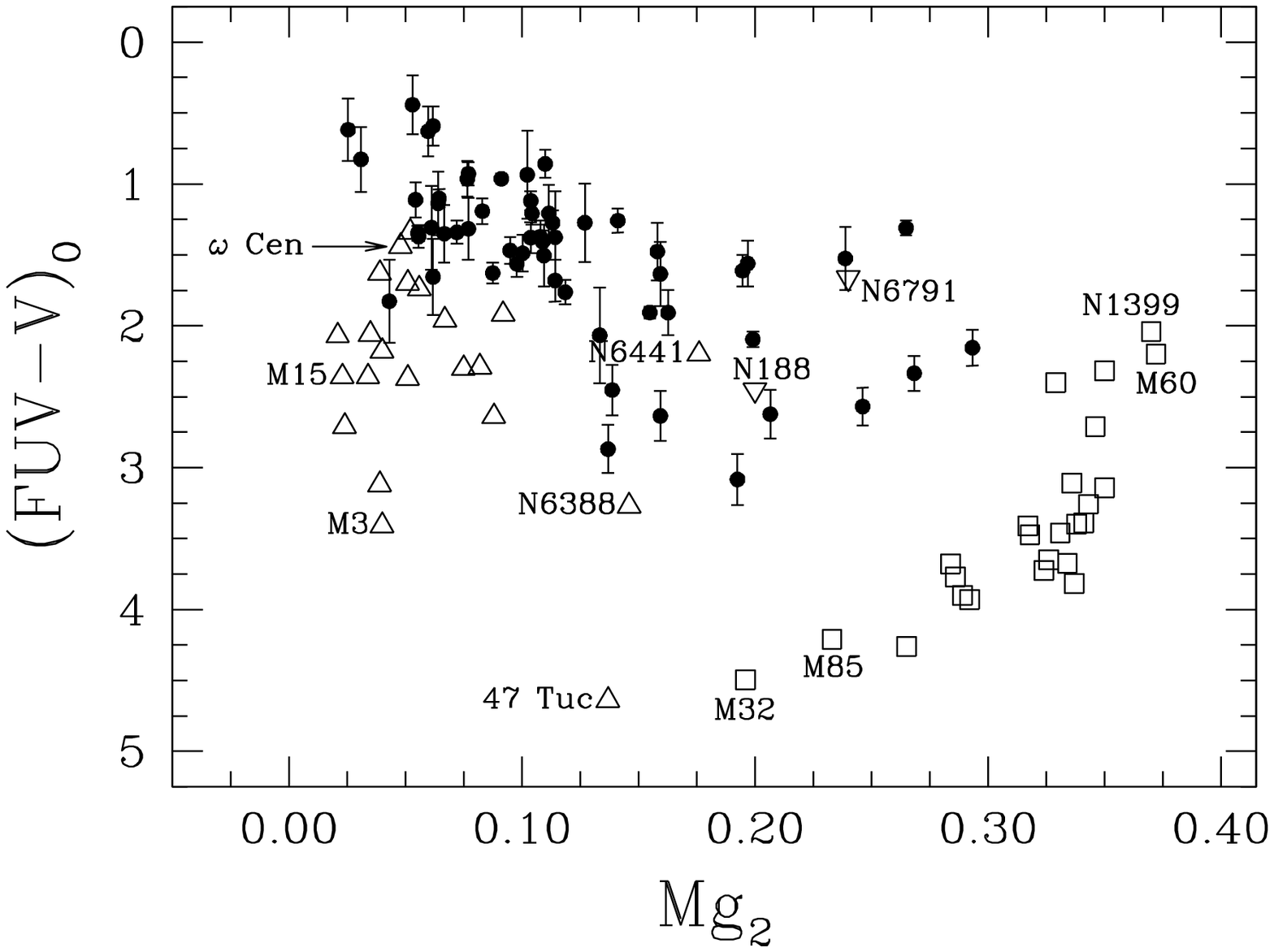} 
\caption{ The metallicity dependence of $(FUV-V)_0$ colors for M87
	  globular clusters ({\it closed circles}), Milky Way GCs
	  ({\it open triangles}), two Milky Way open clusters ({\it inverted
          triangles}), and galaxies ({\it open squares}).
	  Objects with bluer colors lie higher in the diagram. The
	  Mg$_{2}$ indices for M87 clusters were estimated from their
	  $(V-I)_{0}$ colors (see text for details.)  The colors and
	  Mg$_{2}$ indices for Milky Way GCs were adopted from Table
	  \ref{t:mwgc_dor95_dat}, while those for galaxies are from
	  Table 3 of DOR95 updated by Trager et al.\ (1998).  Although
	  some M87 clusters overlap the galaxies in the Mg$_{2}$
	  index, they have distinct FUV properties.
	  \label{f:fuvv_mg2} }
\end{figure}

\begin{figure}
\plotone{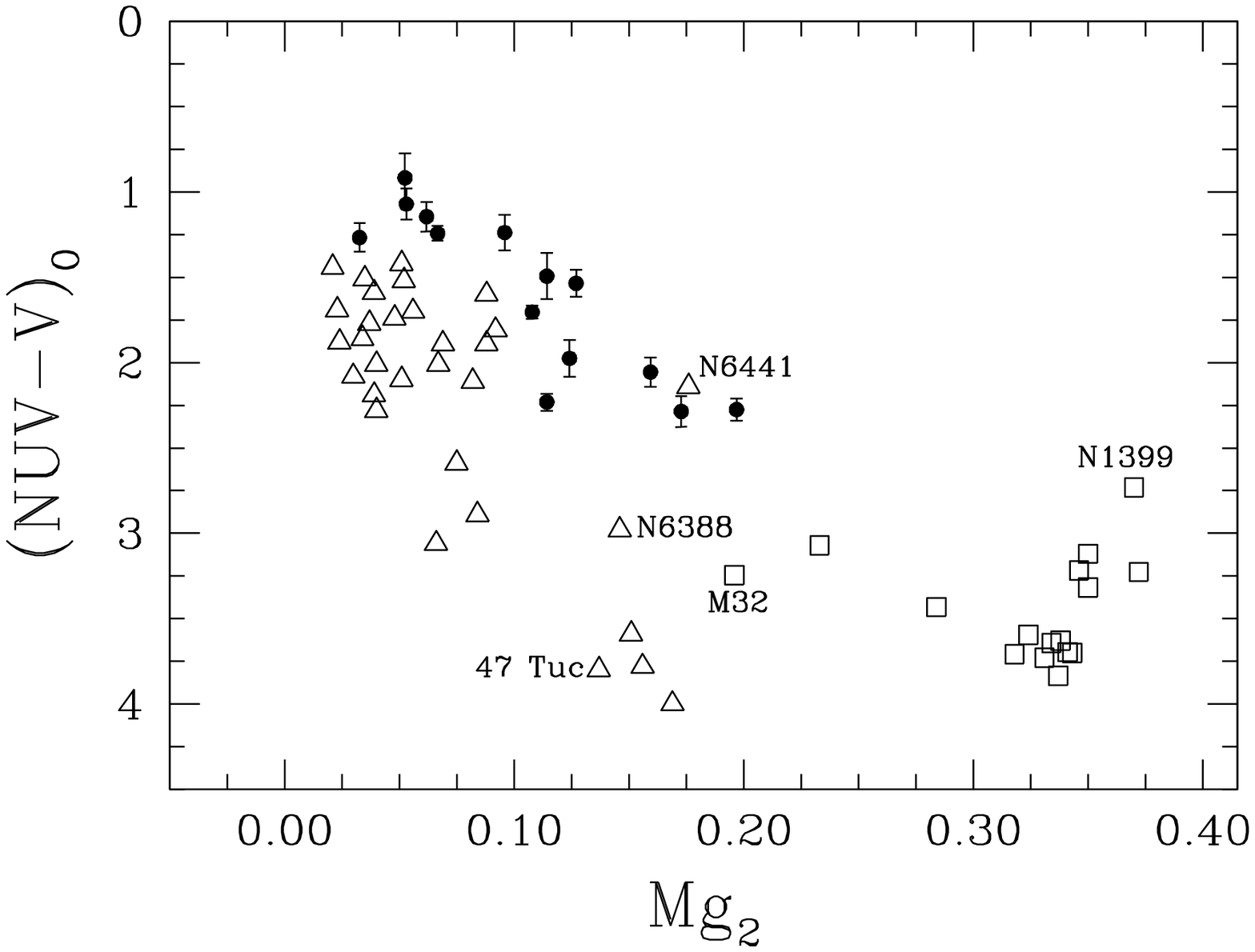} 
\caption{ The metallicity dependence of $(NUV-V)$ colors for Milky Way
	  GCs, galaxies, and M87 GCs in Field 4.  The symbols are same
	  as those of Figure \ref{f:fuvv_mg2}. The Mg$_{2}$ values for
	  M87 GCs are estimated from their $(V-I)_{0}$ colors. Only
	  clusters with $\sigma_{\scriptscriptstyle V-I} < 0.15$ are
	  plotted for clarity.
          \label{f:nuvv_mg2} }
\end{figure}

\begin{figure}
\plotone{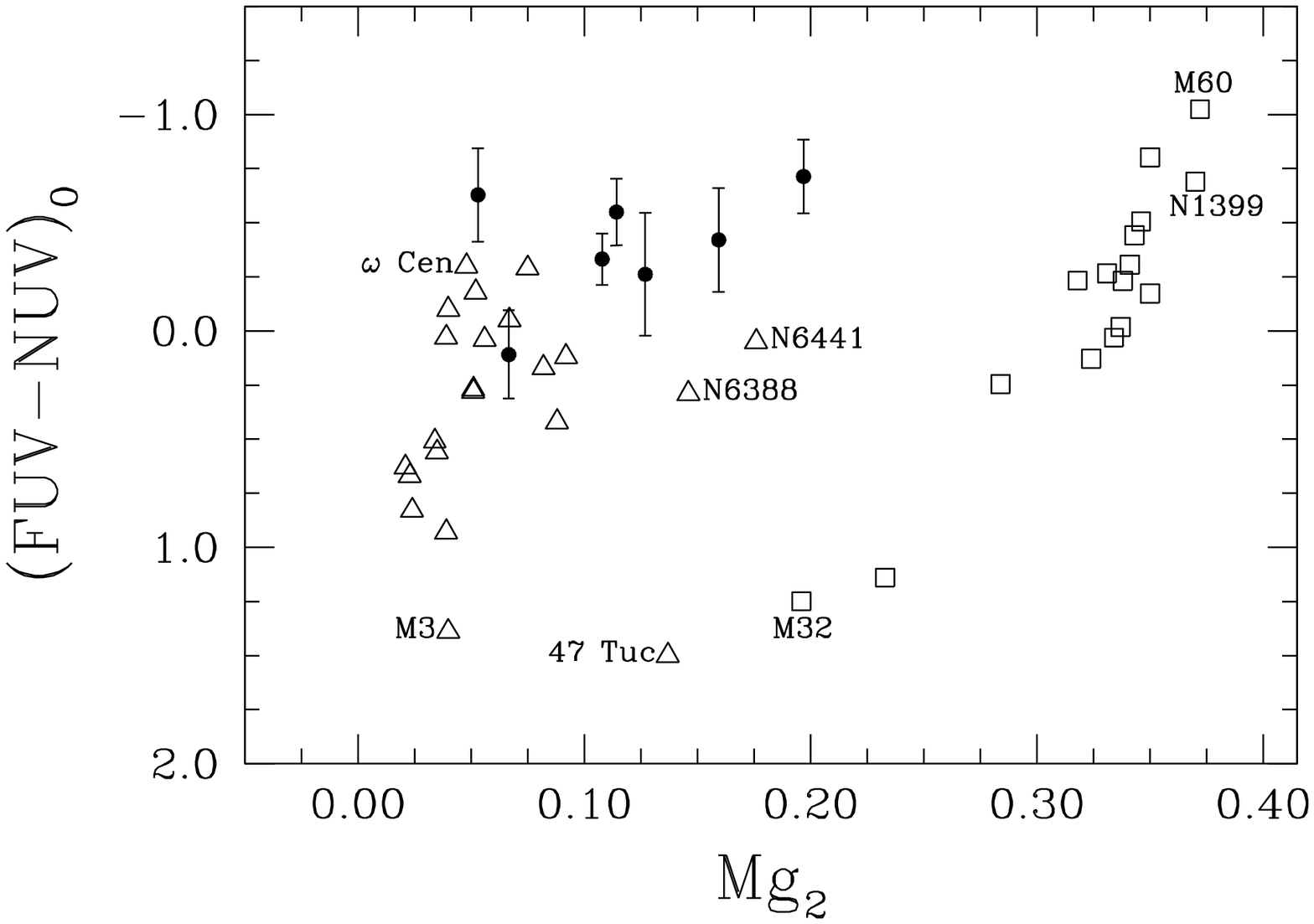} 
\caption{ The metallicity dependence of $FUV-NUV$ colors for Milky Way
	  GCs, galaxies, and M87 GCs.  The symbols are same as those
	  of Figure \ref{f:fuvv_mg2}. The Mg$_{2}$ values for M87 GCs
	  are estimated from their $(V-I)_{0}$ colors.
	  \label{f:fuvnuv_mg2} }
\end{figure}

\begin{figure}
\plotone{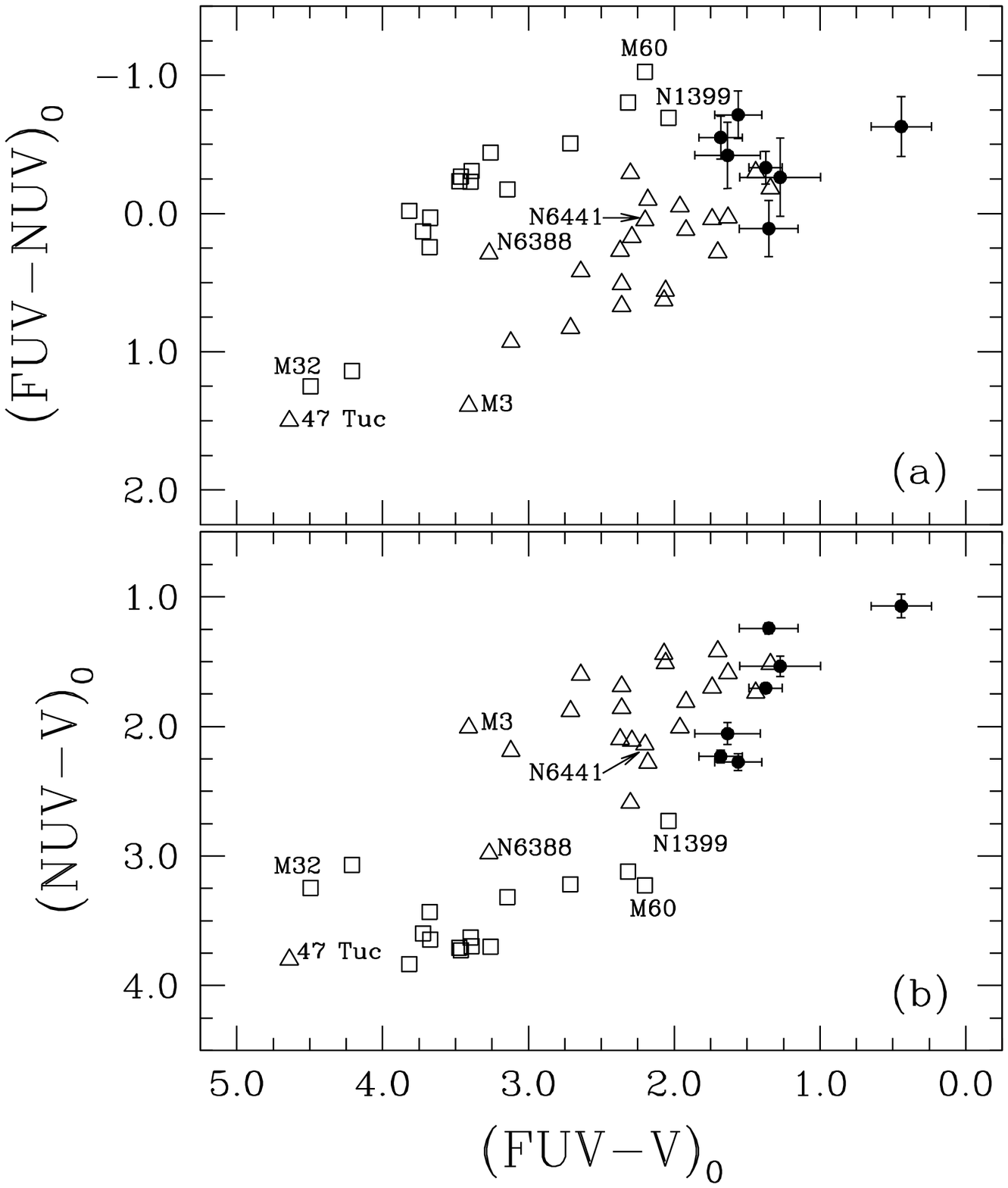} 
\caption{ Ultraviolet color-color diagrams for Milky Way GCs,
	  galaxies, and M87 GCs in (a) $(FUV-V)$ vs. $(FUV-NUV)$ and
	  (b) $(FUV-V)$ vs. $(NUV-V)$.  The symbols are same as those
	  of Figure \ref{f:fuvv_mg2}.  Bluer colors are upward and to
	  the right. The M87 cluster with the exceptionally blue
	  $(FUV-V)$ color is cluster 4002.
	  \label{f:fuv_nuv_v_colorcolor} }
\end{figure}

\begin{figure}
\epsscale{1.0}
\plotone{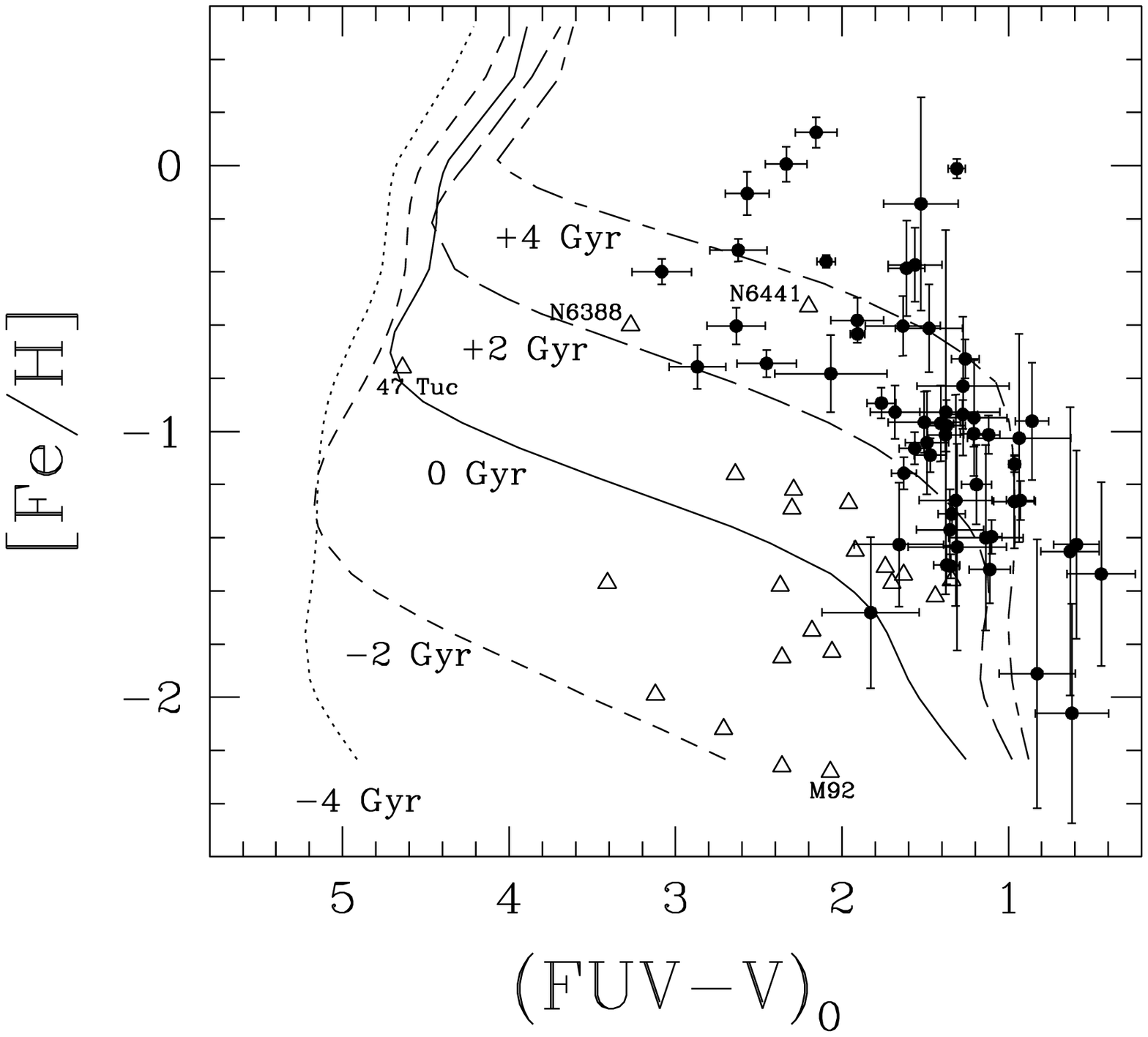} 
\caption{ The observed $(FUV-V)_{0}$ colors versus the [Fe/H] estimated
	  from optical colors for M87 clusters shown in {\it filled
	  circles}.  Milky Way GCs are shown in {\it open triangles}.
	  The theoretical isochrones are from Figure 6 of LLG02.  With
	  current cosmological parameters, the baseline ($\Delta t =
	  0$) is $\sim 12$ Gyr.  \label{f:iso} }
\end{figure}

\begin{figure}
\epsscale{1.0}
\plotone{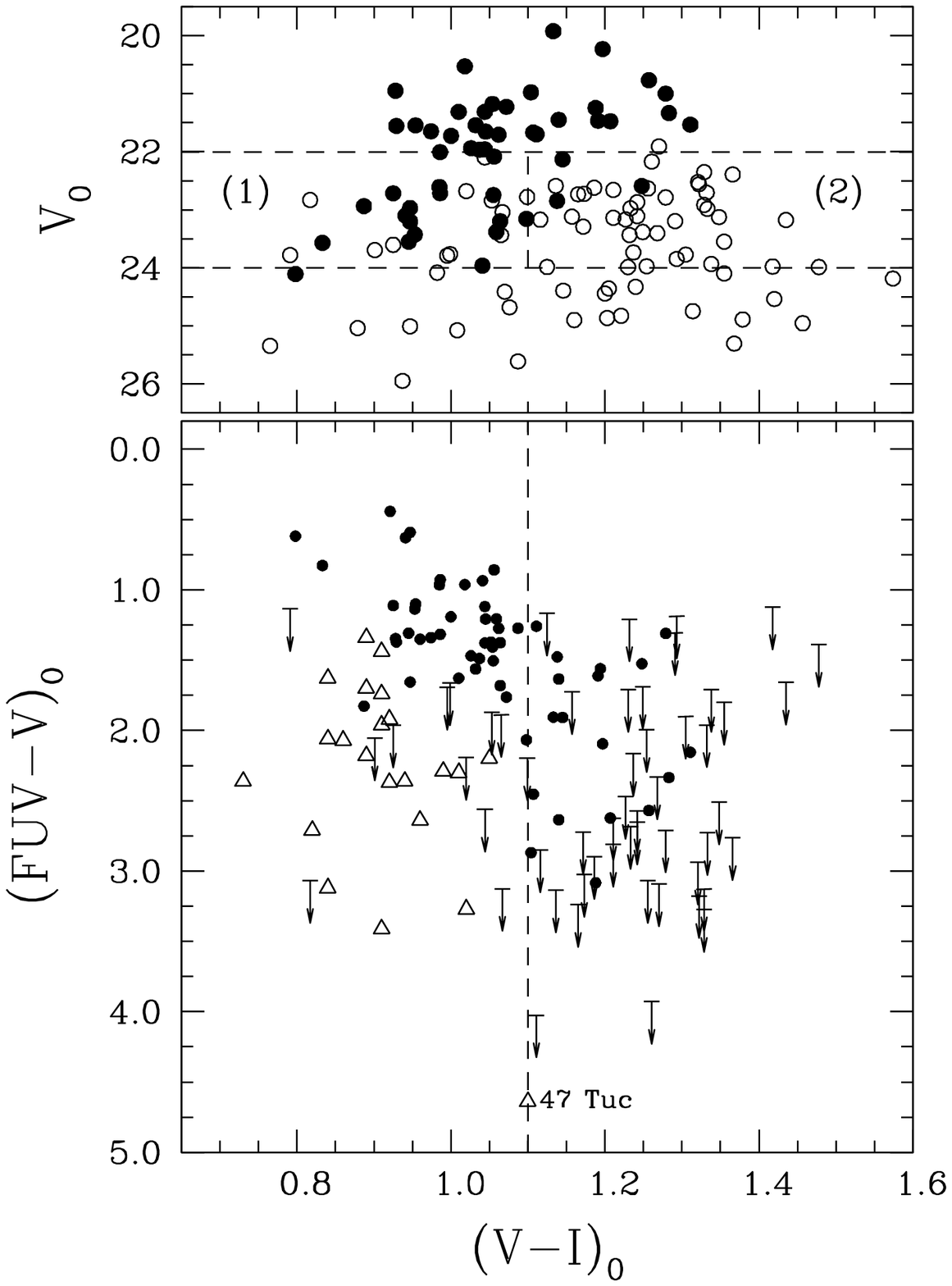} 
\caption{ Selection effects in the M87 and Milky Way samples.  {\it
	  Upper panel}: The optical color-magnitude diagram for M87
	  globular clusters in Fields 1, 2, and 3 as in Figure
	  \ref{f:fuvdetection}. The {\it open circles} are for sources
	  that are detected only in the WFPC2 optical frames while the
	  {\it filled circles} are for sources also detected in the
	  FUV.  {\it Lower panel}:  Color-color diagram for M87
	  clusters with $V_0 < 24.1$ ({\it filled circles} FUV
	  detections; {\it arrows} upper limits) and Milky Way GCs
	  {\it open triangles}.  The dividing lines are explained in
	  the text.  For reference, the 100\% FUV completeness line at
	  $V_0 = 22$ corresponds to $M_{\scriptscriptstyle V, 0} =
	  -9.1$.  \label{f:selection} }
\end{figure}

\begin{figure}
\epsscale{1.0}
\plotone{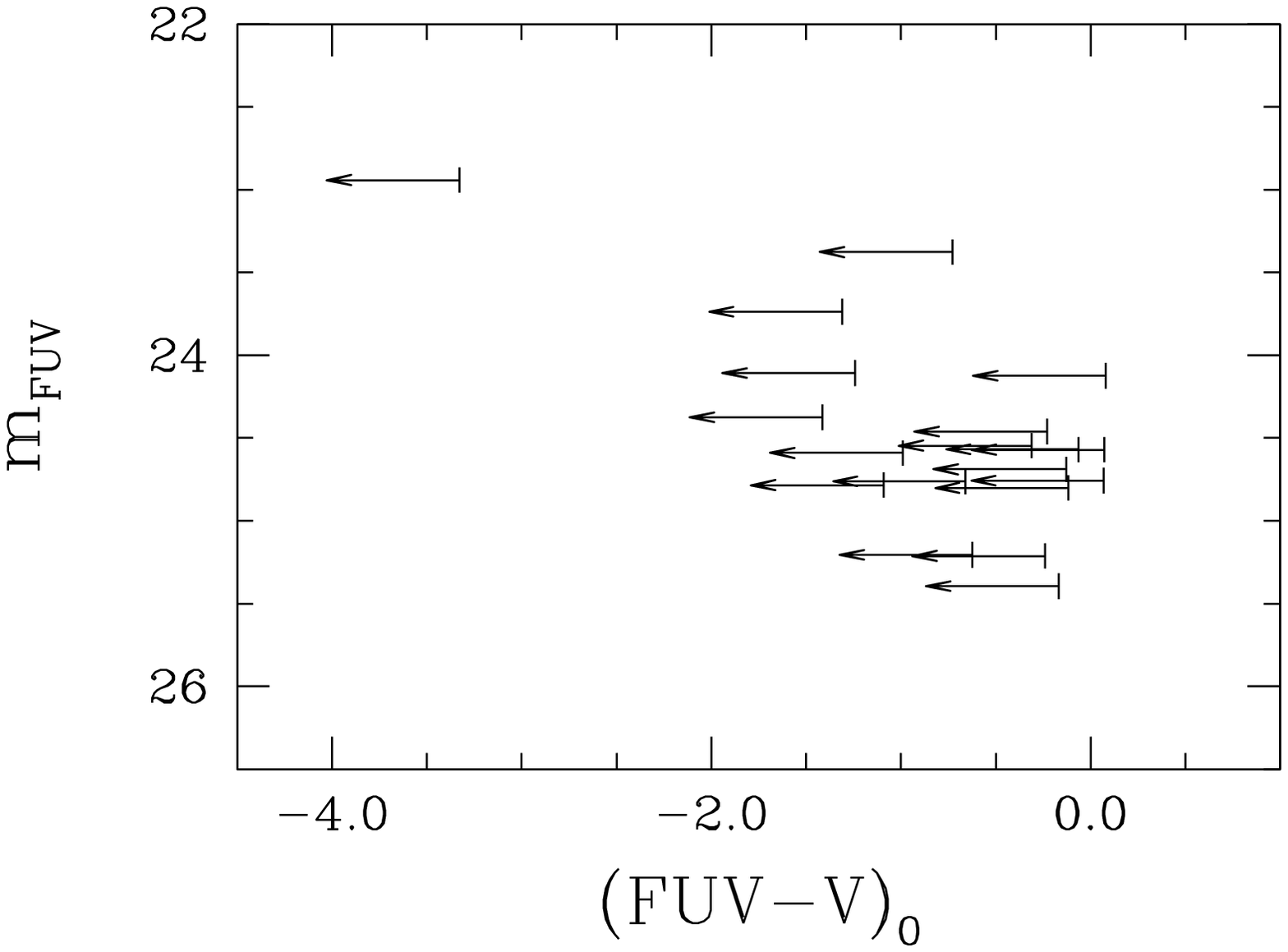}

\caption{ $(FUV-V)$ color limits of sources detected in far-UV frames
	  but not in optical frames for Fields 1, 2, and 3. The upper
	  limit in color for each source is plotted with a vertical
	  bar.  The upper limits to the $V$ magnitudes were determined
	  using the fluxes plus $3 \sigma$ when the measured fluxes
	  are positive and $3 \sigma$ when the measured fluxes are
	  negative.  The bluest color permitted for an old population
	  dominated in the UV by EHB stars is $(FUV-V)_0 \sim 0.1$ (see text).
	  \label{f:fuvonly_colorlimits} }

\end{figure}

\clearpage
\begin{figure}
\figurenum{15a}
\epsscale{1.0}
\plotone{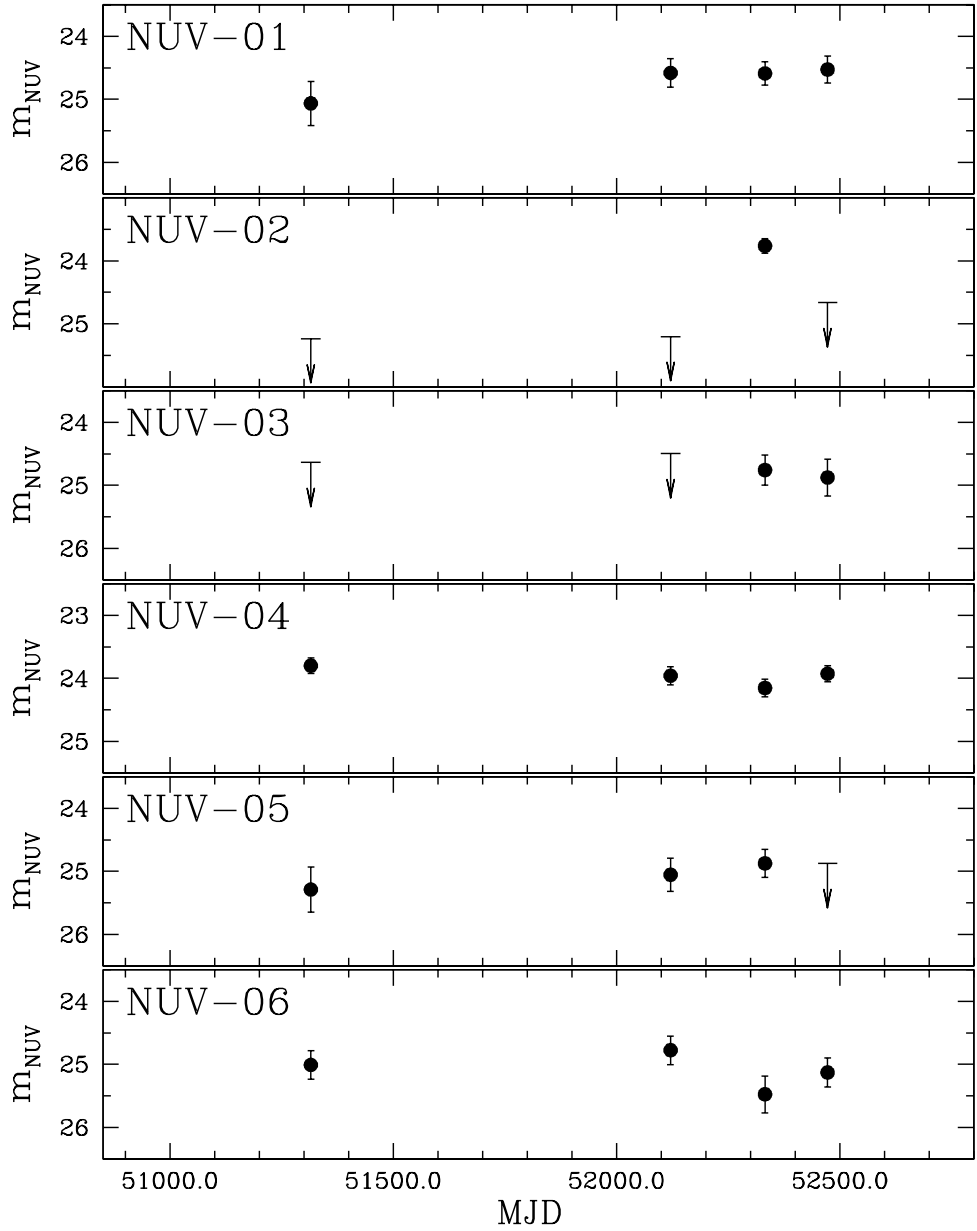} 
\caption{ Near-UV brightness for UV-only sources in Field 4 as a
	  function of modified julian date (MJD).  Brightness limits
	  for the epochs where sources were not detected are shown by
	  a horizontal bar and arrow.
	  \label{f:nuv_var} }
\end{figure}

\begin{figure}
\figurenum{15b}
\plotone{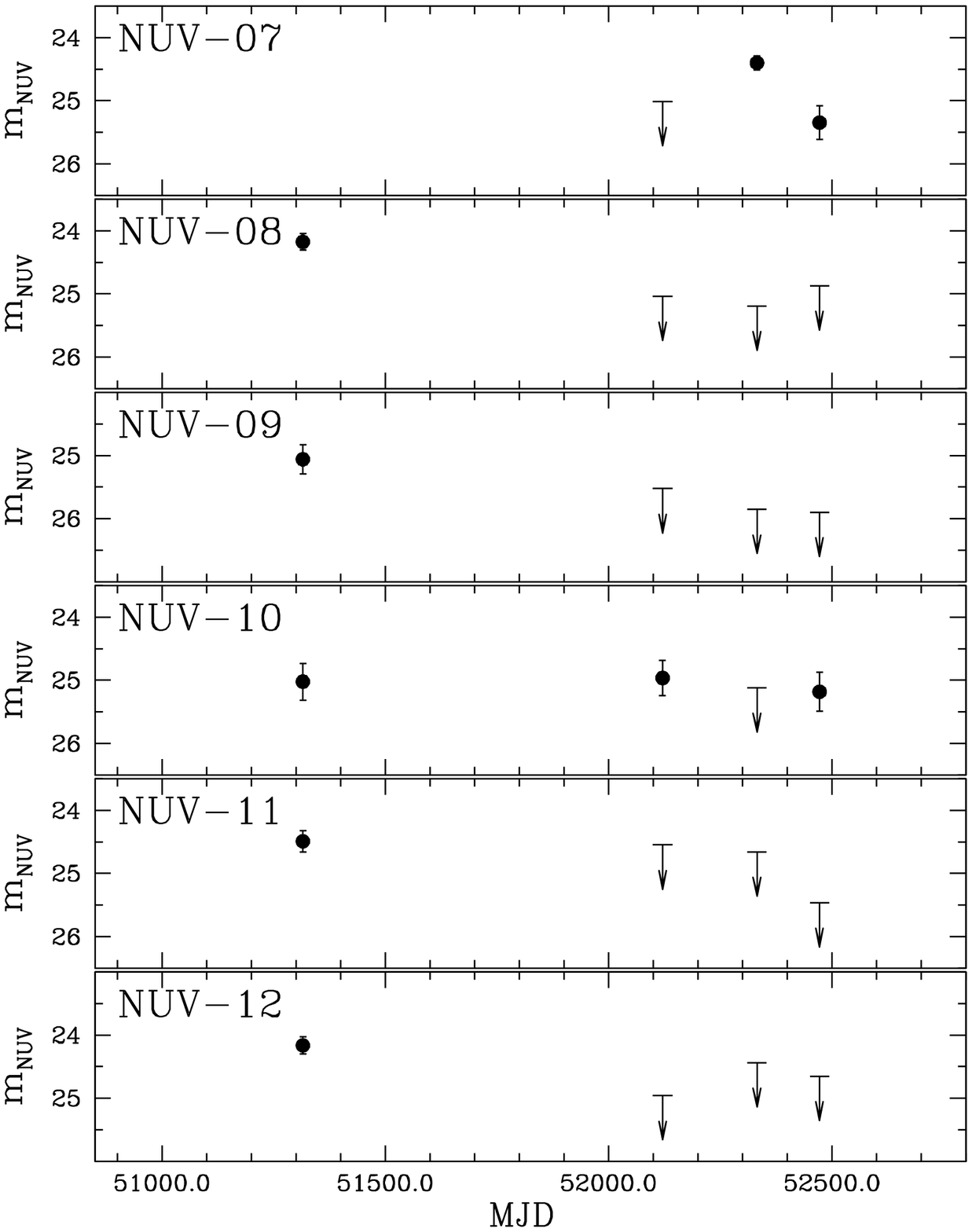} 
\caption{ continued. 
         }
\end{figure}

\begin{figure}
\figurenum{15c}
\plotone{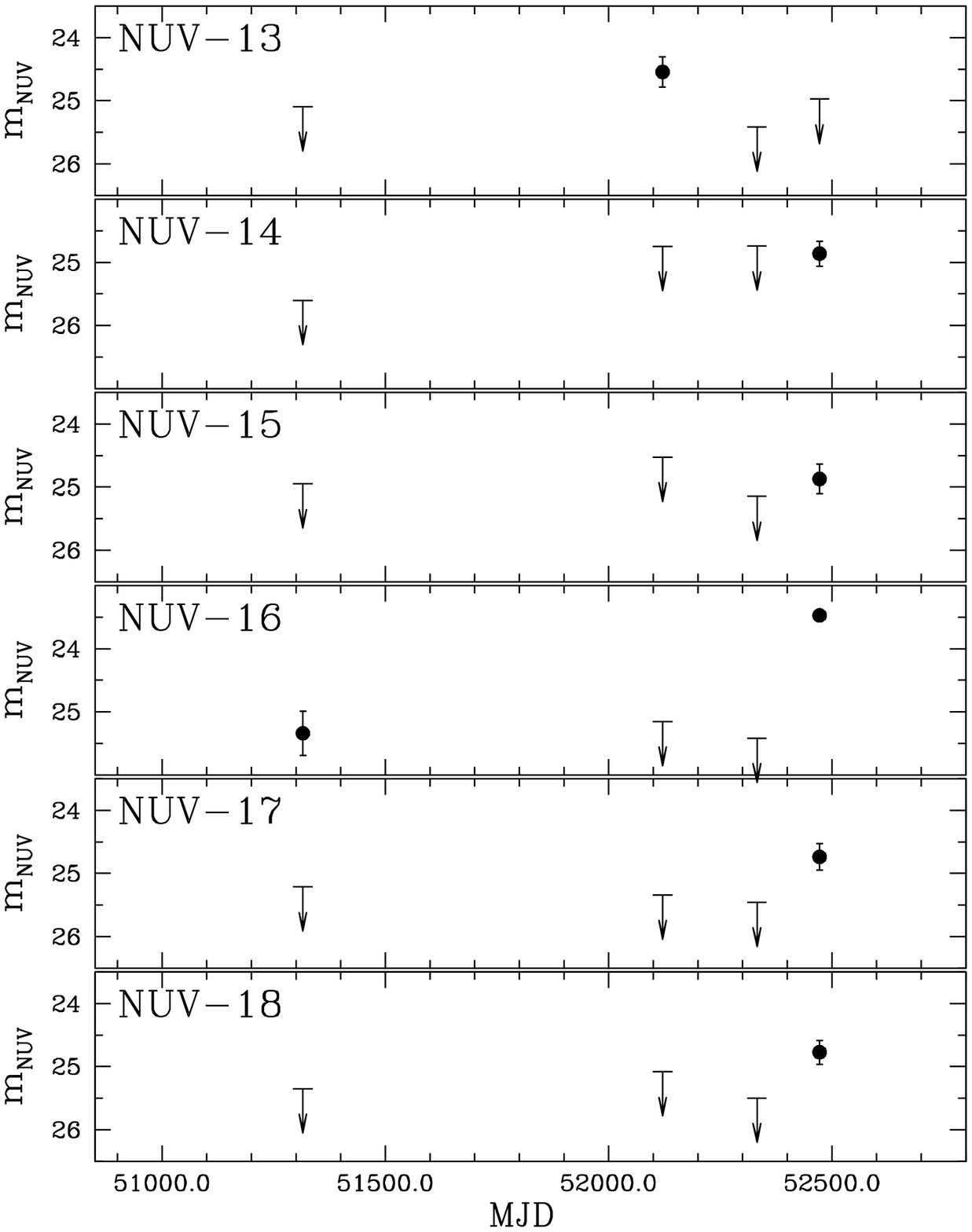} 
\caption{ continued. 
         }
\end{figure}

\begin{figure}
\figurenum{15d}
\plotone{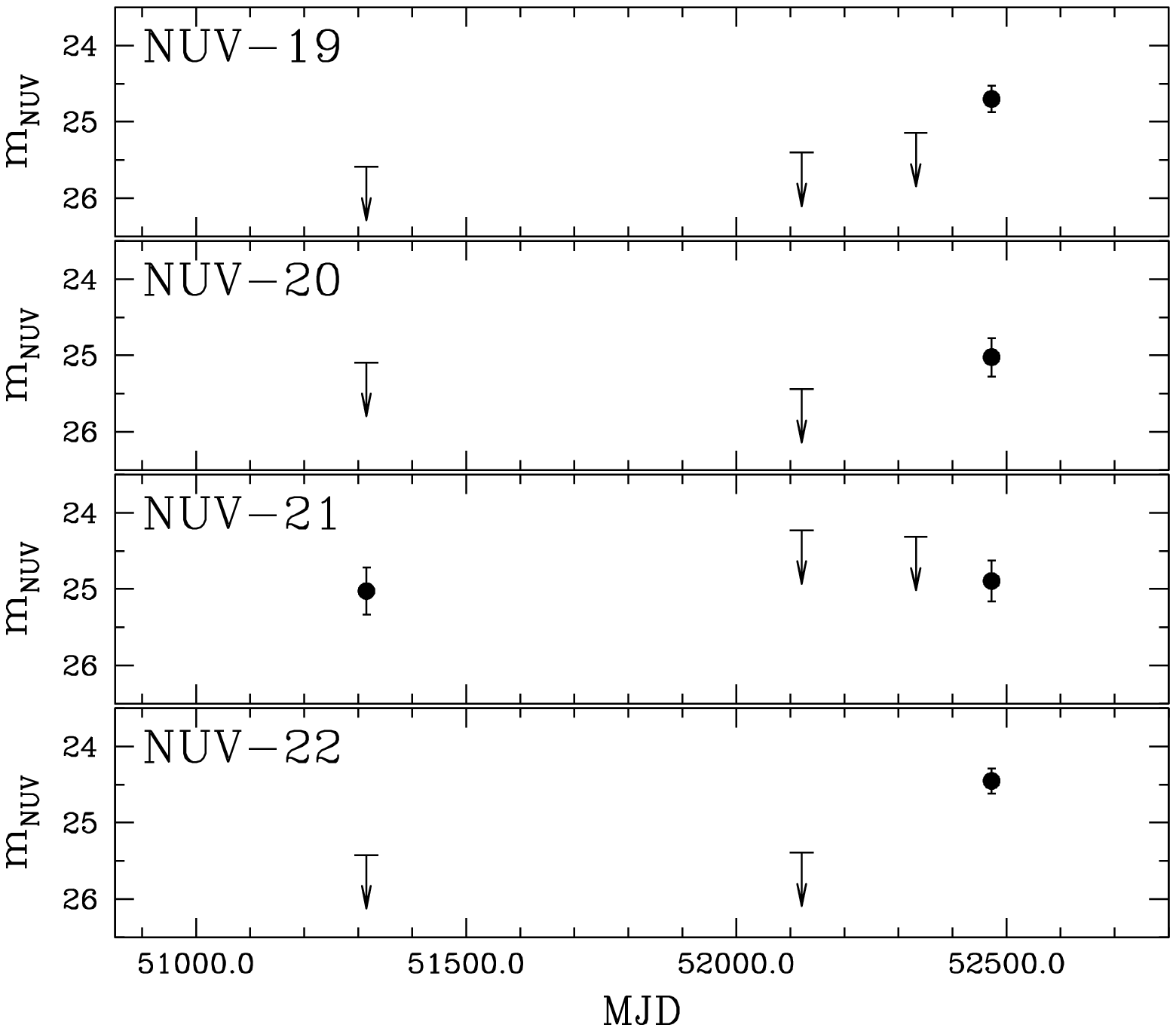} 
\caption{ continued. 
         }
\end{figure}
\setcounter{figure}{15}


\begin{figure}
\figurenum{16a}
\plotone{dummypix.eps}
\caption{ {\Large \bf SEE 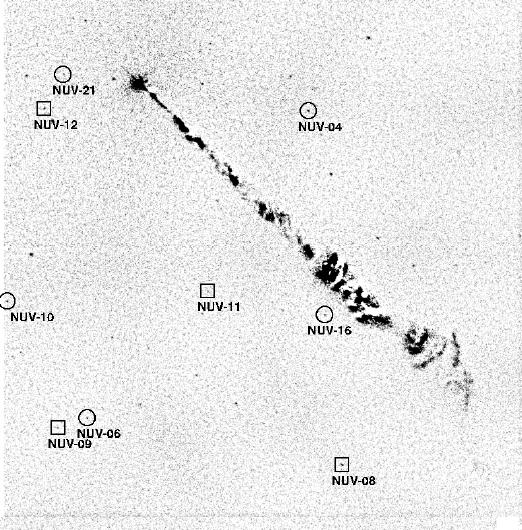}~~~~~~
          Finding chart for the UV-only sources in the near-UV frame
	  of Field 4 for GO-8140.  A median-filtered image has been subtracted
          to suppress the diffuse background.  
          Sources that are detected in only a single
	  epoch are surrounded with {\it squares} while other
	  sources are surrounded with {\it circles}. MJD$ =
	  51315.5633$.  
          \label{f:nuvonly_findingmap} }
\end{figure}

\begin{figure}
\figurenum{16b}
\plotone{dummypix.eps}
\caption{ {\Large \bf SEE 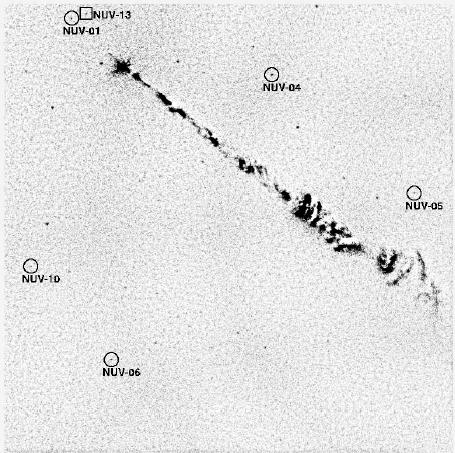}~~~~~~ 
Same as Figure 16(a) but for GO-8780. MJD$ = 52120.9713$.
         }
\end{figure}

\begin{figure}
\figurenum{16c}
\plotone{dummypix.eps}
\caption{ {\Large \bf SEE 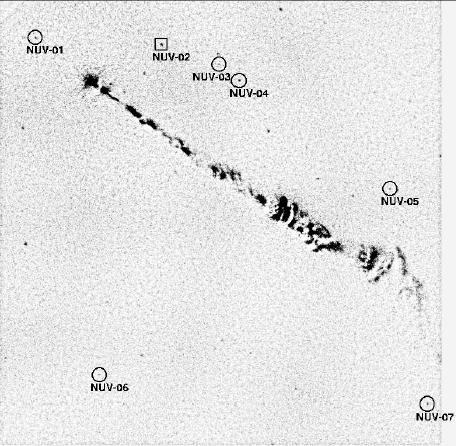}~~~~~~
Same as Figure 16(a) but for GO-8048. MJD$ = 52332.2047$.
         }
\end{figure}

\begin{figure}
\figurenum{16d}
\plotone{dummypix.eps}
\caption{ {\Large \bf SEE 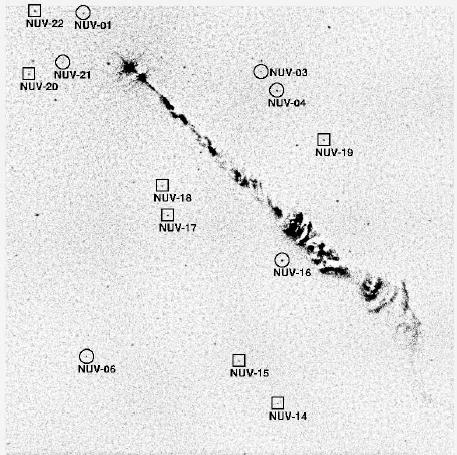}~~~~~~
Same as Figure 16(a) but for GO-9461. MJD$ = 52472.3710$.
         }
\end{figure}
\setcounter{figure}{16}

\begin{figure}
\epsscale{1.0}
\plotone{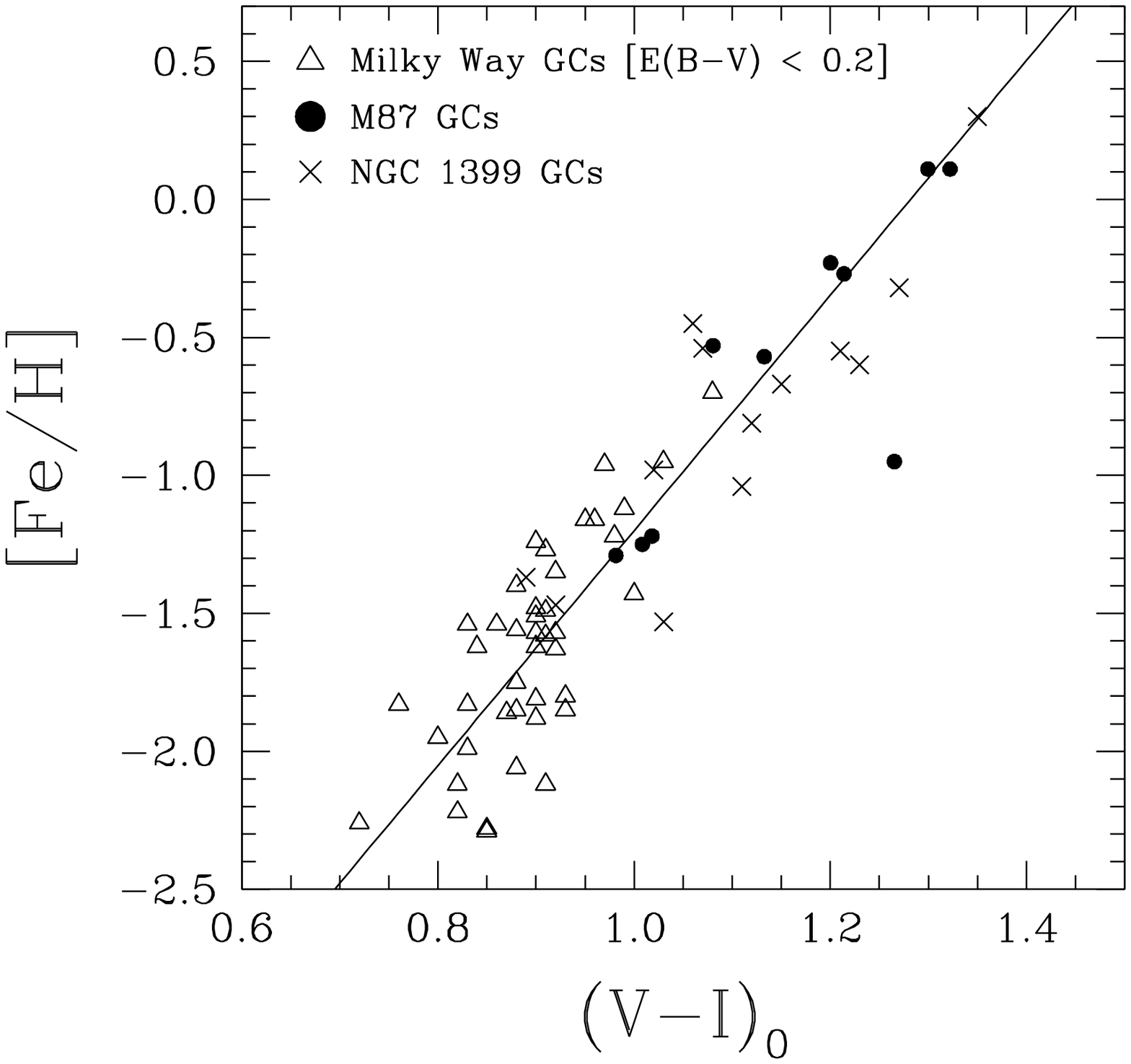} 
\caption{ Correlation between ($V-I$) color and [Fe/H] for Milky Way
	  globular clusters with $E(B-V)<0.2$ ({\it open triangles}),
	  M87 clusters ({\it closed circles}), and NGC 1399 clusters
	  ({\it crosses}).  The solid line is a result of fitting the
	  data points with the bisector method as described in
	  \citet{Isobe90}.  Details given in Appendix B.
	  \label{f:vifeh_conversion} }
\end{figure}



\begin{deluxetable}{cccccc}
\tablewidth{0pt}
\tabletypesize{\footnotesize}
\tablecaption{ Observation log \label{t:obslog}}
\tablehead{ \colhead{} & \colhead{} & \colhead{} & \colhead{} & \colhead{Exp} & \colhead{Start} \\
            \colhead{Dataset} & \colhead{Detector} & \colhead{Filter} & \colhead{Mode\tablenotemark{a}} & \colhead{Time\tablenotemark{b}} & \colhead{Time\tablenotemark{c}}
          }
\startdata
\tableline
\multicolumn{6}{c}{Field 1 -- RA $12^{\rm h}30^{\rm m}49\fs 83$, Dec +12\arcdeg 22\arcmin 50\farcs 38, $r = 38\farcs 1$} \\
\tableline
  o6be01oqq & FUV-MAMA & F25SRF2  & TIMETAG & 2680 & 2001-04-27 ~20:36:28 \\*
  o6be01osq & FUV-MAMA & F25SRF2  & TIMETAG & 2900 & 2001-04-27 ~22:08:09 \\*
  o6be02010 & CCD      & F28X50LP & ACCUM   &  518 & 2001-04-23 ~18:39:19 \\*
  o6be02ceq & FUV-MAMA & F25SRF2  & TIMETAG & 2000 & 2001-04-23 ~18:52:00 \\*
  o6be02cgq & FUV-MAMA & F25SRF2  & TIMETAG & 2900 & 2001-04-23 ~20:12:30 \\*
\tableline
\multicolumn{6}{c}{Field 2 -- RA $12^{\rm h}30^{\rm m}53\fs 12$, Dec +12\arcdeg 24\arcmin 15\farcs 10, $r = 72\farcs 8$} \\*
\tableline
  o6be03xzq & FUV-MAMA & F25SRF2  & TIMETAG & 2680 & 2002-06-12 ~01:10:06 \\*
  o6be03ygq & FUV-MAMA & F25SRF2  & TIMETAG & 2900 & 2002-06-12 ~02:40:57 \\*
  o6be04010 & CCD      & F28X50LP & ACCUM   &  518 & 2002-06-12 ~04:20:04 \\*
  o6be04z4q & FUV-MAMA & F25SRF2  & TIMETAG & 2000 & 2002-06-12 ~04:33:45 \\*
  o6be04znq & FUV-MAMA & F25SRF2  & TIMETAG & 2900 & 2002-06-12 ~05:53:18 \\*
\tableline
\multicolumn{6}{c}{Field 3 -- RA $12^{\rm h}30^{\rm m}50\fs 64$, Dec +12\arcdeg 23\arcmin 40\farcs 84, $r = 22\farcs 4$} \\*
\tableline
  o6be06010 & CCD      & F28X50LP & ACCUM   &  518 & 2002-06-08 ~12:10:52 \\*
  o6be06omq & FUV-MAMA & F25SRF2  & TIMETAG & 2000 & 2002-06-08 ~12:24:33 \\*
  o6be06p3q & FUV-MAMA & F25SRF2  & TIMETAG & 2900 & 2002-06-08 ~13:43:21 \\*
  o6be15m9q & FUV-MAMA & F25SRF2  & TIMETAG & 2680 & 2002-02-14 ~16:47:22 \\*
  o6be15mbq & FUV-MAMA & F25SRF2  & TIMETAG & 2900 & 2002-02-14 ~18:19:24 \\*
\tableline
\multicolumn{6}{c}{Field 4 -- RA $12^{\rm h}30^{\rm m}48\fs 78$, Dec +12\arcdeg 23\arcmin 30\farcs 59, $r = 10\farcs 0$} \\*
\tableline
  o5dc01bpq & FUV-MAMA & F25SRF2  & ACCUM   &  856 & 1999-05-17 ~10:33:00 \\*
  o5dc01bqq & FUV-MAMA & F25SRF2  & ACCUM   &  850 & 1999-05-17 ~10:48:00 \\*
  o5dc01btq & FUV-MAMA & F25SRF2  & ACCUM   &  910 & 1999-05-17 ~12:05:00 \\*
  o5dc01buq & FUV-MAMA & F25SRF2  & ACCUM   &  910 & 1999-05-17 ~12:21:00 \\*
  o5dc01bxq & NUV-MAMA & F25QTZ   & ACCUM   &  614 & 1999-05-17 ~13:31:00 \\*
  o5dc01byq & NUV-MAMA & F25QTZ   & ACCUM   &  600 & 1999-05-17 ~13:42:00 \\*
  o5dc01c0q & NUV-MAMA & F25QTZ   & ACCUM   &  600 & 1999-05-17 ~13:53:00 \\*
  o5dc01c2q & NUV-MAMA & F25QTZ   & ACCUM   &  600 & 1999-05-17 ~14:03:00 \\*
  o6a351evq & NUV-MAMA & F25QTZ   & ACCUM   &  600 & 2001-07-30 ~23:18:00 \\*
  o6a351ewq & NUV-MAMA & F25QTZ   & ACCUM   &  600 & 2001-07-30 ~23:29:00 \\*
  o6a351eyq & NUV-MAMA & F25QTZ   & ACCUM   &  558 & 2001-07-30 ~23:40:00 \\*
  o6a351f0q & NUV-MAMA & F25QTZ   & ACCUM   &  600 & 2001-07-30 ~23:50:00 \\*
  o53c61keq & NUV-MAMA & F25QTZ   & ACCUM   &  600 & 2002-02-27 ~04:54:00 \\*
  o53c61kfq & NUV-MAMA & F25QTZ   & ACCUM   &  600 & 2002-02-27 ~05:53:00 \\*
  o53c61khq & NUV-MAMA & F25QTZ   & ACCUM   &  558 & 2002-02-27 ~06:04:00 \\*
  o53c61kjq & NUV-MAMA & F25QTZ   & ACCUM   &  600 & 2002-02-27 ~06:14:00 \\*
  o6m201w4q & NUV-MAMA & F25QTZ   & ACCUM   &  600 & 2002-07-17 ~08:54:00 \\*
  o6m201w5q & NUV-MAMA & F25QTZ   & ACCUM   &  600 & 2002-07-17 ~09:05:00 \\*
  o6m201w7q & NUV-MAMA & F25QTZ   & ACCUM   &  600 & 2002-07-17 ~09:15:00 \\*
  o6m201w9q & NUV-MAMA & F25QTZ   & ACCUM   &  600 & 2002-07-17 ~09:25:00 
\enddata
\tablenotetext{a}{Refer to text for TIMETAGE mode. ACCUM mode is the normal time-integration mode for imaging.}
\tablenotetext{b}{Exposure times are in seconds.}
\tablenotetext{c}{Start times are in UT and in the format of Year-Month-Day Hour:Minute:Second.}
\end{deluxetable}


\begin{deluxetable}{lcc}
\tablewidth{0pt}
\tablecaption{ Final integration times and PHOTFLAM values used in the photometric conversion \label{t:photflam}}
\tablehead { \colhead{}      & \colhead{Integration Time} & \colhead{PHOTFLAM} \\
             \colhead{Field} & \colhead{(seconds)}        & \colhead{(ergs s$^{-1}$ cm$^{-2}$ \AA$^{-1}$/[counts s$^{-1}$])}
           }
\startdata
1      & 8150. & $4.1153 \times 10^{-17}$ \\
2      & 7650. & $4.2221 \times 10^{-17}$ \\
3      & 6600. & $4.2079 \times 10^{-17}$ \\ 
4 FUV  & 3526. & $3.9437 \times 10^{-17}$ \\
4 NUV  & 2414. & $5.8723 \times 10^{-18}$ \\
4a NUV & 2358. & $6.1125 \times 10^{-18}$ \\
4b NUV & 2358. & $6.0602 \times 10^{-18}$ \\
4c NUV & 2358. & $6.1477 \times 10^{-18}$ 
\enddata
\end{deluxetable}



\begin{deluxetable}{crrccccccccc}
\rotate
\tabletypesize{\footnotesize}
\tablewidth{0pt}
\tablecaption{ Photometric Catalog for Field 1, 2, and 3\label{t:catalog1}}
\tablehead { \colhead{}                    & \colhead{$\Delta$RA\tablenotemark{b}} & \colhead{$\Delta$Dec\tablenotemark{b}} & \colhead{$r$\tablenotemark{c}} & \colhead{}              & \colhead{}                   & \colhead{}                     & \colhead{}                              & \colhead{}                     & \colhead{} \\
             \colhead{ID\tablenotemark{a}} & \colhead{(arcsec)}                     & \colhead{(arcsec)}                      & \colhead{(arcsec)}             & \colhead{$m_{\scriptscriptstyle FUV}$\tablenotemark{d}} & \colhead{$\sigma_{\scriptscriptstyle FUV}$} & \colhead{$(FUV-V)_{0}$\tablenotemark{e}} & \colhead{$\sigma_{FUV-V}$} & \colhead{$(V-I)_{0}$\tablenotemark{e}} & \colhead{$\sigma_{\scriptscriptstyle V-I}$} 
           }
\startdata
  1001 & $-5.15$ &  $-38.36$ &   38.71 & \phm{$>$}22.830 &    0.063 & \phm{$->$}1.101   &    0.064 &    0.954 &    0.015 \\ 
  1002 & $-3.32$ &  $-38.58$ &   38.72 & \phm{$>$}25.040 &    0.267 & \phm{$->$}1.656   &    0.269 &    0.947 &    0.055 \\ 
  1003 & $-2.71$ &  $-35.03$ &   35.14 & \phm{$>$}24.744 &    0.218 & \phm{$->$}1.136   &    0.224 &    0.953 &    0.082 \\ 
  1004 & $-2.61$ &  $-43.85$ &   43.92 & \phm{$>$}22.493 &    0.052 & \phm{$->$}1.310   &    0.052 &    1.279 &    0.008 \\ 
  1005 & $-1.40$ &  $-36.82$ &   36.85 & \phm{$>$}23.121 &    0.074 & \phm{$->$}1.628   &    0.075 &    1.010 &    0.014 \\ 
  1006 &    0.39 &  $-32.06$ &   32.06 & \phm{$>$}22.477 &    0.052 & \phm{$->$}1.347   &    0.052 &    0.928 &    0.011 \\ 
  1007 &    2.12 &  $-47.37$ &   47.42 & \phm{$>$}24.511 &    0.179 & \phm{$->$}3.083   &    0.179 &    1.188 &    0.011 \\ 
  1008 &    4.66 &  $-40.64$ &   40.91 & \phm{$>$}24.505 &    0.201 & \phm{$->$}1.477   &    0.203 &    1.138 &    0.039 \\ 
  1009 &    6.63 &  $-42.78$ &   43.29 & \phm{$>$}24.220 &    0.159 & \phm{$->$}1.908   &    0.160 &    1.145 &    0.020 \\ 
  1010 &    0.01 &  $-29.88$ &   31.51 & \phm{$>$}23.110 &    0.078 & \phm{$->$}1.371   &    0.079 &    0.929 &    0.017 \\ 
  1011 &    0.13 &  $-30.71$ &   32.34 & \phm{$>$}24.948 &    0.290 & \phm{$->$}1.827   &    0.293 &    0.887 &    0.067 \\ 
  1012 &    0.44 &  $-35.13$ &   36.65 & \phm{$>$}23.869 &    0.126 & \phm{$->$}2.155   &    0.126 &    1.311 &    0.013 \\ 
  1013 &   11.31 &  $-25.52$ &   27.91 & \phm{$>$}23.852 &    0.124 & \phm{$->$}2.335   &    0.124 &    1.283 &    0.016 \\ 
  1014 &   11.97 &  $-34.06$ &   36.10 & \phm{$>$}23.138 &    0.083 & \phm{$->$}1.259   &    0.084 &    1.111 &    0.017 \\ 
  1015 &   12.09 &  $-34.96$ &   36.99 & \phm{$>$}22.509 &    0.056 & \phm{$->$}2.095   &    0.056 &    1.197 &    0.006 \\ 
  1016 &   13.23 &  $-41.54$ &   43.59 & \phm{$>$}23.286 &    0.091 & \phm{$->$}1.563   &    0.091 &    1.032 &    0.014 \\ 
  1017 &   13.24 &  $-45.21$ &   47.11 & \phm{$>$}25.078 &    0.304 & \phm{$->$}0.935   &    0.310 &    1.041 &    0.092 \\ 
  1018 &   14.34 &  $-32.94$ &   35.93 & \phm{$>$}21.675 &    0.035 & \phm{$->$}0.963   &    0.035 &    1.018 &    0.007 \\ 
  1101 & $-4.24$ &  $-35.82$ &   36.07 &       $>25.960$ &  \nodata & \phm{$-$}$>2.042$ &  \nodata &    1.237 &    0.091 \\ 
  1102 & $-4.07$ &  $-40.26$ &   40.47 &       $>26.051$ &  \nodata & \phm{$-$}$>2.698$ &  \nodata &    1.116 &    0.049 \\ 
  1103 & $-3.97$ &  $-43.17$ &   43.35 &       $>25.250$ &  \nodata & \phm{$-$}$>0.985$ &  \nodata &    0.982 &    0.111 \\ 
  1104 & $-0.84$ &  $-49.84$ &   49.85 &       $>25.538$ &  \nodata & \phm{$-$}$>2.736$ &  \nodata &    1.186 &    0.031 \\ 
  1105 & $-0.82$ &  $-28.90$ &   28.91 &       $>25.014$ &  \nodata & \phm{$-$}$>2.925$ &  \nodata &    1.270 &    0.023 \\ 
  1106 & $-0.53$ &  $-50.19$ &   50.19 &       $>26.036$ &  \nodata & \phm{$-$}$>1.879$ &  \nodata &    1.254 &    0.088 \\ 
  1107 &    0.08 &  $-35.11$ &   35.11 &       $>25.776$ &  \nodata & \phm{$-$}$>2.870$ &  \nodata &    1.173 &    0.038 \\ 
  1108 &    1.35 &  $-49.49$ &   49.51 &       $>26.113$ &  \nodata & \phm{$-$}$>3.760$ &  \nodata &    1.261 &    0.021 \\ 
  1109 &    1.36 &  $-47.01$ &   47.03 &       $>25.742$ &  \nodata & \phm{$-$}$>2.574$ &  \nodata &    1.333 &    0.040 \\ 
  1110 &    3.35 &  $-44.72$ &   44.84 &       $>25.215$ &  \nodata & \phm{$-$}$>1.047$ &  \nodata &    1.125 &    0.093 \\ 
  1111 &    3.63 &  $-38.11$ &   38.28 &       $>25.745$ &  \nodata & \phm{$-$}$>0.554$ &  \nodata &    0.947 &    0.287 \\ 
  1112 &    5.54 &  $-41.31$ &   41.68 &       $>25.523$ &  \nodata & \phm{$-$}$>2.554$ &  \nodata &    1.279 &    0.035 \\ 
  1113 &    5.59 &  $-44.50$ &   44.85 &       $>25.490$ &  \nodata & \phm{$-$}$>2.650$ &  \nodata &    1.211 &    0.031 \\ 
  1114 &    7.93 &  $-34.81$ &   35.70 &       $>25.938$ &  \nodata & \phm{$-$}$>1.345$ &  \nodata &    1.070 &    0.168 \\ 
  1115 &    8.58 &  $-32.27$ &   33.39 &       $>25.692$ &  \nodata & \phm{$-$}$>2.535$ &  \nodata &    1.233 &    0.048 \\ 
  1116 &    8.74 &  $-41.19$ &   42.11 &       $>25.742$ &  \nodata & \phm{$-$}$>2.973$ &  \nodata &    1.136 &    0.030 \\ 
  1117 &    9.16 &  $-35.31$ &   36.48 &       $>24.982$ &  \nodata & \phm{$-$}$>1.021$ &  \nodata &    0.791 &    0.134 \\ 
  1118 &   10.67 &  $-37.76$ &   39.24 &       $>25.724$ &  \nodata & \phm{$-$}$>2.909$ &  \nodata &    1.256 &    0.030 \\ 
  1119 &   11.27 &  $-29.87$ &   31.93 &       $>25.760$ &  \nodata & \phm{$-$}$>3.020$ &  \nodata &    1.322 &    0.035 \\ 
  1120 &   11.34 &  $-23.98$ &   26.52 &       $>25.690$ &  \nodata & \phm{$-$}$>2.341$ &  \nodata &    1.227 &    0.081 \\ 
  1121 &   11.35 &  $-33.28$ &   35.16 &       $>25.464$ &  \nodata & \phm{$-$}$>0.419$ &  \nodata &    1.203 &    0.282 \\ 
  1122 &   11.97 &  $-34.06$ &   36.10 &       $>25.733$ &  \nodata & \phm{$-$}$>3.854$ &  \nodata &    1.111 &    0.017 \\ 
  1123 &   14.32 &  $-27.90$ &   31.36 &       $>25.807$ &  \nodata & \phm{$-$}$>2.487$ &  \nodata &    1.211 &    0.062 \\ 
  1124 &   14.52 &  $-31.03$ &   34.26 &       $>25.471$ &  \nodata & \phm{$-$}$>2.422$ &  \nodata &    1.242 &    0.047 \\ 
  1125 &   14.74 &  $-37.36$ &   40.16 &       $>25.170$ &  \nodata & \phm{$-$}$>1.010$ &  \nodata &    1.418 &    0.091 \\ 
  1126 &   18.45 &  $-39.05$ &   43.19 &       $>25.807$ &  \nodata & \phm{$-$}$>1.930$ &  \nodata &    0.901 &    0.102 \\ 
  1127 &   18.69 &  $-42.26$ &   46.21 &       $>25.779$ &  \nodata & \phm{$-$}$>1.205$ &  \nodata &    1.146 &    0.155 \\ 
  1128 &   19.18 &  $-41.89$ &   46.07 &       $>25.403$ &  \nodata & \phm{$-$}$>1.671$ &  \nodata &    1.355 &    0.069 \\ 
  1201 & $-4.34$ &  $-38.83$ &   39.07 & \phm{$>$}24.944 &    0.253 & $<-0.658$         &  \nodata &  \nodata &  \nodata \\ 
  1202 & $-4.25$ &  $-36.69$ &   36.93 & \phm{$>$}25.395 &    0.336 & $<-0.239$         &  \nodata &  \nodata &  \nodata \\ 
  1203 & $-3.26$ &  $-44.09$ &   44.21 & \phm{$>$}24.940 &    0.241 & \phm{$-$}$<0.071$ &  \nodata &  \nodata &  \nodata \\ 
  1204 & $-0.91$ &  $-41.06$ &   41.07 & \phm{$>$}24.773 &    0.214 & $<-0.989$         &  \nodata &  \nodata &  \nodata \\ 
  1205 &    2.28 &  $-31.90$ &   31.98 & \phm{$>$}23.921 &    0.124 & $<-1.311$         &  \nodata &  \nodata &  \nodata \\ 
  1206 &    4.20 &  $-49.31$ &   49.49 & \phm{$>$}25.388 &    0.328 & $<-0.624$         &  \nodata &  \nodata &  \nodata \\ 
  1207 &    9.06 &  $-31.39$ &   32.67 & \phm{$>$}24.752 &    0.246 & $<-0.062$         &  \nodata &  \nodata &  \nodata \\ 
  1208 &   12.72 &  $-34.03$ &   36.33 & \phm{$>$}24.729 &    0.264 & $<-0.312$         &  \nodata &  \nodata &  \nodata \\ 
  1209 &   14.22 &  $-28.78$ &   32.10 & \phm{$>$}24.291 &    0.170 & $<-1.241$         &  \nodata &  \nodata &  \nodata \\ 
\\
\hline
\\
  2001 &   44.34 &     45.39 &   63.45 & \phm{$>$}24.012 &    0.121 & \phm{$->$}1.112   &    0.122 &    0.925 &    0.030 \\ 
  2002 &   50.64 &     55.81 &   75.36 & \phm{$>$}22.011 &    0.042 & \phm{$->$}1.906   &    0.042 &    1.133 &    0.004 \\ 
  2003 &   52.51 &     35.20 &   63.21 & \phm{$>$}24.771 &    0.197 & \phm{$->$}1.207   &    0.200 &    1.059 &    0.051 \\ 
  2004 &   53.77 &     48.32 &   72.29 & \phm{$>$}23.168 &    0.081 & \phm{$->$}1.340   &    0.081 &    0.974 &    0.013 \\ 
  2005 &   56.01 &     59.09 &   81.42 & \phm{$>$}23.591 &    0.095 & \phm{$->$}1.469   &    0.096 &    1.026 &    0.015 \\ 
  2006 &   57.50 &     56.72 &   80.76 & \phm{$>$}25.405 &    0.337 & \phm{$->$}2.067   &    0.338 &    1.098 &    0.034 \\ 
  2007 &   57.63 &     63.33 &   85.63 & \phm{$>$}24.908 &    0.213 & \phm{$->$}0.618   &    0.219 &    0.798 &    0.097 \\ 
  2008 &   58.52 &     38.44 &   70.01 & \phm{$>$}24.303 &    0.178 & \phm{$->$}2.453   &    0.178 &    1.107 &    0.012 \\ 
  2009 &   59.68 &     53.04 &   79.84 & \phm{$>$}24.278 &    0.171 & \phm{$->$}2.623   &    0.171 &    1.207 &    0.010 \\ 
  2010 &   60.29 &     42.14 &   73.56 & \phm{$>$}23.521 &    0.108 & \phm{$->$}1.378   &    0.108 &    1.044 &    0.015 \\ 
  2011 &   62.53 &     42.94 &   75.86 & \phm{$>$}23.160 &    0.088 & \phm{$->$}1.274   &    0.088 &    1.062 &    0.013 \\ 
  2012 &   63.68 &     46.28 &   78.72 & \phm{$>$}23.044 &    0.080 & \phm{$->$}1.208   &    0.080 &    1.045 &    0.011 \\ 
  2013 &   63.86 &     43.62 &   77.34 & \phm{$>$}24.434 &    0.216 & \phm{$->$}1.505   &    0.217 &    1.055 &    0.027 \\ 
  2014 &   65.49 &     49.50 &   82.09 & \phm{$>$}23.117 &    0.082 & \phm{$->$}0.928   &    0.083 &    0.986 &    0.017 \\ 
  2101 &   44.97 &     52.04 &   68.78 &       $>26.072$ &  \nodata & \phm{$-$}$>0.587$ &  \nodata &    1.368 &    0.214 \\ 
  2102 &   46.17 &     41.83 &   62.30 &       $>26.224$ &  \nodata & \phm{$-$}$>1.599$ &  \nodata &    1.200 &    0.104 \\ 
  2103 &   46.46 &     54.90 &   71.92 &       $>25.987$ &  \nodata & \phm{$-$}$>3.073$ &  \nodata &    1.165 &    0.025 \\ 
  2104 &   48.05 &     43.02 &   64.49 &       $>25.602$ &  \nodata & \phm{$-$}$>1.817$ &  \nodata &    0.925 &    0.065 \\ 
  2105 &   49.29 &     45.24 &   66.90 &       $>26.150$ &  \nodata & \phm{$-$}$>1.069$ &  \nodata &    1.160 &    0.170 \\ 
  2106 &   49.45 &     37.71 &   62.19 &       $>25.931$ &  \nodata & \phm{$-$}$>1.070$ &  \nodata &    1.076 &    0.133 \\ 
  2107 &   49.67 &     41.46 &   64.69 &       $>25.444$ &  \nodata & \phm{$-$}$>0.373$ &  \nodata &    1.379 &    0.143 \\ 
  2108 &   49.84 &     37.69 &   62.49 &       $>26.224$ &  \nodata & \phm{$-$}$>0.966$ &  \nodata &    1.008 &    0.190 \\ 
  2109 &   49.93 &     41.95 &   65.22 &       $>25.459$ &  \nodata & $>-0.334$         &  \nodata &    1.087 &    0.354 \\ 
  2110 &   51.23 &     39.52 &   64.71 &       $>25.901$ &  \nodata & \phm{$-$}$>0.763$ &  \nodata &    1.457 &    0.150 \\ 
  2111 &   52.53 &     51.91 &   73.86 &       $>26.042$ &  \nodata & \phm{$-$}$>2.568$ &  \nodata &    1.172 &    0.040 \\ 
  2112 &   53.40 &     36.46 &   64.66 &       $>26.150$ &  \nodata & \phm{$-$}$>1.869$ &  \nodata &    1.355 &    0.065 \\ 
  2113 &   54.04 &     56.46 &   78.15 &       $>26.064$ &  \nodata & \phm{$-$}$>2.967$ &  \nodata &    1.329 &    0.027 \\ 
  2114 &   54.39 &     52.36 &   75.50 &       $>25.658$ &  \nodata & \phm{$-$}$>2.352$ &  \nodata &    1.348 &    0.034 \\ 
  2115 &   55.39 &     58.38 &   80.48 &       $>26.064$ &  \nodata & \phm{$-$}$>1.524$ &  \nodata &    1.205 &    0.093 \\ 
  2116 &   55.55 &     36.30 &   66.36 &       $>25.613$ &  \nodata & \phm{$-$}$>0.087$ &  \nodata &    0.765 &    0.296 \\ 
  2117 &   56.66 &     54.46 &   78.59 &       $>25.920$ &  \nodata & \phm{$-$}$>2.906$ &  \nodata &    0.817 &    0.031 \\ 
  2118 &   56.93 &     43.15 &   71.43 &       $>25.655$ &  \nodata & \phm{$-$}$>0.937$ &  \nodata &    1.420 &    0.102 \\ 
  2119 &   56.96 &     62.77 &   84.77 &       $>26.189$ &  \nodata & \phm{$-$}$>2.965$ &  \nodata &    1.067 &    0.031 \\ 
  2120 &   57.96 &     41.65 &   71.37 &       $>25.691$ &  \nodata & \phm{$-$}$>1.573$ &  \nodata &    1.338 &    0.062 \\ 
  2121 &   58.08 &     62.43 &   85.27 &       $>25.800$ &  \nodata & $>-0.329$         &  \nodata &    0.937 &    0.407 \\ 
  2122 &   60.14 &     58.64 &   84.00 &       $>26.077$ &  \nodata & $>-0.450$         &  \nodata &    1.655 &    0.437 \\ 
  2123 &   60.35 &     37.09 &   70.84 &       $>25.979$ &  \nodata & \phm{$-$}$>0.971$ &  \nodata &    1.221 &    0.138 \\ 
  2124 &   64.90 &     49.68 &   81.73 &       $>25.171$ &  \nodata & \phm{$-$}$>2.596$ &  \nodata &    1.366 &    0.019 \\ 
  2125 &   66.74 &     41.56 &   78.63 &       $>25.766$ &  \nodata & \phm{$-$}$>0.546$ &  \nodata &    0.879 &    0.193 \\ 
  2201 &   45.12 &     52.96 &   69.58 & \phm{$>$}25.578 &    0.347 & $<-0.168$         &  \nodata &  \nodata &  \nodata \\ 
  2202 &   53.09 &     44.46 &   69.25 & \phm{$>$}24.558 &    0.201 & $<-1.414$         &  \nodata &  \nodata &  \nodata \\ 
  2203 &   55.06 &     44.59 &   70.85 & \phm{$>$}24.968 &    0.275 & $<-1.090$         &  \nodata &  \nodata &  \nodata \\ 
  2204 &   67.12 &     46.96 &   81.92 & \phm{$>$}23.125 &    0.082 & $<-3.327$         &  \nodata &  \nodata &  \nodata \\ 
\\
\hline
\\
  3001 &    6.85 &     16.78 &   18.13 & \phm{$>$}22.765 &    0.077 & \phm{$->$}1.407   &    0.080 &    1.054 &    0.033 \\ 
  3002 &   11.66 &     17.01 &   20.62 & \phm{$>$}24.756 &    0.309 & \phm{$->$}1.376   &    0.324 &    1.064 &    0.161 \\ 
  3003 &   12.06 &     21.49 &   24.65 & \phm{$>$}23.099 &    0.087 & \phm{$->$}1.192   &    0.089 &    1.000 &    0.035 \\ 
  3004 &   13.36 &     14.08 &   19.41 & \phm{$>$}24.297 &    0.214 & \phm{$->$}1.525   &    0.223 &    1.248 &    0.094 \\ 
  3005 &   13.81 &     16.00 &   21.13 & \phm{$>$}23.125 &    0.093 & \phm{$->$}0.858   &    0.099 &    1.056 &    0.052 \\ 
  3006 &   15.16 &     10.83 &   18.63 & \phm{$>$}23.519 &    0.132 & \phm{$->$}2.569   &    0.133 &    1.257 &    0.019 \\ 
  3007 &   17.44 &      5.25 &   18.21 & \phm{$>$}23.258 &    0.108 & \phm{$->$}1.612   &    0.111 &    1.191 &    0.042 \\ 
  3008 &   19.22 &      8.68 &   21.09 & \phm{$>$}24.210 &    0.211 & \phm{$->$}1.316   &    0.219 &    0.986 &    0.093 \\ 
  3009 &   19.35 &     24.44 &   31.17 & \phm{$>$}24.579 &    0.217 & \phm{$->$}0.827   &    0.228 &    0.833 &    0.119 \\ 
  3010 &   19.58 &     13.69 &   23.89 & \phm{$>$}23.912 &    0.155 & \phm{$->$}0.629   &    0.174 &    0.941 &    0.127 \\ 
  3011 &   21.28 &     26.49 &   33.98 & \phm{$>$}23.752 &    0.122 & \phm{$->$}0.964   &    0.125 &    0.985 &    0.041 \\ 
  3012 &   21.93 &     15.66 &   26.95 & \phm{$>$}23.743 &    0.129 & \phm{$->$}0.591   &    0.138 &    0.947 &    0.083 \\ 
  3013 &   22.48 &      8.25 &   23.94 & \phm{$>$}23.634 &    0.127 & \phm{$->$}1.489   &    0.130 &    1.037 &    0.046 \\ 
  3014 &   22.63 &     22.07 &   31.61 & \phm{$>$}23.169 &    0.086 & \phm{$->$}1.763   &    0.086 &    1.072 &    0.013 \\ 
  3015 &   22.76 &      4.98 &   23.30 & \phm{$>$}24.025 &    0.170 & \phm{$->$}2.869   &    0.170 &    1.104 &    0.019 \\ 
  3016 &   26.85 &     13.34 &   29.98 & \phm{$>$}22.612 &    0.065 & \phm{$->$}1.119   &    0.066 &    1.044 &    0.017 \\ 
  3017 &   27.24 &     20.19 &   33.91 & \phm{$>$}25.041 &    0.290 & \phm{$->$}1.308   &    0.295 &    0.945 &    0.091 \\ 
  3018 &   28.12 &     16.76 &   32.73 & \phm{$>$}24.268 &    0.175 & \phm{$->$}2.635   &    0.175 &    1.140 &    0.016 \\
  3101 &    4.02 &     18.15 &   18.59 &       $>24.793$ &  \nodata & \phm{$-$}$>1.177$ &  \nodata &    1.232 &    0.223 \\ 
  3102 &    7.04 &     11.73 &   13.68 &       $>25.046$ &  \nodata & \phm{$-$}$>1.686$ &  \nodata &    1.435 &    0.289 \\ 
  3103 &    8.06 &     19.93 &   21.50 &       $>24.727$ &  \nodata & \phm{$-$}$>1.840$ &  \nodata &    1.332 &    0.089 \\ 
  3104 &    9.35 &     18.19 &   20.45 &       $>24.922$ &  \nodata & \phm{$-$}$>1.620$ &  \nodata &    1.157 &    0.134 \\ 
  3105 &    9.91 &     17.21 &   19.86 &       $>25.445$ &  \nodata & \phm{$-$}$>1.828$ &  \nodata &    1.065 &    0.200 \\ 
  3106 &   11.68 &      8.12 &   14.22 &       $>24.986$ &  \nodata & \phm{$-$}$>2.125$ &  \nodata &    1.020 &    0.198 \\ 
  3107 &   14.05 &     23.59 &   27.46 &       $>25.740$ &  \nodata & \phm{$-$}$>0.812$ &  \nodata &    1.314 &    0.299 \\ 
  3108 &   14.11 &     20.36 &   24.77 &       $>25.602$ &  \nodata & \phm{$-$}$>1.631$ &  \nodata &    0.995 &    0.203 \\ 
  3109 &   15.38 &      5.77 &   16.43 &       $>24.714$ &  \nodata & \phm{$-$}$>2.434$ &  \nodata &    1.044 &    0.088 \\ 
  3110 &   15.86 &     19.23 &   24.92 &       $>25.161$ &  \nodata & \phm{$-$}$>1.133$ &  \nodata &    1.293 &    0.178 \\ 
  3111 &   18.06 &      3.84 &   18.47 &       $>24.801$ &  \nodata & \phm{$-$}$>1.776$ &  \nodata &    1.053 &    0.148 \\ 
  3112 &   18.91 &      6.18 &   19.90 &       $>24.599$ &  \nodata & \phm{$-$}$>1.221$ &  \nodata &    1.291 &    0.137 \\ 
  3113 &   19.08 &     20.78 &   28.21 &       $>25.392$ &  \nodata & \phm{$-$}$>0.884$ &  \nodata &    1.240 &    0.210 \\ 
  3114 &   19.17 &     13.22 &   23.29 &       $>25.024$ &  \nodata & \phm{$-$}$>2.064$ &  \nodata &    1.099 &    0.088 \\ 
  3115 &   20.34 &     11.55 &   23.39 &       $>25.498$ &  \nodata & \phm{$-$}$>2.795$ &  \nodata &    1.321 &    0.059 \\ 
  3116 &   20.72 &     15.91 &   26.13 &       $>25.643$ &  \nodata & \phm{$-$}$>1.274$ &  \nodata &    1.574 &    0.230 \\ 
  3117 &   21.77 &     17.85 &   28.15 &       $>25.653$ &  \nodata & \phm{$-$}$>3.119$ &  \nodata &    1.329 &    0.038 \\ 
  3118 &   22.10 &      5.69 &   22.82 &       $>25.563$ &  \nodata & \phm{$-$}$>1.616$ &  \nodata &    0.999 &    0.229 \\ 
  3119 &   22.53 &      7.63 &   23.79 &       $>25.535$ &  \nodata & \phm{$-$}$>1.366$ &  \nodata &    1.478 &    0.208 \\ 
  3120 &   24.37 &     20.00 &   31.53 &       $>25.763$ &  \nodata & \phm{$-$}$>1.807$ &  \nodata &    1.305 &    0.124 \\ 
  3121 &   25.02 &     21.20 &   32.79 &       $>25.806$ &  \nodata & \phm{$-$}$>2.512$ &  \nodata &    1.242 &    0.065 \\ 
  3122 &   25.96 &     14.92 &   29.94 &       $>25.137$ &  \nodata & \phm{$-$}$>1.573$ &  \nodata &    1.249 &    0.088 \\ 
  3123 &   26.12 &     15.91 &   30.58 &       $>25.798$ &  \nodata & \phm{$-$}$>2.214$ &  \nodata &    1.268 &    0.097 \\ 
  3124 &   26.41 &     23.34 &   35.25 &       $>25.790$ &  \nodata & \phm{$-$}$>1.621$ &  \nodata &    1.230 &    0.137 \\ 
  3201 &    8.69 &     20.86 &   22.60 & \phm{$>$}23.560 &    0.124 & $<-0.727$         &  \nodata &  \nodata &  \nodata \\ 
  3202 &   18.15 &     23.79 &   29.93 & \phm{$>$}24.307 &    0.182 & \phm{$-$}$<0.081$ &  \nodata &  \nodata &  \nodata \\ 
  3203 &   21.28 &     19.63 &   28.95 & \phm{$>$}24.754 &    0.261 & \phm{$-$}$<0.072$ &  \nodata &  \nodata &  \nodata \\ 
  3204 &   22.76 &     11.15 &   25.34 & \phm{$>$}24.872 &    0.306 & $<-0.127$         &  \nodata &  \nodata &  \nodata \\ 
  3205 &   23.89 &      8.30 &   25.29 & \phm{$>$}24.645 &    0.262 & $<-0.229$         &  \nodata &  \nodata &  \nodata \\ 
  3206 &   25.09 &     16.86 &   30.22 & \phm{$>$}24.984 &    0.299 & $<-0.119$         &  \nodata &  \nodata &  \nodata
\enddata


\tablenotetext{a}{ID: The identification number. Each object has a 4
digit ID number.  For fields 1, 2, and 3, the first digit is the field
number, the second is a flag concerning whether an object meets the
$3\sigma$ detection criteria {\it both in optical and far-UV} (0),
{\it only in optical} (1), or {\it only in far-UV} (2); and the last
two digits are serial numbers for each detection flag type sorted in
order of ascending $\Delta$ RA.  For Field 4, we use a prefix of
``NUV-'' followed by two digits numbers for the clusters detected in
the near-UV but not in the optical.}

\tablenotetext{b}{$\Delta$RA, $\Delta$Dec: RA and Dec offset in
arcseconds from the center of M87 at RA = $12^{\rm h}30^{\rm m}49\fs
42$ and Dec = +12\arcdeg 23\arcmin 28\farcs 0 (ICRS; Ma et al. 1998).
Note that K99 used different coordinates for the center of M87.}

\tablenotetext{c}{$r$: Angular radial distance in arcseconds from the center of M87.}

\tablenotetext{d}{$m_{\scriptscriptstyle FUV}$,
$\sigma_{\scriptscriptstyle FUV}$: Far-UV magnitudes in the
STMAG system and photometric errors. Errors include random photon
noises, scatter in sky pixel values, and uncertainties in mean sky
measurements.  For optical-only sources in Fields 1, 2, and 3, we
determine upper limits to the far-UV magnitudes using the fluxes plus
$3 \sigma$ when the measured fluxes are positive and $3 \sigma$ when
the measured fluxes are negative.}

\tablenotetext{e}{$(FUV-V)_{0}$, $(V-I)_{0}$:
Dereddened colors and errors.  We used the adopted extinction to correct for
reddening. $V$ and $I$ magnitudes and errors are from a complete
catalog of K99. For UV-only sources, we determine upper limits to the
$V$ magnitudes using the fluxes plus $3 \sigma$ when the measured
fluxes are positive and $3 \sigma$ when the measured fluxes are
negative.}

\tablenotetext{e}{Data taken from K99.}

\end{deluxetable}


\begin{deluxetable}{crrcccccccccccc}
\setlength{\tabcolsep}{0.025in}
\rotate
\tabletypesize{\footnotesize}
\tablewidth{0pt}
\tablecaption{ Photometric Catalog for Field 4\tablenotemark{a}\label{t:catalog2}}
\tablehead { \colhead{}                    & \colhead{$\Delta$ RA} & \colhead{$\Delta$ Dec} & \colhead{$r$} & \colhead{}              & \colhead{}                   & \colhead{}              & \colhead{}                   & \colhead{}                                       & \colhead{}                         & \colhead{}                     & \colhead{}                         & \colhead{}                             & \colhead{}               & \colhead{}  \\
             \colhead{ID} & \colhead{(arcsec)}                     & \colhead{(arcsec)}                      & \colhead{(arcsec)}             & \colhead{$m_{\scriptscriptstyle FUV}$} & \colhead{$\sigma_{\scriptscriptstyle FUV}$} & \colhead{$m_{\scriptscriptstyle NUV}$\tablenotemark{b}} & \colhead{$\sigma_{\scriptscriptstyle NUV}$} & \colhead{$(FUV-V)_{0}$\tablenotemark{d}} & \colhead{$\sigma_{FUV-V}$} & \colhead{$(NUV-V)_{0}$\tablenotemark{c}} & \colhead{$\sigma_{NUV-V}$} & \colhead{$(V-I)_{0}$} & \colhead{$\sigma_{\scriptscriptstyle V-I}$} & \colhead{K99 ID\tablenotemark{d}} 
           }
\startdata
  4001 & $-14.92$&  $-0.91$&   14.95 &  \nodata &  \nodata &   24.531 &    0.099 &  \nodata &  \nodata &    1.238 &    0.106 &    1.028 &    0.065 &   62\\ 
  4002 & $-13.06$&  $-4.93$&   13.96 &  \nodata &  \nodata &   24.356 &    0.076 &  \nodata &  \nodata &    1.145 &    0.086 &    0.947 &    0.066 &   56\\ 
  4003 & $-12.47$&    9.20 &   15.50 &   23.849 &    0.224 &   24.265 &    0.084 &    1.634 &    0.225 &    2.055 &    0.086 &    1.140 &    0.026 &   60\\ 
  4004 & $-10.76$&  $-3.46$&   11.30 &  \nodata &  \nodata &   24.965 &    0.122 &  \nodata &  \nodata &    1.493 &    0.136 &    1.064 &    0.095 &   44\\ 
  4005 & $-10.61$&    8.23 &   13.42 &   23.671 &    0.201 &   24.295 &    0.078 &    0.442 &    0.206 &    1.071 &    0.091 &    0.921 &    0.081 &   50\\ 
  4006 &  $-9.29$&  $-4.52$&   10.33 &  \nodata &  \nodata &   24.938 &    0.121 &  \nodata &  \nodata &    0.832 &    0.169 &    1.345 &    0.154 &   39\\ 
  4007 &  $-6.89$&    7.56 &   10.23 &   23.321 &    0.147 &   23.866 &    0.047 &    1.681 &    0.148 &    2.231 &    0.049 &    1.064 &    0.023 &   26\\ 
  4008 &  $-5.84$&    2.31 &    6.28 &  \nodata &  \nodata &   24.605 &    0.096 &  \nodata &  \nodata &    1.975 &    0.107 &    1.082 &    0.069 &   19\\ 
  4009 &  $-4.37$&  $-6.52$&    7.84 &   22.915 &    0.114 &   23.242 &    0.035 &    1.372 &    0.115 &    1.704 &    0.038 &    1.052 &    0.023 &   24\\ 
  4010 &  $-3.73$&   13.34 &   13.85 &  \nodata &  \nodata &   24.992 &    0.137 &  \nodata &  \nodata &    0.981 &    0.171 &    0.645 &    0.201 &   47\\ 
  4011 &  $-1.86$&    7.82 &    8.04 &   23.792 &    0.275 &   24.049 &    0.069 &    1.273 &    0.277 &    1.535 &    0.078 &    1.087 &    0.061 &   20\\ 
  4012 &  $-1.74$&   11.86 &   11.99 &   23.540 &    0.201 &   23.426 &    0.037 &    1.351 &    0.202 &    1.242 &    0.043 &    0.960 &    0.036 &   30\\ 
  4013 &  $-0.43$&    4.00 &    4.02 &  \nodata &  \nodata &   24.094 &    0.084 &  \nodata &  \nodata &    2.286 &    0.090 &    1.160 &    0.048 &    3\\ 
  4014 &  $-0.13$&   12.75 &   12.76 &  \nodata &  \nodata &   24.193 &    0.133 &  \nodata &  \nodata &    0.917 &    0.143 &    0.919 &    0.095 &   35\\ 
  4015 &    0.96 &  $-2.99$&    3.14 &   23.190 &    0.160 &   23.899 &    0.060 &    1.561 &    0.162 &    2.275 &    0.064 &    1.194 &    0.033 &    8\\ 
  4016 &    3.41 &  $-3.17$&    4.66 &  \nodata &  \nodata &   24.011 &    0.068 &  \nodata &  \nodata &    1.267 &    0.084 &    0.842 &    0.086 &   11
\enddata

\tablenotetext{a}{Explanations for columns with headings as in Table 3 are given
in the notes to Table 3.}
\tablenotetext{b}{$m_{\scriptscriptstyle NUV}$,
$\sigma_{\scriptscriptstyle NUV}$: Near-UV magnitudes in the
STMAG system and photometric errors. Errors include random photon
noises, scatter in sky pixel values, and uncertainties in mean sky
measurements.}
\tablenotetext{c}{$(NUV-V)_{0}$: Dereddened colors and errors.}
\tablenotetext{d}{ID number in Table 2 of \citet{K99}.}
\end{deluxetable}


\begin{deluxetable}{crrrrrrrrrrrrrrcc}
\rotate
\setlength{\tabcolsep}{0.04in}
\tablewidth{0pt}
\tabletypesize{\footnotesize}
\tablecaption{ Photometry of NUV-Only Sources in Field 4\label{t:catalog3} }
\tablehead{ \colhead{}  & \colhead{$\Delta$ RA} & \colhead{$\Delta$ Dec}  & \multicolumn{2}{c}{MJD 51315.56} & & \multicolumn{2}{c}{MJD 52120.97} & & \multicolumn{2}{c}{MJD 52332.20} & & \multicolumn{2}{c}{MJD 52472.37} & & \multicolumn{2}{c}{Mean Mag}\\
            \cline{4-5} \cline{7-8}  \cline{10-11}  \cline{13-14} \cline{16-17}
            \colhead{ID}& \colhead{(arcsec)} & \colhead{(arcsec)} & \colhead{$m_{NUV}$} & \colhead{$\sigma_{NUV}$} & & \colhead{$m_{NUV}$} & \colhead{$\sigma_{NUV}$} & & \colhead{$m_{NUV}$} & \colhead{$\sigma_{NUV}$} & & \colhead{$m_{NUV}$} & \colhead{$\sigma_{NUV}$} & & \colhead{$m_{NUV}$} & \colhead{$\sigma_{NUV}$}
            }
\startdata
NUV-01 &   $4.68$ &  $0.34$ & \phm{$>$}25.063  &    0.351 & & \phm{$>$}24.581  &    0.227 & & \phm{$>$}24.590  &    0.186 & & \phm{$>$}24.527  &    0.215  & &  24.568  &    0.120  \\
NUV-02 &   $0.39$ &  $5.61$ &        $>25.238$ &  \nodata & &        $>25.203$ &  \nodata & & \phm{$>$}23.763  &    0.117 & &        $>24.664$ &  \nodata  & &  23.763  &    0.117  \\
NUV-03 &  $-2.34$ &  $7.51$ &        $>24.638$ &  \nodata & &        $>24.498$ &  \nodata & & \phm{$>$}24.757  &    0.240 & & \phm{$>$}24.875  &    0.293  & &  24.804  &    0.186  \\
NUV-04 &  $-3.67$ &  $7.84$ & \phm{$>$}23.799  &    0.122 & & \phm{$>$}23.958  &    0.146 & & \phm{$>$}24.153  &    0.139 & & \phm{$>$}23.926  &    0.130  & &  23.947  &    0.067  \\
NUV-05 & $-13.30$ & $11.00$ & \phm{$>$}25.288  &    0.356 & & \phm{$>$}25.053  &    0.265 & & \phm{$>$}24.874  &    0.220 & &        $>24.877$ &  \nodata  & &  25.010  &    0.153  \\
NUV-06 & $-12.55$ & $-7.42$ & \phm{$>$}25.008  &    0.226 & & \phm{$>$}24.775  &    0.226 & & \phm{$>$}25.475  &    0.293 & & \phm{$>$}25.129  &    0.229  & &  25.054  &    0.120  \\
NUV-07 & $-24.12$ &  $6.01$ &         \nodata  &  \nodata & &        $>25.015$ &  \nodata & & \phm{$>$}24.399  &    0.114 & & \phm{$>$}25.346  &    0.266  & &  24.546  &    0.105  \\
NUV-08 & $-19.68$ &  $2.26$ & \phm{$>$}24.175  &    0.132 & &        $>25.040$ &  \nodata & &        $>25.189$ &  \nodata & &        $>24.870$ &  \nodata  & &  24.205  &    0.132  \\
NUV-09 & $-12.38$ & $-8.87$ & \phm{$>$}25.059  &    0.230 & &        $>25.525$ &  \nodata & &        $>25.851$ &  \nodata & &        $>25.898$ &  \nodata  & &  25.059  &    0.230  \\
NUV-10 &  $-5.89$ & $-8.48$ & \phm{$>$}25.022  &    0.293 & & \phm{$>$}24.963  &    0.279 & &        $>25.116$ &  \nodata & & \phm{$>$}25.182  &    0.307  & &  25.049  &    0.169  \\
NUV-11 &  $-9.45$ &  $0.12$ & \phm{$>$}24.491  &    0.168 & &        $>24.548$ &  \nodata & &        $>24.656$ &  \nodata & &        $>25.462$ &  \nodata  & &  24.491  &    0.168  \\
NUV-12 &   $1.66$ & $-3.12$ & \phm{$>$}24.165  &    0.136 & &        $>24.959$ &  \nodata & &        $>24.438$ &  \nodata & &        $>24.652$ &  \nodata  & &  24.165  &    0.136  \\
NUV-13 &   $4.45$ &  $1.19$ &        $>25.091$ &  \nodata & & \phm{$>$}24.543  &    0.240 & &        $>25.415$ &  \nodata & &        $>24.975$ &  \nodata  & &  24.543  &    0.240  \\
NUV-14 & $-19.32$ &  $0.67$ &        $>25.599$ &  \nodata & &        $>24.748$ &  \nodata & &        $>24.742$ &  \nodata & & \phm{$>$}24.861  &    0.198  & &  24.861  &    0.198  \\
NUV-15 & $-16.33$ & $-0.16$ &        $>24.950$ &  \nodata & &        $>24.526$ &  \nodata & &        $>25.146$ &  \nodata & & \phm{$>$}24.871  &    0.234  & &  24.871  &    0.234  \\
NUV-16 & $-12.28$ &  $4.16$ & \phm{$>$}25.338  &    0.350 & &        \nodata   &  \nodata & &         \nodata  &  \nodata & & \phm{$>$}23.885  &    0.125  & &  23.885  &    0.125  \\
NUV-17 &  $-7.40$ & $-0.26$ &        $>25.212$ &  \nodata & &        $>25.340$ &  \nodata & &        $>25.455$ &  \nodata & & \phm{$>$}24.736  &    0.207  & &  24.736  &    0.207  \\
NUV-18 &  $-5.79$ &  $0.16$ &        $>25.350$ &  \nodata & &        $>25.082$ &  \nodata & &        $>25.497$ &  \nodata & & \phm{$>$}24.772  &    0.190  & &  24.772  &    0.190  \\
NUV-19 &  $-7.24$ &  $8.96$ &        $>25.591$ &  \nodata & &        $>25.403$ &  \nodata & &        $>25.146$ &  \nodata & & \phm{$>$}24.700  &    0.174  & &  24.700  &    0.174  \\
NUV-20 &   $2.89$ & $-3.70$ &        $>25.095$ &  \nodata & &        $>25.442$ &  \nodata & &         \nodata  &  \nodata & & \phm{$>$}25.022  &    0.250  & &  25.022  &    0.250  \\
NUV-21 &   $2.70$ & $-1.80$ & \phm{$>$}25.026  &    0.307 & &        $>24.231$ &  \nodata & &        $>24.315$ &  \nodata & & \phm{$>$}24.894  &    0.268  & &  24.951  &    0.202  \\
NUV-22 &   $5.89$ & $-1.95$ &        $>25.420$ &  \nodata & &        $>25.394$ &  \nodata & &         \nodata  &  \nodata & & \phm{$>$}24.452  &    0.164  & &  24.452  &    0.164
\enddata
\end{deluxetable}

\begin{deluxetable}{lcc}
\rotate
\tablewidth{0pt}
\tablecaption{ Various calibrations of [Fe/H] against $(V-I)_{0}$ colors.\tablenotemark{1} \label{t:vifeh_relation}}
\tablehead { \colhead{Reference} & \colhead{$a$} & \colhead{$b$} 
           }
\startdata
This study                            & $-5.46$  & 4.26  \\
\tableline
Couture, Harris, \& Allwright (1990) & $-6.10$  & 5.05  \\
Kundu \& Whitmore (1998)              & $-5.89$  & 4.72  \\
Kissler-Patig et al. (1998)           & $-4.50$  & 3.27  \\
Harris et al. (2000)                  & $-6.76$  & 5.88
\enddata
\tablenotetext{1}{[Fe/H] = $a$ + $b(V-I)_{0}$.}
\end{deluxetable}

\begin{deluxetable}{llcccccccrcccc}
\setlength{\tabcolsep}{0.02in}
\rotate
\tabletypesize{\scriptsize}
\tablewidth{0pt}
\tablecaption{ Ultraviolet Data for Milky Way Globular Clusters \label{t:mwgc_dor95_dat}}
\tablehead { \colhead{NGC} & \colhead{Name} & \colhead{$(FUV-V)_{0}$} & \colhead{$\sigma_{FUV-V}$} & \colhead{$(FUV-NUV)_{0}$} & \colhead{$\sigma_{FUV-NUV}$} & \colhead{$(NUV-V)_{0}$} & \colhead{$\sigma_{NUV-V}$} & \colhead{$(V-I)_{0}$} & \colhead{$M_{V}$} & \colhead{[Fe/H]} & \colhead{Mg$_{2}$} & \colhead{$\sigma_{Mg_{2}}$} & \colhead{$E(B-V)$}
           }
\startdata
 104\phn...... & 47 Tuc       &    4.64 & \nodata & \phm{$-$}1.50 & \nodata &    3.80 &    0.02 &    1.10 &  $-9.42$ &  $-0.76$ &    0.137 &    0.014 &    0.04\phantom{$^{\rm a}$} \\*
 362\phn...... &              & \nodata & \nodata &       \nodata & \nodata &    1.89 &    0.08 &    0.95 &  $-8.41$ &  $-1.16$ &    0.088 &    0.014 &    0.05$^{\rm a}$ \\*
 1261......    &              & \nodata & \nodata &       \nodata & \nodata &    1.89 &    0.08 &    0.91 &  $-7.81$ &  $-1.35$ &    0.069 &    0.014 &    0.01$^{\rm a}$ \\*
 1851......    &              &    2.29 & \nodata & \phm{$-$}0.17 & \nodata &    2.11 &    0.08 &    0.99 &  $-8.33$ &  $-1.22$ &    0.082 &    0.014 &    0.02$^{\rm a}$ \\*
 1904......    & M79          &    1.70 & \nodata & \phm{$-$}0.28 & \nodata &    1.42 & \nodata &    0.89 &  $-7.86$ &  $-1.57$ &    0.051 &    0.014 &    0.01$^{\rm a}$ \\*
 2298......    &              &    2.36 &    0.07 & \phm{$-$}0.51 &    0.06 &    1.86 &    0.09 &    0.94 &  $-6.30$ &  $-1.85$ &    0.034 &    0.014 &    0.14$^{\rm a}$ \\*
 5024......    & M53          &    3.12 &    0.07 & \phm{$-$}0.93 &    0.06 &    2.19 &    0.09 &    0.84 &  $-8.70$ &  $-1.99$ &    0.039 &    0.010 &    0.02$^{\rm a}$ \\*
 5139......    & $\omega$ Cen &    1.44 &    0.06 &       $-0.30$ &    0.10 &    1.74 &    0.08 &    0.91 & $-10.29$ &  $-1.62$ &    0.048 &    0.014 &    0.12$^{\rm a}$ \\*
 5272......    & M3           &    3.41 &    0.04 & \phm{$-$}1.39 &    0.02 &    2.01 &    0.05 &    0.91 &  $-8.93$ &  $-1.57$ &    0.040 &    0.008 &    0.01\phantom{$^{\rm a}$} \\*
 5897......    &              & \nodata & \nodata &       \nodata & \nodata &    1.77 &    0.08 &    0.93 &  $-7.21$ &  $-1.80$ &    0.037 &    0.014 &    0.09$^{\rm a}$ \\*
 5904......    & M5           &    1.96 & \nodata &       $-0.05$ & \nodata &    2.01 &    0.08 &    0.91 &  $-8.81$ &  $-1.27$ &    0.067 &    0.010 &    0.03\phantom{$^{\rm a}$} \\*
 6093......    & M80          &    2.18 &    0.22 &       $-0.10$ &    0.18 &    2.28 &    0.29 &    0.89 &  $-8.23$ &  $-1.75$ &    0.040 &    0.014 &    0.18$^{\rm a}$ \\*
 6121......    & M4           & \nodata & \nodata &       \nodata & \nodata &    2.89 &    0.08 &    0.98 &  $-7.20$ &  $-1.20$ &    0.084 &    0.014 &    0.36$^{\rm a}$ \\*
 6205......    & M13          &    1.63 &    0.00 & \phm{$-$}0.03 &    0.00 &    1.59 &    0.00 &    0.84 &  $-8.70$ &  $-1.54$ &    0.039 &    0.005 &    0.02\phantom{$^{\rm a}$} \\*
 6266......    & M62          &    2.30 &    0.07 &       $-0.29$ &    0.07 &    2.59 &    0.10 &    1.01 &  $-9.19$ &  $-1.29$ &    0.075 &    0.014 &    0.47$^{\rm a}$ \\*
 6341......    & M92          &    2.07 &    0.04 & \phm{$-$}0.63 &    0.02 &    1.44 &    0.05 &    0.86 &  $-8.20$ &  $-2.28$ &    0.021 &    0.006 &    0.02\phantom{$^{\rm a}$} \\*
 6356......    &              & \nodata & \nodata &       \nodata & \nodata &    4.00 &    0.12 &    1.11 &  $-8.52$ &  $-0.50$ &    0.169 &    0.008 &    0.28$^{\rm a}$ \\*
 6388......    &              &    3.27 &    0.22 & \phm{$-$}0.29 &    0.16 &    2.98 &    0.27 &    1.02 &  $-9.42$ &  $-0.60$ &    0.146 &    0.006 &    0.37$^{\rm a}$ \\*
 6402......    & M14          & \nodata & \nodata &       \nodata & \nodata &    3.06 &    0.08 &    0.91 &  $-9.12$ &  $-1.39$ &    0.066 &    0.014 &    0.60$^{\rm a}$ \\*
 6441......    &              &    2.20 &    0.26 & \phm{$-$}0.05 &    0.18 &    2.14 &    0.32 &    1.05 &  $-9.64$ &  $-0.53$ &    0.176 &    0.006 &    0.47$^{\rm a}$ \\*
 6541......    &              &    2.06 &    0.07 & \phm{$-$}0.56 &    0.05 &    1.51 &    0.09 &    0.84 &  $-8.37$ &  $-1.83$ &    0.035 &    0.014 &    0.14$^{\rm a}$ \\*
 6624......    &              & \nodata & \nodata &       \nodata & \nodata &    3.78 &    0.05 &    1.08 &  $-7.49$ &  $-0.44$ &    0.156 &    0.008 &    0.28$^{\rm a}$ \\*
 6626......    & M28          &    1.92 &    0.41 & \phm{$-$}0.12 &    0.29 &    1.81 &    0.50 &    0.92 &  $-8.18$ &  $-1.45$ &    0.092 &    0.014 &    0.40$^{\rm a}$ \\*
 6637......    &              & \nodata & \nodata &       \nodata & \nodata &    3.59 &    0.05 &    1.08 &  $-7.64$ &  $-0.70$ &    0.151 &    0.005 &    0.16$^{\rm a}$ \\*
 6681......    & M70          &    1.74 &    0.04 & \phm{$-$}0.04 &    0.04 &    1.70 &    0.05 &    0.91 &  $-7.11$ &  $-1.51$ &    0.056 &    0.014 &    0.07\phantom{$^{\rm a}$} \\*
 6715......    & M54          &    2.37 &    0.04 & \phm{$-$}0.27 &    0.04 &    2.10 &    0.05 &    0.92 & $-10.01$ &  $-1.58$ &    0.051 &    0.014 &    0.15$^{\rm a}$ \\*
 6752......    &              &    1.34 &    0.04 &       $-0.18$ &    0.02 &    1.52 &    0.05 &    0.89 &  $-7.73$ &  $-1.56$ &    0.052 &    0.014 &    0.04\phantom{$^{\rm a}$} \\*
 6779......    & M56          & \nodata & \nodata &       \nodata & \nodata &    2.08 &    0.05 &    0.92 &  $-7.38$ &  $-1.94$ &    0.030 &    0.014 &    0.20$^{\rm a}$ \\*
 6864......    & M75          &    2.64 &    0.22 & \phm{$-$}0.42 &    0.16 &    1.60 &    0.08 &    0.96 &  $-8.55$ &  $-1.16$ &    0.088 &    0.014 &    0.16$^{\rm a}$ \\*
 7078......    & M15          &    2.36 &    0.07 & \phm{$-$}0.67 &    0.05 &    1.69 &    0.09 &    0.73 &  $-9.17$ &  $-2.26$ &    0.023 &    0.007 &    0.10\phantom{$^{\rm a}$} \\*
 7099......    & M30          &    2.71 &    0.04 & \phm{$-$}0.83 &    0.04 &    1.88 &    0.05 &    0.82 &  $-7.43$ &  $-2.12$ &    0.024 &    0.014 &    0.03$^{\rm a}$  
\enddata
\tablenotetext{a}{Clusters with new reddenings.}
\end{deluxetable}

\end{document}